\documentclass[12pt,preprint]{aastex}
\usepackage[english]{babel} 
\usepackage {graphics}
\usepackage{latexsym}
\usepackage {graphicx}
\usepackage {amsmath} 
\usepackage {rotating} 
\usepackage{natbib}
\usepackage{amssymb}
\usepackage[a4paper]{geometry}
\usepackage{longtable}
\usepackage{xcolor}
\usepackage{epstopdf}
\usepackage{multirow}

\def\cm3{cm$^{-3}$\/}
\def\hbbc{{\sc{H}}$\beta_{\rm BC}$\/}
\def\hbnc{{\sc{H}}$\beta_{\rm NC}$\/}
\def\mgii{{Mg \sc{ii}} $\lambda$2800\/}
\def\fe{{\sc{Fe}}\/}

\def\kms{km~s$^{-1}$}

\def\lya{{ Ly}$\alpha$}
\def\nc{$N_{\rm c}$\/}
\def\lledd{$L/L_{\rm Edd}$}
\def\nh{$n_{\mathrm{H}}$\/}
\def\hb{H$\beta$}
\def\ca2{Ca {\sc ii}}
\def\fe2{Fe {\sc ii}}
\def\oi{O {\sc i}}
\def\oil{O {\sc i} $\lambda$8446\/}
\def\o3{{\sc{[O iii]}} $\lambda\lambda$4959, 5007\/}
\def\s3{{\sc{[S iii]}}}
\def\pa9{Pa9 $\lambda$9229\/}
\def\rfe{$R_{\rm FeII}$}
\def\feiiq{Fe {\sc ii}$\lambda$4570\/}
\shorttitle{\caii }
\shortauthors{Mart\'{i}nez--Aldama et al.}

\begin{document}
\title{\oi\ and \ca2\ observations in intermediate redshift quasars\altaffilmark{1}}
\author{Mary Loli Mart\'{i}nez--Aldama\altaffilmark{2}, Deborah Dultzin\altaffilmark{2}, Paola Marziani\altaffilmark{3}, Jack W. Sulentic\altaffilmark{4}, Alessandro Bressan\altaffilmark{5}, Yang Chen\altaffilmark{5} and Giovanna M. Stirpe\altaffilmark{6}}
\altaffiltext{1}{Based on observations collected at the European Organisation for Astronomical Research in the Southern Hemisphere, Chile, under programme ID 085.B--0158(A).}
\altaffiltext{2}{Instituto de Astronom\'{\i}a, Universidad Nacional Aut\'{o}noma de M\'{e}xico, M\'{e}xico}
\altaffiltext{3}{INAF, Osservatorio Astronomico di Padova, Italia}
\altaffiltext{4}{IAA-CSIC, Granada, Espa\~na}
\altaffiltext{5}{Scuola Internazionale Superiore di Studi Avanzati (SISSA), Trieste, Italy}
\altaffiltext{6}{INAF, Osservatorio Astronomico di Bologna, Italy}

\begin{abstract}
We present an unprecedented spectroscopic survey of the \ca2\ triplet + \oi\  for a sample of 14 luminous ($ -26 \ga M_V   \ga -29$), intermediate redshift (0.85 $\la z \la$ 1.65) quasars. The ISAAC spectrometer at ESO VLT allowed us to cover the \ca2\ NIR spectral region redshifted into the H and K windows. We describe in detail  our data analysis which enabled us to detect \ca2\ triplet emission in all 14 sources (with the possible exception of HE0048--2804) and  to retrieve accurate line widths and fluxes of the triplet and \oil. The new measurements show trends consistent with previous lower $z$\ observations, indicating that \ca2\ and optical \fe2\ emission are probably closely related. The ratio between the \ca2\ triplet and the optical \fe2\ blend at $\lambda$4570 \AA\ is apparently systematically larger in our intermediate redshift sample relative to a low--$z$\ control sample. Even if this result needs a larger sample for adequate interpretation, higher \ca2/optical \fe2\ should be  associated with 
recent episodes of star formation in the intermediate redshift quasars and, at least  in part, explain an apparent correlation of  \ca2\ triplet equivalent width  with $z$ and $L$.  The  \ca2\ triplet measures yield significant constraints on the emitting region  density and  ionization parameter, implying \ca2\ triplet emission from $\log$ \nh\ $ \ga$ 11 [\cm3] and ionization parameter $\log U \la -1.5$. Line width and intensity ratios suggest properties consistent with emission from the outer part of a high density broad line region (a line emitting accretion disk?).
\end{abstract}


\section{Introduction}

Explaining the origin of Fe emission in quasar spectra is a long--standing problem in studies of active galactic nuclei  (AGN). Recent observations and models fail to reach a consensus on the ionization mechanism even in the case of moderate \fe2\ emitters. Purely collisional models are unable to account for the spectral energy distribution of \fe2\ emitters \citep{KUE08}. At the same time, the ratio UV \fe2\  to optical \fe2\ emission is more easily explained in the context of collisional ionization \citep{SAM11}.  Akn  564 \fe2\ reverberates in response to  continuum changes \citep{SHA12}, providing support for the role of  photoionization. At the same time   Akn 120 \fe2\ does not show any response \citep{KUE08}.  The extreme complexity of the \fe2\ ion makes theoretical model calculations very difficult and line blending makes estimation of \fe2\ width and strength parameters uncertain.

Given the difficult interpretation and prediction of the \fe2\ spectrum,  the study of simpler ionic species like \ca2\ (emitting the IR triplet $\lambda$8498, $\lambda$8542, $\lambda$8662, hereafter CaT) and \oi\ is more straightforward. The \ca2\ ion is much simpler. The ionization potential of neutral Calcium ($\approx$6.1 eV) leads us to expect that \ca2\ ions exist where Hydrogen is not fully ionized. Several lines of evidence suggest that the CaT and optical \fe2\ are produced in the same region. Data from \citet{PER88} and photoionization calculations \citep{JOL89} found that CaT is emitted by gas at low temperature (8000 K), high density ($>$10$^{11}$ cm$^{-3}$) and high column density ($>$10$^{23}$ cm$^{-2}$) similar to optical \fe2. \citet{MAT07, MAT08} computed photoionization models using the \oil\ and $\lambda$11287 lines and the CaT and found that  high density ($\sim$ 10$^{11.5}$ cm$^{-3}$) and low--ionization parameter ($U \sim$ 10$^{-2.5}$) are needed to reproduce flux ratios, consistently 
with the physical conditions expected for optical \fe2\ emission. Density and ionization parameters are in agreement with the ones found by \citet{JOL89}.

\citet{FERPER89} improved the photoionization models including physical processes like H${^0}$ free-free, H${^-}$\ bound-free and Compton recoil ionization. These authors demonstrated the need for very large column densities ($N_\mathrm{c} >  10 ^{24.5}$ cm$^{-2}$) to reproduce the \ca2\ spectrum. Such large column densities could be provided by an accretion disk. A similar behavior to the one described by \citet{JOL89} for CaT/\hb\ and \fe2/\hb\ was found by \citet{DUL99}. These authors also suggested that the line emitting region could be associated with the outer part of an accretion disk. The width measured on the unblended \fe2\ $\lambda$11126 line (that belongs to the so-called 1$\mu$m \fe2\ lines), is very similar to the ones of CaT and \oil\, and narrower than the one of the Hydrogen lines. In a purely kinematical interpretation, this means that low ionization lines like \fe2, CaT and \oil\ are emitted in the outer zone of the broad line region \citep{ROD02, ROD02b, MAT07, MAT08}, probably associated 
 with the accretion disk, with physical conditions different from the region emitting most of the high ionization lines.

Physical modelling (via a photoionization code like {\sc cloudy}, \citealt{FER13}) requires input from measurements that are retrieved  through model fits of the observed spectra. The \ca2\ triplet and \oil\ lines  can be easier to model than \fe2\ but they  are not as strong as  \fe2\ optical features which, however, are heavily blended. They are located in a wavelength range where the putative accretion disk continuum  and the high energy tail of hot dust emission form a sort of valley in the spectral energy distribution. The old stellar population associated with the quasar host galaxy peaks at $\approx 1 \mu m$\ \citep{LAN11}. In many low--luminosity quasars CaT and \oil\ are confused with the underlying absorption of the host galaxy. 

The near-infrared region (NIR) has not been easy to observe due to the lack of high quality IR spectrometers. Only in relatively recent times the development of new instruments has made it possible to obtain high S/N spectra for sources with redshift larger than $z \ga$ 0.1. It is not surprising that  the \ca2\ and \oi\ features remain relatively unstudied even in sources for which a wealth of optical data exist. 

The present work extends the study of the \ca2\ triplet in emission to include 14 high luminosity and intermediate redshift quasars (\S \ref{sample}). Whenever possible they are analyzed within the 4D eigenvector 1 context \citep[4DE1;][]{BOR92, SMD2000a, SUL2000, MAR01, MAR03a, MAR03b, SUL07}, which serves as a spectroscopic unifier/discriminator of the emission lines properties for type 1 AGN. The parameters defining the four dimensions of the E1 space involve: (1) FWHM(\hb$_{BC}$), (2) the ratio between the equivalent width (W) of the optical \fe2\ $\lambda$ 4570 blend and \hb, \rfe = W(\fe2 $\lambda$4570)/W(\hb) $\approx I$(\fe2 $\lambda$4570)/$I$(\hb), (3) the soft X-ray photon index ($\Gamma_\mathrm{soft}$), and (4) the centroid line shift of high--ionization C {\sc iv} $\lambda$1549, with the strongest correlations involving parameters 2, 3 and 4  \citep{SUL07}. In the 4DE1  framework the broad line AGN can be divided into two populations, A and B \citep{SUL02}. Considering the broad component of the \
hb\ line, populations A and B can be separated at FWHM(\hb$_{BC}$) = 4000 km s$^{-1}$. Spectra of sources above and below this limit look substantially different. It is possible to introduce a finer subdivision that will not  be used in this paper because of its  small sample size: we will distinguish sources as Pop. A or B only. Quasars can be divided into bins of $\Delta$FWHM(\hb) and $\Delta$\rfe. The bins of Pop. A (A1, A2, A3, A4), are defined in terms of increasing \rfe, while Pop. B bins (B1, B1$^{+}$, B1$^{++}$) are defined in terms of increasing FWHM(\hb). Population A shows: (1) a scarcity of RL sources, (2) strong/moderate \fe2\ emission, (3) a soft X-ray excess, (4) high--ionization broad lines (HIL) with blueshift/asymmetry and (5) low--ionization broad line profiles (LIL) best described by Lorentz fits. Population B: (1) includes the large majority of  RL sources, (2) shows weak/moderate \fe2\ emission, includes sources (3) with less prominent or no soft X-ray excess \citep{SUL07} and (4) with 
HIL blueshift/asymmetry or  no blue shifts at all. Last, (5) Pop. B shows LIL Balmer lines best fit with double Gaussian models. The physical drivers that change along the 4DE1 sequence have been identified: number density appears to increase from Pop. B to A, and black hole mass increases with large scatter from A to B. The principal driver of source occupation in the 4DE1 space is probably Eddington ratio which increases from B to A \citep{MAR01, BOR02}. The 4DE1 parameter space is, to a first approximation, not strongly dependent on luminosity. The same basic distinction between Pop. A and B is recognizable also for very luminous sources \citep{MAR09}.

This paper presents new observations and data reduction of spectra of the \ca2\ IR triplet, as a representative LIL, as well as of the \oil\ line, in very luminous quasars of the Hamburg-ESO  survey (\S \ref{sample}, \S \ref{obs} and \S \ref{red}). We discuss in some detail the identified line and continuum components  for a proper data analysis (\S \ref{meas}). Basic results obtained from the measurements are briefly reported in \S \ref{res}. In \S\ref{dis} we compare observational data and photoionization predictions, the role of CaT and \oil\ within the 4DE1 context, possible implications for star formation, and a preliminary geometry for the broad line region (BLR). Conclusions are presented in \S \ref{conclusion}.

\section[]{Sample selection}
\label{sample}

Our original sample of Hamburg-ESO  (HE) sources was defined with the aim of testing/extending the 4DE1 correlation at relatively high--$z$ and at high luminosity \citep{MAR09}. The HE original sample of \hb\ ISAAC observations was therefore flux limited and unbiased in terms of \fe2\ strength. Our \ca2\ subsample  was extracted from  the 52 sources ISAAC sample of  \citet{MAR09}. The only  selection criterion was that IR atmospheric absorptions were not affecting the \ca2\ and \oi\ lines. All targets in the present investigation have therefore optical spectra around \hb\ already obtained with the same instrument. 

The \ca2\  + \oi\ sample contains 15 high luminosity Hamburg-ESO (HE) quasars, {with $M_B$ $<$ -26, in the redshift range 0.846 $< z <$ 1.638.} They were observed  at the Very Large Telescope (VLT-UT3) equipped with the Infrared Spectrometer And Array Camera (ISAAC) that yielded a spectral resolution $R_\mathrm{S} \approx$1000. Sample characteristics and the log of the observations are reported in the Table \ref{table:obs}. The first column lists the coordinate name from the Hamburg ESO survey for each object of our sample. In  columns 2, 3, 4 and 5 we report redshift, apparent and absolute magnitude, and Kellerman's  for radio--loudness parameter, as previously given by \citet{SUL04} and \citet{MAR09}. Column 6 lists the population in the Eigenvector 1 sequence \citep{SUL02}.  One of the sources (HE2259--5524) was excluded from the analysis because of poor S/N ratio.

The median absolute magnitude of the selected 14 HE sources is $\approx$ --28, which makes them among the most luminous quasars known. The sample presented in this paper is small; however, \ca2\ and \oi\ observations of quasars have been until now obtained for only 66 sources \citep{PER88,MAT05,MAT07,MAT08,LAN08}, of which only one at $z >$ 1. Therefore the present sample represents the first systematic study of the \ca2\ and \oi\ lines in quasars at redshift $>$ 1. 

In the following we will consider as a reference low--$z$, low--luminosity  \ ``control'' sample the sources of \citet{PER88} that were observed with very high S/N and spectral resolution similar to the one achieved in the present paper. This previous \ca2\ sample was sources with strong \fe2\ emission at low redshift, as are other, more recent ones \citep{MAT07}. We stress again that our sample is not selected on the basis of \fe2\ intensity. The selected ISAAC sources do not show very strong optical \fe2\ emission: 10 of them are Pop. B and 4 Pop. A (Table \ref{table:obs}). The Pop. A sources are of spectral type A1 following \citet{SUL02}; for all sources we therefore have \rfe\ $\la$ 0.5. This allows us to explore a domain in \fe2\ emission and in the 4DE1 parameter space where the study of \ca2\ could give new insights on the origin of low--ionization lines for the majority of quasars with moderate \fe2\ emission. At the same time, the preferential selection of high \rfe\ sources at low--$z$ should be 
taken into account  when  possible luminosity and $z$ effects are studied. 

\section{Observations}
\label{obs}

Observations were aimed at obtaining spectra of the \oil\ and CaT lines for the sample described in the previous section. Considering the rather low efficiency of IR spectrometers available until a few years ago, and the relatively low equivalent width of the features we wanted to observe ($\approx 10 - 20 $ \AA), the collecting area of a large telescope was needed to obtain sufficient signal with a moderate dispersion. 

The VLT includes four telescopes of 8.2 m diameter located on top of Cerro Paranal, and instruments are mounted at the foci of the four telescopes. ISAAC was\renewcommand{\thefootnote}{\fnsymbol{footnote}}\footnote{ISAAC has been decommissioned in 2013.} one of them, and has been for many years one of the few instruments available worldwide for moderate resolution IR spectroscopy of faint sources like high--$z$ quasars. ISAAC was able to obtain images and spectra in the wavelength range 1 -- 5 $\mu$m  \citep{MOO98}. It was equipped with gratings for high ($R_\mathrm{S}$ $\sim$ 3000) and low spectral resolution (R${_S}$ $\sim$ 1000). 

Our spectra were collected on the Rockwell CCD detector (pixel size 18.5 $\mu$m yielding a pixel scale of 0.148 ''/pixel) at low resolution but with a slit width of $0.6$'', that ensured R${_S} \approx$ 1000, as measured on sky and arc lamp lines. The second part of Table \ref{table:obs} provides date of observation, photometric band of the covered ranges, detector integration time (DIT), and number of sequential exposures with fixed DIT in the spectral band covering the CaT + \oi\ blend. Seeing values, reported in the penultimate column of Table \ref{table:obs}, are averages over 10 -- 15  measures at Paranal obtained during each quasar exposure. The S/N was measured on the continuum at 8300 to 8400 \AA. The instrumental spectral resolution at FWHM measured on sky lines was $\approx$ 330 \kms.

\section{Data reduction}
\label{red}

We used the {\sc iraf} software to reduce our spectra. Frames were obtained alternating  exposures of two or three DITs in two different positions (A and B) along the slit, following a sequence like A--B--B--A. The frames in each position were averaged and then subtracted one from the other. This procedure cancelled the bias and dark current present on each frame, assuming that they remained constant during the sequence, and also subtracted most of the sky background, facilitating the spectrum extraction. After the subtraction, frames were divided by the corresponding flat fields. We extracted the 1D spectra using the {\sc iraf} task {\sc apsum}, {where a low--order polynomial function was fitted and subtracted to eliminate any residual background}. 1D spectra were wavelength calibrated using an Ar-Xe lamp. The calibration was achieved using a 3rd order polynomial, with a typical rms of 0.3 \AA\ in the H band and 0.4 \AA\ in the K band. This calibration, if applied to the sky spectrum yielded a slight 
displacement that was corrected using measures of OH sky line wavelengths. We applied the offset--corrected wavelength calibration to the quasar spectra. Finally 1D quasar spectra were combined to get one median spectrum. The spectra of the standard stars (observed right before or after the quasar spectrum) were extracted and wavelength calibrated in  the same way. 

We corrected for the effect of telluric lines using the {\sc telluric iraf} task. This procedure involved a change in the quasar continuum shape since the task {\sc telluric} carried out a division between the quasar and the standard star. To return to the original quasar continuum after the elimination of the telluric bands, we divided the quasar spectrum by the standard star continuum (i.e., by the standard star divided by a synthetic atmospheric transmission spectrum). 

\begin{figure}[ht]
  \begin{center}
      \includegraphics[width=6cm,keepaspectratio=true]{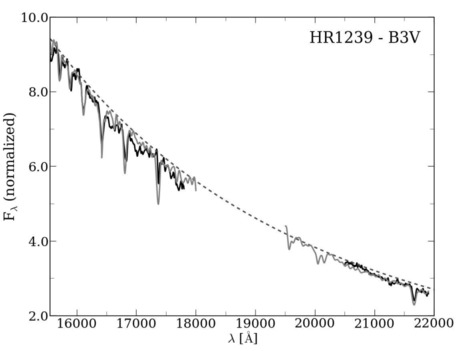}\\
      \includegraphics[width=6cm,keepaspectratio=true]{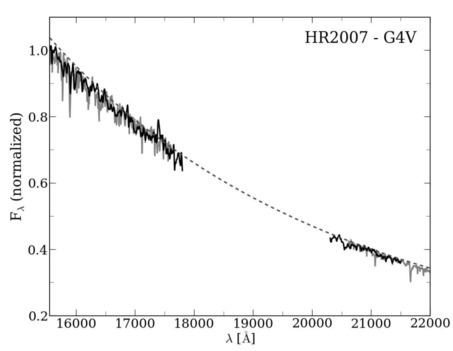}\\
      \includegraphics[width=6cm,keepaspectratio=true]{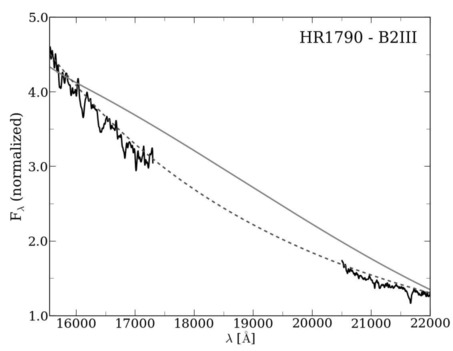}
      \caption{Comparison between Pickles stellar theoretical and observed emission of main sequence B (upper panel) and G (middle panel) stars and a giant B (bottom panel) star with a black--body spectral energy distribution. The black line is the observed emission, the grey line is the theoretical emission and the dashed line is the black--body emission.}
     \label{fig:bb}
  \end{center}
\end{figure}

Fourteen standard stars were spectral type B and one G. The B stars belonged to different luminosity classes (supergiants, giants and main sequence),  while the star G is a main sequence star. Comparing the NIR stellar spectral library from \citet{RAN04} and \citet{RAN07} with theoretical libraries in the H and K bands \citep{PICK98}, we found that the theoretical spectra could reproduce the observed spectra of main sequence stars, but not those of giants and supergiant stars. For the giants and supergiants stars we considered a spectral energy distribution provided by black--body emission at the temperature tabulated for the star \citep{AMM06}. Black--body emission fits very well the spectra after erasing the stellar absorptions lines with the tool {\sc xydraw} of the {\sc splot iraf} task  (see Figure \ref{fig:bb}). In order to render our approach uniform we fitted all the stars using a black--body, because there is practically no difference between the simple black--body fit and the continuum spectra of 
the  stars. Finally, the standard star spectra were normalized to H and K 2MASS magnitudes. 

The sensitivity function was obtained  dividing the standard star spectra by the black--body model. The final relative flux calibration was achieved dividing the spectrum by the sensitivity function. Spectra were corrected for Galactic extinction following \citet{SFD98} but no internal extinction correction was applied. Quasar spectra were normalized to 2MASS magnitudes as done for the standard stars in order to minimize the effect in the continuum flux change possibly induced by intrinsic variability (see next section). Redshift correction was performed using the redshift values reported in \citet{SUL04} and \citet{MAR09}. These redshifts were measured on H$\beta_{NC}$, H$\gamma_{NC}$ and \o3\ with an uncertainty usually  $cz <$ 150 km s$^{-1}$\ in the rest frame. Figure \ref{fig:cont} shows the rest-frame  flux and wavelength calibrated spectra before continuum subtraction.

\subsection{Variability}
\label{variability}
The \hb\ spectra were taken between 2001 and 2005, while the near-infrared spectra were collected in 2010. There is a difference of 5 -- 10 years between the \hb\ and NIR observations. In order to estimate a possible broad line flux variation in response to continuum changes, we used the BLR radius vs. luminosity relation of \citet{BEN13}. Comparing the expected response time with the difference in date between the optical and IR observations, we found that eight objects could have varied. Extracting the light curves from the Catalina Real-Time Transient Survey \citep[CRTS;][]{DRA09}, we could identify eight cases of actual variation. Five of the eight objects show a systematic variation $\lesssim$ 0.1 mag in a period of 6 -- 8 years when they were monitored by the CRTS. The remaining three objects, HE0048--2804, HE0203--4627 and HE2147--3212, show a ``high frequency'' variation  which is apparent even from a visual inspection of the light curve. The amplitude is modest, $\approx$ 0.2 magnitudes, and does 
not hint at any systematic variation over longer periods. We normalized the \ca2\ spectra to H and K 2MASS magnitudes, and the \hb\ spectra to J and H magnitudes, in order to obtain an accurate flux scale as well as to remove any variability effect on line and continuum fluxes. This normalization was not possible for two objects, HE0048--2804 and HE2340--4443, since 2MASS does not have information on the I or Z photometric bands that cover redshifted \hb\ (see Appendix \ref{appex} for a brief discussion).

\section{Measurements}
\label{meas}

The fits and data analysis were carried out within the Eigenvector 1 context \citep{BOR92, SMD2000a, SUL2000, MAR01, MAR03a, MAR03b, SUL07}. We performed multicomponent fits including all the relevant components that we presume are contributing to the spectral ranges under investigation. The objects analyzed in this paper belong to a sample of intermediate redshift quasars for which \hb\ is available from previous ISAAC observations. The \hb\ line profile parameters can guide us in the interpretation of the  NIR region. The optical spectra were previously analyzed by \citet{SUL04} and \citet{MAR09}. The analysis of this paper adds an \hb\ multicomponent {\sc specfit} modelling that was not carried out for individual sources  or, if carried out, was not presented in previous papers.  The optical and the NIR spectra were modeled with the task {\sc specfit} of {\sc iraf}, a routine that carries out simultaneous fits of several line and continuum components to 1D spectra \citep{KRI94}. Each  component is 
described by a set of parameters that are varied according to the initial guess between a   maximum and a minimum value. The best fit is achieved via $\chi{^2}$ minimization. The multicomponents fits are shown in the Figure \ref{fig:fits}. The left panel shows the optical spectra and  the right panel the NIR spectra.

Few parameters were actually free to vary in the NIR spectral range. The \hb\ profile was taken as a proxy for the \pa9\ and \oil, and the peak shifts of these two lines were set as  measured on \hb. The flux ratio between the broad component and the very broad component of Pa9 is assumed to be the same of \hb. The  width and shift of \o3\ set the  width and shift  of \s3 $\lambda$9531, so that only the intensity of the latter line is left free to vary. The centroid of CaT was set in the quasar rest-frame.  {The only remaining free parameters were therefore: the specific flux scale and the index of the power-law continuum (2), the intensity scaling of the \fe2\ template (2), the intensity and width of \oil\ and CaT (4), the intensity scaling of the high order Paschen lines (1) and the intensity and temperature of Paschen continuum (2), giving a  total number of 11 free parameters. However, not all these parameters refer to   features blended with \oil\ and CaT: the Paschen continuum properties do not affect 
the blend \oi\ + CaT; the intensity scalings of the high order Paschen lines and of the \fe2\ template are also set by features outside of the blend. This makes the fits more robust, and ensures that  \oil\ + CaT  intensity and width are the parameters mostly affecting the $\chi^2_\nu$ in the wavelength range they cover.} The following sections describe each one of the components used for the fits and the error estimates. \\

\subsection{Continuum}

The main continuum emission present from the UV to the NIR spectral region of quasars is due to an accretion disk \citep{MAL82, MAL83}. Accretion disk emission is thermal (a ``stretched black--body'') but can be modeled locally with a power law with a slope of $-2.4$. Optical and NIR spectra should share  in part the same underlying continuum  i.e., the low--energy tail of the accretion disk emission. However, we could not consistently fit  the continuum with a single slope from the optical to the NIR. In the majority of the sources the NIR spectrum is $\sim$ 30 -- 80\% above   the NIR extrapolation of the optical   continuum. Therefore, a local power-law continuum was fitted for each quasar assuming a free power-law index. The optical continuum was defined at 4750 and 5100 \AA. The IR continuum was set at 8100, 8800 and/or 9400 \AA, depending on the wavelength range covered by each NIR spectrum.

A possible origin of the disagreement between  optical and NIR continuum is a contribution from two additional NIR components \citep{LAN11}. The first is a diffuse continuum (due to thermal emission and scattering) that may flatten the NIR spectrum \citep{KOR01}. The second component is due the finite albedo of the same dusty clouds emitting the thermal far-IR continuum \citep{LAN11}. The small range covered by our spectra makes it impossible to measure their contributions. The continuum can be also affected by the host galaxy (this seems to be case for only one quasar in our sample, see Section \ref{host_gal}).

\subsection{\fe2\ template}  
Over the years, there have been several efforts to model the ultraviolet and optical \fe2\ contribution \citep{SIG03, SIG04, BRU08}. In the NIR, a semi--empirical  \fe2\ template based on the I Zw 1 spectrum was produced  by \citet{GAR12} who kindly made it available to us. A theoretical template based on photoionization was also computed by the same authors (Table 3 of \citealt{GAR12}). A comparison of the two templates is shown in the upper left panel of Figure \ref{fig:fe2_templates}. The agreement between the two templates is fair; both indicate significant \fe2\ emission at around 8400 \AA, 9200 \AA, and 1$\mu$m. These features are expected to be produced through \lya\ pumping of high energy levels ($\approx$ 11 eV) in the Fe$^+$ ion, although \lya\ pumping may not be the only production mechanism, and additional mechanisms are still debated \citep{SIG98,RUD2000,ROD02,GAR12}. \lya\ fluorescence makes the difference between optical and NIR emission. At present we do not have evidence that this process 
could induce a  difference between the NIR \fe2\ spectrum of Pop. A and B sources. We considered  \fe2\ measures of Table 6 of \citet{LAN08} for Pop. A (13) and B (4) sources. The median \fe2\ relative intensities measured in the range 8000 -- 9810 \AA\ are the almost identical. 

The upper right side of Figure \ref{fig:fe2_templates} shows an example of  fits to the same spectrum  (HE1349+0007) using different  templates. The lower panels of Figure \ref{fig:fe2_templates} show the HE1349+0007 NIR spectral range after background subtraction, to illustrate the effect of the \fe2\ template on \oil\ and CaT measures.

In our spectra with high S/N  we  observe a rather symmetrical bump due to Pa9 + \fe2\ at  $\approx$ 9200 \AA. The emission on the blue side of this bump cannot be reproduced with the semi--empirical template. The width of  Pa9 is limited by the assumption that it should be consistent with the one measured for the Balmer lines in the optical spectra. For this reason   we could not model Pa9 with a  profile broad enough to fill  the observed excess emission. In addition, higher order Paschen lines constrain the Pa9 intensity. Therefore, the excess of emission with respect to the semi--empirical template has to be ascribed to \fe2. Given the problem with the semi--empirical template, the \fe2\ emission of our spectra was fitted with both the semi--empirical and the theoretical template. 

Between 8600 and 8800 \AA\ the semi--empirical template presents a contribution that is not modeled by the theoretical one.  This feature affects the flux and equivalent width of the line \ca2\ $\lambda$ 8662 \AA. Because the three lines have the same intensity, if the semi--empirical template is used, the equivalent width and flux of the CaT and \oil\ decrease. In our sample the \oil\ and CaT flux and equivalent widths is affected by less than 20\%, save for: two sources where the difference reaches $\approx$ 20 -- 35\% for \oil\ (HE2340--4443 and HE0203--4627), and four sources where the CaT differs by 20 -- 35\%\ (HE0203--4627,  HE2352--4010, HE0248--3628, HE2340--4443). 

We then choose the template that gave the minimum $\chi^2_\nu$ value. We found that the theoretical template suits better in the majority of the cases. In only three cases (HE2147--3212, HE2349--3800  and HE2352--4010) the semi--empirical template was used; in the remaining sources the theoretical template was applied.

In the IR, the strongest, most isolated \fe2\ emission features are around 1 $\mu$m, but this wavelength range is not covered by our spectra. In the wavelength range covered by our spectra we have no truly isolated \fe2\ features. However,  we could still accurately measure the \fe2\ contribution in the range of 9000 -- 9400 \AA\ from the  scaled and broadened template, and consider it as an estimate of the NIR \fe2\ emission. In the majority of our spectra, we were able to neatly fit the \fe2\ and Paschen 9 line contribution.

For the \fe2\ optical contribution in the \hb\ region we used the template previously employed by \citet{MAR09}, mainly based on a high S/N optical spectrum of I Zw 1. Iron emission gives rise to a fairly isolated bump at $\sim$ 4450 -- 4680 \AA. Optical \fe2\ emission was defined by the total emission in this range (conventionally indicated with \feiiq), a standard practice \citep{BOR92}.

\subsection{Broad component} 

Following the 4DE1 approach \citep{SMD2000a, SUL2000, MAR01, MAR03a, MAR03b, SUL07}, we assume that low--ionization lines in Pop. A and B sources have profiles that are best fitted by Lorentzian (Pop. A) or double Gaussian (Pop. B). All the broad components were modeled with this assumption. We note that the Lorentzian approach is roughly equivalent to fitting two  Gaussians with the same shifts (usually $\la 200$ \kms\ for Pop. A sources, \citealt{SUL12}), one broader and one narrower. However, this is strictly true if the S/N is low: at very high S/N (not the case of our data) a Lorentz function yields a lower $\chi^2_\nu$\ \citep{MAR03b}. Since the best data suggest that Pop. A \hb\ profiles are Lorentzian \citep{VER01,MAR03b,ZAM10,SHA12}, we fit Lorentzian functions to the broad line profiles of all Pop. A sources in our sample. With a double Gaussian fit, the difference between Pop. A and B would be that the broader Gaussian is usually redshifted in Pop. B. 

The \hb\ line profile was taken as a reference for modelling  the CaT and \oil\ blend, since its profile is well defined and well understood. CaT was modeled with three BCs of equal intensity and the same FWHM since it is believed that the \ca2\ triplet components are optically thick \citep{PER88, FERPER89}. As \ca2\ lines are completely blended, it is difficult to measure a shift, so that the \ca2\ lines were held fixed at rest-frame wavelength. The maximum shifts possible for \oil\ were the ones measured on \hb, due to the relation between the two lines (see also section \ref{pho}). FWHM and shifts of \hb\ and high order Paschen lines were assumed to be identical within the uncertainties.

\subsection{Very broad component} 
A VBC was fitted to \hb\ and the most intense high order Paschen lines  (Pa9, Pa$\epsilon$ $\lambda$9546 and Pa$\delta$ $\lambda$10049) for all Pop. B sources. The shift and FWHM were assumed equal to those of \hb\ within the uncertainties. A VBC was not fit to \ca2\ triplet, since the VBC is believed to be associated with a very high ionization region (the very broad line region, VBLR  e. g. \citealt{SNE07} and \citealt{MARZ10}) that is expected to emit negligible \fe2\ and \ca2. Since there is no clear spectroscopic evidence of the presence of a VBC for \oi, only a BC was first  considered also for this line (although the possibility of an \oi\ VBC  is debatable  and discussed in section \S\ \ref{oivbc}).

\subsection{Narrow components and narrow lines} 
We considered the \hb\ narrow component (\hbnc) and \o3. In the NIR, we were able to detect a hint of the NC for \oi\  only in HE1349+0007, a source in which \hbnc\ is especially strong. Also, in all the spectra covering the region around $\sim$ 9500 \AA, we could detect \s3 $\lambda$9531.

\subsection{High order Paschen lines} 
Modelling the NIR spectral range with only the  strongest high order Paschen lines visible in our spectra (\pa9, Pa$\epsilon$ $\lambda$9546 and Pa$\delta$ $\lambda$10049) results in an emission deficit between 8700 and 9100 \AA. Therefore we decided to include   even higher order Paschen lines: from  \pa9\ up to Pa24 $\lambda$8334. Since Pa9 is clearly seen in the majority of our spectra, it was taken as a reference to scale the intensity of the other Paschen lines. The  intensity ratios were computed from  case B values from {\sc cloudy} simulations \citep[version 08.00;][]{FER98,FER13}  which have an exponential decay with increasing order number. This set of lines forms a pseudo-continuum that cannot be neglected in some cases. Previously, \citet{PER88} found that Paschen lines contribute $\sim$ 12$\%$ of the flux of \ca2\ $\lambda$8662 in Mrk 42, so that they were ignored. Comparing the flux of Pa13 $\lambda8665$ and the \ca2\ $\lambda$8662, we found that Paschen lines do contribute by a fraction $\
gtrsim$ 30\% in ten sources; in two sources the contribution is $\sim$ 10$\%$\ (HE0203--4627, HE2147--3212). In two cases (HE0048--2804 and HE0058--3231) the high order Paschen lines dominate the fit.

\subsection{Paschen continuum} 
We could detect a hint of Paschen continuum (PaC) in HE0048--2804 and HE2202--2557, so we included an optically thin model of the Paschen continuum at 8204 \AA. The inclusion has little effect on the \oil\ and CaT measures, since the Paschen series head is separated in wavelength from the \oi\ + CaT\ blend. A {\sc cloudy} simulation with $\log U = -2.5$\ and $\log n_\mathrm{H}$ = 12 provided us with predictions on the relative intensity of Pa9 and of the integrated PaC \citep[c.f.][]{OF06}. We then estimated the continuum specific intensity at the Paschen edge, and assumed an exponential decrease toward shorter wavelengths appropriate for an optically thin case. Our estimates appear to be in excess with respect to the observations since the best fits require a PaC smaller than that predicted. There are three main possible explanations: (1)  {\sc cloudy} computations over--predict the recombination continuum (a distinctly possible case since the assumed conditions are unlikely to be found in optically thin 
gas), (2) the quasar continuum level is not correct, and the actual continuum is lower, or (3) PaC is not optically thin. We cannot test these options on our data since the spectra cover only a small wavelength range. To properly define PaC we need to cover a spectral range spanning from the optical to the NIR. We will try to address the issue of the PaC intensity in the future.

\subsection{Contribution of the host galaxy} 
\label{host_gal}
A new stellar population synthesis model (Chen et al. in preparation), based on the code of \citet{BGS98} with updated stellar evolutionary tracks \citep{BRE12} and with stellar atmospheres was used to account for the possible contribution of the quasar host galaxy. These computations are suitable for the analysis of stellar absorption lines in the optical and NIR spectral regions \citep{SAN06a, SAN06b, SAN06c, RAY09}. From the lookback time at the quasar redshift we estimated an upper limit to the host age. Then we computed the black hole mass, and estimated the host mass assuming the $M_\mathrm{bulge} / M_\mathrm{BH}$ ratio \citep{MAG98, MER10} appropriate for the quasar redshift. The 0.9 $\mu$m specific flux was then derived from the stellar population synthesis models that are provided as a function of mass. We found that the underlying stellar absorption of the host galaxy is significant only in HE2202-2557, with a luminosity contribution of $\approx$ 50$\%$, while the rest of the sample is affected 
only by $\lesssim$ 10$\%$. In the  HE2202--2557 case, we  assumed a spheroid mass $\approx$ 1.1 $\cdot \ 10^{12} M_\odot$, an age of 2.4 Gyr and a metallicity of 2Z$_\odot$. These properties are consistent with massive ellipticals expected to host very luminous quasars at intermediate redshift \citep{TRA00, KUK01, SAN04, FAL04, KOT07, SPO09, FLOY13}.

In order to test the behavior of several stellar templates, we performed several models taking different spheroid masses (10$^{10}$, 10$^{11}$, 10$^{12}$, 10$^{13}$ M$_\odot$, but keeping {  metallicity equal to 2Z$_\odot$ and host age  2.4 Gyr}). Model spectra with different spheroid masses give a similar total spectrum, but spectra  differ in the CaT range where the stellar absorptions  have  different depths and widths, as  shown in Figure \ref{fig:HG}. The CaT intensity in absorption decreases with increasing  galaxy mass and/or velocity dispersion \citep{CEN03, VAN12} while its width increases. These changes affect the emission lines measurements. The flux and equivalent width of  CaT in emission and of \oil\  increase with galaxy mass. The effect induced by a 3 order of magnitude mass change   is, in general, $\lesssim$ 50\%. {In the HE2202--2557 case, Figure \ref{fig:HG} shows that  host masses lower than $1 \cdot 10^{12}$ $M_\odot$\ would yield  significant residuals and a worse $\chi^2_\nu$, while 
for $1 \cdot10^{13} M_\odot$\ the $\chi^2_\nu$\ will not change significantly. We assume a bulge mass  $1.1 \cdot 10^{12}$ $M_\odot$, consistent with the value expected from   $M_\mathrm{bulge}$/$M_\mathrm{BH}$\ ratio at intermediate redshift following \citet{MER10}.

\section{Results}
\label{res}

Figure \ref{fig:fits} shows the multicomponent fits after continuum subtraction for the \hb\ and \oil\ + CaT spectral regions. Results of  multicomponent fits are provided in Tables \ref{tab:hb+pa9} and \ref{tab:oi+ca}. Table \ref{tab:hb+pa9} reports \hb\ and \pa9\ measurements. The first column lists the name of the object and the second column is the continuum flux measured at the line's rest-frame. The next columns list equivalent width (W), flux (F) and full--width at half maximum (FWHM) for the broad component (BC), narrow component (NC) and very broad component (VBC). The last column reports the  FWHM of the full profile i. e., the sum of BC and VBC. Table \ref{tab:oi+ca} reports the \oil\ and CaT $\lambda$8498, $\lambda$8542 and $\lambda$8662 measurements. As for Table \ref{tab:hb+pa9}, the second column lists continuum flux at the \oil\ wavelength and the other columns report  W, flux and FWHM for the BCs. The NC was detected  for \oi\ only in the case of HE1349+0007. Error estimates were obtained as 
described in Appendix \ref{error}. The equivalent width and flux for optical and NIR \fe2\ are reported in Table \ref{tab:fe2}. The last column reports which NIR template (i.e., theoretical or semi--empirical) was used in the fits. Shifts are, at a first guess, assumed to be consistent with those of \hb, and are found to be in general close to 0 (i.e., consistent with the quasar  rest frame). They are therefore not reported. Only in one case, HE1409+0101, the \hb$_\mathrm{BC}$ shows a significant blueshift, --700 \kms, and a similar shift was applied to the \oi\ peak to make it consistent with \hb. The \ca2\ triplet was assumed to be always at rest frame. The VBC of Hydrogen lines was fitted with a broad Gaussian that has a large shift to the red, $\sim$ 1000 -- 3000 \kms,  assumed to be the same for optical and NIR Hydrogen lines.

\subsection{\ca2\ triplet detection}
\label{ca2_det}
The \ca2\ triplet has been observed in sources with strong \fe2\ emission, as in the majority of the \citet{PER88} sample. CaT could be observed in many \fe2\ strong sources because they are Narrow Line Seyfert 1, and emission lines are therefore relatively narrow (FWHM \hb\ $\la 2000$ \kms) and sharply peaked. In our sources the CaT lines are often broad and totally blended with \oil\ and high order Paschen lines making it necessary to confirm that CaT is really detected. The $\lambda$8498 and $\lambda$8542 lines do not provide conclusive evidence, because they are heavily blended among themselves and with \oil. On the converse, \ca2\ $\lambda$8662 is in a region relatively free from strong contaminating lines, and in some cases this line could be detected by a simple visual inspection of the spectra (HE0248--3628 is an example). In two cases of sources with weak CaT, like  HE0048--2804 and HE2202--2557, a visual inspection is inconclusive. To test that  detections are real, we performed a fit with no CaT 
emission. In HE0048--2804 we could observe a slightly positive residual around 8600 \AA; the CaT emission is  very weak, if any  (Figure \ref{fig:sCa}, left). The positive residual is strong in HE2202--2557 and without any doubt CaT emission is present (Figure \ref{fig:sCa}, right). We conclude that \ca2\ triplet is detected in all of our sources with the possible exception of HE0048--2804, even if detection of CaT emission was not a priori expected on the basis of \fe2\ -- \ca2\ correlation \citep{PER88, JOL89, FERPER89, DUL99}. We also note that W(CaT)  can be as large as $\approx 50$ \AA.

\subsection{Relations between \ca2\ triplet and \oil}

The equivalent width distributions of CaT and \oil\ for \citet{PER88} and our sample are shown in the upper panels of Figure \ref{fig:histo}. The spread of \oil\ and CaT values is similar in the two samples. In our sample W(CaT) is distributed over a wide range, whereas the W(\oi) distribution is peaked around $\approx$ 15 -- 20 \AA. The Persson sample similarly shows a narrower distribution of W(\oil). The difference in the equivalent width distributions is reflected in the wide range of the CaT/\oil\ ratio: from almost 0 to CaT significantly stronger than \oil. 

{If we add the 11 upper limits in the sample of \citet{PER88} to the distribution, the CaT seems to be significantly more prominent in our sample. Two sample tests that include  censored data indicate a confidence level between 0.92 (Gehan's Generalized Wilcoxon test) and 0.98 (Logrank test). The \oil\ distributions (no upper limit; lower left) are instead consistent. \citet{PER88} fails to detect CaT emission if CaT/\oil\ $\la$ 0.3, with W(\oil) $\sim$ 10 -- 20 \AA. These sources all show \rfe\ $\la$ 0.5, and are of spectral type B1 or A1.  This finding leaves open the possibility that there could be a dependence on $L$ and/or $z$\  (\S \ref{lumz}) specially for CaT.}

\subsection{Line widths of \hb, \oil\ and \ca2 triplet}

Persson's and our data are complementary because our sample is mainly made of Pop. B sources (Table \ref{table:obs}; column 6). Twelve of the \citet{PER88} sources are Pop. A  for which FWHM(\hb) $\le$ 4000 \kms\ and 2 are Pop. B. ({\sc ii} Zw 136 and Mrk 376).  If we compare the FWHM of \hb\ (measured on the full profile, without separating VBC from BC; top panel of Figure \ref{fig:fwhm}) we see an overall consistency of FWHM(\hb) with FWHM(CaT) and FWHM(\oil) for Pop. A sources, while for Pop.B sources FWHM(\hb) is significantly larger than the FWHM of the NIR lines. This is the effect of the VBC that broadens the full profile of Pop. B by 20 -- 30\%\ \citep{MAR13b}.

If we restrict the attention to the broad component of \hb, there is an obvious correlation between FWHM(H$\beta_{BC}$) $\gtrsim$ FWHM(CaT) $\sim$ FWHM(\oil) (middle panel of Figure \ref{fig:fwhm}), which confirms and extends the trend found by \citet{PER88}. In the middle panel of Figure \ref{fig:fwhm} Pop. A sources are located around the equality line, while Pop. B sources are located below. The Pop. B trend is significant at a 2--$\sigma$ confidence level accord into to a Wilcoxon signed rank test \citep{WIL45}, and may indicate that there is a  systematic difference between the widths of \hb\ BC only and \oil\ or CaT in Pop. B sources. A discrepancy could also be due to different quality of data and/or to different measurement techniques, since  Pop. A sources are mainly from the Persson sample and Pop. B sources are from our sample and the H$\beta$\ profiles need a BC/VBC decomposition. If broadening is  predominantly due to virial motions, and if FWHM(H$\beta_{BC}$) is really systematically larger 
than FWHM(CaT) and FWHM(\oil),   the \hb\ emission associated with the  BC only may still preferentially come from regions closer to the central black hole than CaT and \oil. 

It is also interesting to compare the FWHM of \oil\ and CaT (bottom panel of Figure \ref{fig:fwhm}). At least to a first approximation, \oil\ and CaT show consistent FWHM values. The Persson data suggest that CaT is slightly systematically broader than \oil. This effect is not seen in our data that have however much larger uncertainties.  A small CaT and \oil\ FWHM difference may be explained in the context of the different emissivity of the two lines as a function of ionization parameter and density. If the two lines are not emitted  in a coexistence region \citep{ROD02,MAT05, MAT07}, they can be emitted in two regions with different physical conditions, but kinematically coupled. Considerations on ionization parameter and density may support this possibility (discussed in \S \ref{pho}). 

\subsection{Relation to \fe2}

Photoionization models performed by \citet{JOL89} have shown that the relation between the ratios CaT/\hb\ and \feiiq/\hb\ provides  evidence of a common origin for CaT and \feiiq: CaT/\hb\ increases at high density and low temperature as does \feiiq/\hb\ \citep{JOL87}. Our sample follows the same trend found by Joly (Figure \ref{fig:ratio_fe+ca}), with a Spearman correlation coefficient $\rho \approx 0.73$,  yielding a negligible probability ($P<10^{-4})$ that the correlation   arises from statistical fluctuations. The correlation is consistent with the idea that \feiiq\ and CaT are both emitted in a similar region, since the two lines are expected to be emitted under very similar physical conditions. If not, a hidden variable (metallicity or even orientation: Dultzin-Hacyan, Taniguchi \& Uranga 1999) could give rise to the observed correlation.

A least-squares fit including censored data gives the following relation: $\log$ (CaT/\hb) $\approx$ (1.33 $\pm$ 0.23) $\log$(\feiiq/\hb) -- (0.63 $\pm$ 0.07). The relation between \oil/\hb\ and \feiiq/\hb\ is, on the contrary, much more shallow, even if the slope is significantly different from 0: $ \log$(\oil/\hb) = (0.56 $\pm$ 0.14) $\log$(\feiiq/\hb) -- (0.76 $\pm$ 0.05), implying that  the systematic  \oil/\hb\ change  is a factor $\approx$ 3.5 for an order-of-magnitude increase in \rfe.

\subsection{Correlation with redshift and luminosity}
\label{lumz}

Figure \ref{fig:lumz} shows that W(\oil) does not correlate with luminosity over a range covering 4 dex in luminosity, as defined by our sample and Persson's data, with the addition of intermediate luminosity  sources  from \citet{MAT07}. The present data do not support earlier claims of a low--ionization Baldwin effect in \oi, neither as a function of luminosity nor as a function of redshift (i.e., an ``evolutionary Baldwin effect''). 

The data shown in Figure \ref{fig:lumz} are statistically biased in terms of both $z$\ and $L$\ (i.e., they are not uniformly distributed over the  covered range). In addition, there is a less-visible, difficult to quantify ``Eddington bias'' (see \citealt{SUL14} for a discussion on this topic): our high--$z$\ Pop. B sources may be those with the largest \lledd\ possible for Pop. B. Therefore, great care should be exerted in interpreting the bottom left panel of Figure \ref{fig:lumz} before claiming a  correlation, even if the correlation coefficient is --0.564, with a probability $<$10$^{-4}$ of a chance correlation. Accepted at face value, the trend between $M_\mathrm{B}$\ and W(CaT) would imply an increase in prominence of CaT with $L$\ and $z$. We see two possible explanations for this result, in addition to an intrinsic luminosity effect: (a) a larger ratio CaT/\feiiq\ in the high--$L$\ sample; (b) a systematic difference in the \feiiq/\hb\ distribution between our sample and the sample of \citet{PER88}.
 
The 
correlation between CaT/\hb\ and \feiiq/\hb\ implies that a strong CaT emission is associated with a strong \feiiq\ emission. The bottom left panel  of Figure \ref{fig:histo} shows the \rfe\ distribution for our sample and Persson's. They are not significantly different according to generalized Wilcoxon's tests (repeated also without censored data). Therefore, the trend of Figure \ref{fig:lumz}  cannot be ascribed to a systematic difference in \rfe. The bottom right panel  of Figure \ref{fig:histo} shows the CaT/\hb\  distributions. In this case, the same two-sample tests yield a significant difference. Figure \ref{fig:ca2fe2} shows the CaT/\feiiq\ distributions (only for detected data) that are probably least affected by selection effects\footnote{{ There is no a-priori reason why we should select sources with a biased distribution of CaT/\feiiq\ since we are able to measure the ratio CaT/\feiiq\ over a broad range of values ($\sim 0.05 - 0.7$) and we have   undistinguishable \rfe\ distributions for the 
Persson and ISAAC sample.}} and they are found significantly different with a confidence above the 2--$\sigma$\ level: the CaT/\feiiq\ ratio average is larger by a factor $\approx 1.7$\ for the ISAAC sample. This effect can account at least in part for the $L$ and $z$ trends of Figure \ref{fig:lumz}, and is a potentially important result  that will be discussed in \S \ref{ca2-fe2} and \S \ref{alphafe}, even if a confirmation from a larger sample and a set of observations obtained with the same instrument including low-- and high--$z$ sources is needed.

\section{Discussion}
\label{dis}

\subsection{Comparison with photoionization models}
\label{pho}
We computed photoionization models using the code {\sc Cloudy} \citep{FER98,FER13} in order to constrain the physical conditions of the region where CaT and \oil\ are emitted. The simulations include a 371--level  \fe2\ ion\ that allows some limited comparison with the observations of optical and IR \fe2\ emission. Simulations span the density range 7.00 $ < \log n_\mathrm{H}  <  14.00$\ cm$^{-3}$ and ionization parameter range --4.50 $< \log U <$ 00.00, in intervals of 0.25 dex assuming plane--parallel geometry. The spectral energy distribution of the ionizing continuum is the one of \citet{MATFER87} that is considered a standard for quasars. Open geometry, line thermal broadening and solar metallicity are  assumed. }  We consider  column density values   10$^{23}$ and 10$^{25}$ cm$^{-2}$.  Figure \ref{fig:sim_ca+oi} shows isopleths of the predicted CaT/\oil\  ratio as a function of $U$\ and \nh\ for the two values of column density.  Figure \ref{fig:sim} shows the panels isopleths in the plane ($U, n_\mathrm{H}$) for CaT/\hb, \oil/\hb, CaT/Pa9 and \oil/Pa9, again for the two values of  column density. 

From a first comparison between the photoionization prediction at $Z = Z_\odot$\ and the observed line ratios we can draw the following preliminary conclusions:

\begin{itemize}
\item The geometrical depth of the fully-ionized zone does not exceed the geometrical depth of the gas slab in the case of $N_\mathrm{c} = 10 ^{25}$\ cm$^{-2}$. There is a depth range where the gas is only partially ionized, even at the highest $U$\ ($\log U \sim 0$). In other words, the ionizing photon flux is not able to make the cloud optically thin.  In a photoionization context, \oil\ and especially CaT  emission will be observed only if the emitting gas remains optically thick to the Lyman continuum. Therefore, in the high ionization parameter area  of the plane (\nh,$U$) we still have  some emission of \oil\ and CaT that it is not predicted when $N_\mathrm{c}$ is smaller, 10$^{23}$ cm$^{-2}$. 

\item Constraints on the density of the emitting regions follow from maximum and minimum measured values of CaT/\hb, \oil/\hb, CaT/Pa9 and \oil/Pa9. Regardless of \nc, the ratios involving CaT\ (when CaT emission is detected at a high significance) suggest that emission occurs only at relatively high density $\log n_\mathrm{H} \gtrsim$ 11.00, with an upper limit to the ionization parameter $\log U \approx$ --1.5, in agreement with the results of \citet{MAT07}.

\item The median values of the  CaT/\hb\  ($\approx -0.6$) and CaT/\oi\ ($\approx$ 0.2) ratios do not favor the area in the plane ($U$, \nh) at very low--ionization and high--density ($\log$        \nh $\gtrsim$ 12, $\log U \lesssim -2.5$), unless density becomes $\log$ \nh\ $\sim 13$. In this region significant emission of CaT --  but not of \oil\ --  can occur. 

\item The ratios of CaT and \oil\ normalized to \hb\ and Pa9 look strikingly similar (Figure \ref{fig:sim}). \hb\ and Pa9 are practically interchangeable. The upper half of Table \ref{tab:ratios} lists the logarithm of the observed flux ratios for  \hb\ and Pa9 (BC only), \oil\ and CaT. 
\end{itemize}

The observed equivalent width of CaT can be $\approx 50$ \AA, and W(\oil) $\lesssim $ 25 \AA. These large equivalent widths are observed in both our sample and the Persson sample. Reaching values that high for CaT is possible in the framework of photoionization within the broad \nh\ lower limit  and $U$\ upper limit set above. The W(CaT) is strongly dependent on the column density in the area centered at (\nh,$U$) $\approx$ (11.5, --2.5): the larger column density case yields an almost 10--fold increase in W(\ca2), while the effect is much lower for \oil. 

If we want to impose the conditions that \oil\ and CaT are emitted mainly within the same range of physical conditions, then a high metallicity (5 $Z_\odot$) seems to be required in order to account for the largest observed W(\oil) with a reasonable covering factor ($\approx$ 0.2).  Metallicity  solar or higher than solar is a very likely condition for  HE sources. They are among the most luminous quasars known  and there is a well defined correlation between $Z$ and luminosity \citep[e.g.,][]{SHIN13}. Among Pop. B sources the HE quasars show large  \rfe. Therefore, strong CaT and \oil\ emitters (the values of Table \ref{tab:oi+ca} are loosely correlated) would benefit from large column density and high--$Z$.  We remark that the condition of high--$Z$\ is needed to explain especially the \oi\ strength if we require that \oil\ and CaT are emitted under similar conditions. This might not be necessarily the case. 

At variance with CaT, \oil\ can be emitted over a wide range of densities: 9.00 $\lesssim \log n_\mathrm{H}$ $\lesssim$ 12.00. We interpret the difference in behavior as due to:  1) the similarity of neutral Hydrogen and Oxygen ionization potentials ($\approx$ 13.6 eV) and 2) the Bowen fluorescence mechanism that is strongly influenced by the ionizing photon flux. The Bowen mechanism has indeed been found to be the major contributor to the \oi\ intensity in most AGN studied by \citet{ ROD02} and \citet{LAN08}. Therefore \oil\ emission can originate in deep regions exposed to a large ionizing photon flux provided that the column density is high $N_\mathrm{c}$ $\sim$ 10$^{23-25}$ cm$^{-2}$. These properties are consistent with those of the VBLR seen in Pop. B sources \citep{MARZ10}.

To frame these results on  physical conditions in a broader scenario encompassing kinematics, we have to consider  inferences that come from analysis of internal line shifts and profile differences.  A basic result is that the BLR can be separated into two main regions: one emitting mostly LILs and the other emitting  HILs.   In Pop. A sources a blue shift and  asymmetry (as in the prototypical HIL \ion{C}{4}$\lambda$1549) indicates that the HILs are emitted within a partially-obscured flow \citep{SUL07, RICH11, WANG11}. Photoionization models further suggest that the HIL emitting region has  a relatively low density ($n_\mathrm{H} \sim$ 10$^9$ cm$^{-3}$), column density of $10^{21-23}$ cm$^{-2}$, and a high ionization parameter, $U$ $\sim 10^{-1}$, more than  10 times the value appropriate for LIL emission \citep{NET13}.

The hydrogen lines (typically \hb) have been used as representative lines of the LILs. Photoionization models   show that   LIL  emission is associated with a density that can be as high as $\sim 10^{11.5 - 12}$ cm$^{-3}$,  column density $\gtrsim$ $10^{23}$ cm$^{-2}$ and a rather low--ionization parameter $\log U \approx -2$ \citep{MARZ10, NET13}. However, even under these conditions, it is not possible to reproduce the intensity of  very low--ionization lines, like the \ion{Fe}{2}\ features \citep{WIL85}. \ion{Fe}{2}\ lines need a higher column density ($N_\mathrm{c} \sim$ $10^{24}$ cm$^{-2}$) that makes possible an extended partially ionized zone at a  relatively low electron temperature  \citep[T $\sim$ 8000 K;][]{COL80, COL87, JOL87, FERPER89, MAT07}.

In Pop. B the blue shifted component seems to be less prominent.  The HILs are  emitted    closer to the central source than the LILs \citep{PET99}, implying a radial stratification of the emitting region. The distinction BC/VBC is the most simple parameterization of the ionization stratification, with the VBC being associated with highly ionized ($\log U \sim -0.1$), large column density gas ($\log$ \nc\ $\gtrsim 23$), probably with a broad range of density \citep{SNE07, MARZ10}).

\subsection{A VBC for \oil?}
\label{oivbc}
We observe  a similarity between the \oil\ and \hb\ profiles \citep[c.f.][]{LAN08}. If we take into account the results of photoionization models it is reasonable to investigate the possibility that  \oil\ shows significant VBC emission similar to redshifted VBC emission observed in \hb\  of Pop. B  sources. 

In order to test the  presence of a VBC in \oil\ we added this component to the fits  for Pop. B sources. The fits for HE0035--2853 and HE1349+0007 are shown in Figure \ref{fig:bc+vbc}. Results  are basically the same with or without inclusion of a VBC. Even if FWHM \oil\ and CaT\ are the same using models with and without a VBC, the flux values changed. Inclusion of a VBC reduces the intensity in \oil\ and CaT by 10 -- 30$\%$ and 20 -- 60$\%$, respectively.  This reduction implies only slightly different physical conditions when the intensity ratios are entered in the $(U, n_\mathrm{H})$\ contour diagrams. The lower half of Table \ref{tab:ratios} reports values for the CaT/\hb, \oil/\hb, CaT/Pa9, \oil/Pa9, CaT/\oil\ ratios including a VBC for \oil.  Comparing the ratios with and without VBC, we see that there is no difference within the errors. For example, if we only assume a BC, the range of the ratio CaT/\oil\ is [-0.59, 0.53] while including the \oil\ VBC implies [--0.62, 0.40]. Therefore, the inclusion 
of an  \oil\ VBC does not significantly modify inferences about density  in the CaT and \oil\ emitting region: $\log n_\mathrm{H} \gtrsim$ 11.00 and 9.00 $\lesssim$ log n$_{H}$ $\lesssim$ 12.00, respectively.

The fits carried out without an \oil\ VBC indicate that there is no empirical requirement for significant VBC emission to obtain acceptable residuals. However the photoionization models indicate that some \oil\ VBC contribution could be present. Also the shape of the CaT + \oil\ blend bears a striking similarity to the \hb\ profile in sources where the  \hb\ VBC is unambiguously present. The S/N of our data are not high enough to constrain any VBC emission in  $\chi^2_\nu$ solutions. We consequently decided to consider a  model involving a single Gaussian for \oil\ with eventual consideration for a possible  VBC contribution.  

\subsection{\ca2\ triplet and \oil\ (and \fe2) in the 4DE1 context}
\label{ca2-fe2}

Several important aspects emerge from the previous analysis: in the context of a photoionized gas, CaT\ can be mainly emitted in a low--ionization, dense medium of high column density. In the last decades, there has been considerable progress in  modelling \fe2\ emission in the context of photoionization. As mentioned, the same physical conditions suited for \ca2\ are also suggested for \fe2\ emission.  The relation between \rfe\ and \ca2/\hb\ implies that the same physical mechanisms should be operating in the emission of both \fe2\ and \ca2.

The present study confirms that \ca2\ and \fe2\ emissions are correlated even in very high luminosity objects (Figure \ref{fig:ratio_fe+ca}). Figure \ref{fig:fwhm} shows that our sample and the low--luminosity sample of Persson partly overlap and smoothly merge. This is another confirmation that the 4DE1 correlations involving low--ionization emission line ratios are also orthogonal to luminosity.

The 4DE1 sequence is, at least in part, a sequence of increasing optical \fe2\ prominence \citep[and references therein]{MAR09}, and our sample contains mainly Pop. B sources; the \ca2\ detection confirms that a dense and low--ionization region has to be present also in Pop. B sources. Therefore, the change along the 4DE1 sequence should be ascribed to a systematic variation contribution of low--ionization region: from a low contribution to \fe2\ and \ca2\ emission in extreme Pop. B sources, to a maximum contribution in extreme Pop. A sources \citep{NEG12}.  

Recent results on the reverberation time lag measured for three Seyfert 1 galaxies (NGC 4593, Mrk 1511 and Akn 564) strongly support the conclusion that \fe2\ emission originates in photoionized gas located predominantly in the outer portion of the BLR \citep{SHA12, BAR13}. Our analysis indicates that CaT\ and  \oil\ can be interpreted as emission from photoionizated gas, with CaT strongly favoured at low--ionization and high density. The conditions we suggest are similar to those identified for \fe2\ \citep{BRU08, VER99, SIG03, SIG04}. Therefore, photoionization models are probably  appropriate at least in moderate \fe2\ emitters. Refinements in modelling \fe2\ emission are needed to reconcile observations with theory, in order to reproduce the observed multiplet ratios and specially the \fe2$_{UV}$/\fe2$_{opt}$ \citep{SAM11}.

Can also the strongest \fe2\ emitters be explained in a pure photoionization scheme?  According to \citet{COL86} and \citet{JOL87}, a different source of heating is needed. {The strongest \fe2\ emitters show continuity with fainter \fe2\ emitters in the 4DE1 sequence \citep{MARSUL12,ZAM10}.} By Occam's razor, invoking an additional mechanism should be avoided if no discontinuity is observed. {By virtue of the correlation in Figure \ref{fig:ratio_fe+ca}, strong CaT emitter  should be also strong \fe2\ emitters. If $\log$ CaT/\hb\ $\approx$ --0.3,   we expect \rfe\ $\approx 2$. }The results of this paper indicate that  the strongest observed CaT can be accounted for in a photoionization scenario, suggesting that also strong \fe2\ emitters with \rfe\ $\lesssim$ 2 could be explained by photoionization, if CaT is a valid tracer of the \fe2\ emitting gas.

\subsection{Implications for Star Formation}
\label{alphafe}

We pointed out (\S \ref{lumz}) a possible increase of W(CaT) with luminosity. This trend is prone to selection effects and should be viewed with caution. The CaT/\feiiq\ ratio is, on average, a factor $\approx$ 1.7 larger in our sample than in \citet{PER88}. We regard CaT/\feiiq\ as a reliable measurement. It is unlikely that the difference stems from different analysis techniques. In fact our measures are corrected for \fe2\ and high order Paschen lines, while Persson's were not. There could be several other  factors affecting the  CaT/\feiiq\ ratio, for example effects associated with microturbulence and spectral energy distribution  \citep{VER03,MAT07}. The effect of microturbulence on UV \fe2\ could be especially relevant if  $v_\mathrm{turb}$\ is as high as 100 \kms. Its effect on the optical \fe2\ emission could be much less, as tested by  {\sc cloudy} simulations for plausible values of $U$\ and \nh\ ($-2.5$, $12.00$). The possibility of a softer SED has been checked  by comparing a simulation with 
the \citet{MATFER87} SED  and with the NLSy1 SED of \citet{MAR14}. The effect of a softer X--ray spectrum is relatively modest, yielding an $\approx$ 25\%\ decrease in  CaT/\feiiq\ ratio at a fixed $U$. This last effect may influence our data since Persson sources are mainly NLSy1 while our sample includes mainly Pop. B, but is unable  to fully  account for the difference between two samples. 

If 1) CaT/\feiiq\ traces the calcium abundance as well as the abundance ratio of calcium and iron and 2) if calcium abundance  scales with other $\alpha $ elements, then a CaT/\feiiq\ increase at higher $z$\ could be associated with circumnuclear star formation or host galaxy evolution because the ratio [$\alpha$/Fe] represents a sort of  ``chemical clock"  with [$\alpha$/Fe] higher at earlier cosmic times \citep[e.g.,][and references therein]{MATT12}. An overabundance, relative to Fe, of $\alpha$\ elements such as calcium is expected if there has been enrichment from  a recent burst of star formation: the $\alpha$ elements are produced through core collapse supernovae, on timescales $\la 3 \cdot 10^7$ yr, while iron is due to type Ia supernov\ae\ associated with much longer timescales, $\sim 10^9$\ yr \citep{WYS88,MAT03}. The sources with largest W(CaT) in the \citet{PER88} sample (e.g. the ultra--luminous IR galaxy (ULIRG) Mrk 231, \citealt[][]{BRA04}) are also those known to be associated with strong star 
formation and large \rfe. These low--$z$\ sources however show CaT/\feiiq\ ratio that is close to the average of the low--$z$ sample.  In the simplest scenario of a circumnuclear burst of star formation, enhanced CaT/\feiiq\ ratio may imply that star formation in intermediate--$z$ quasars has been sustained over a shorter timescale than in the case of the large \rfe\ sources in the low--$z$\ sample or, more likely, that massive stars are  continuously forming. It is generally accepted that black hole growth and star formation are closely related \citep[e.g.][ for a review]{SAN10,kormendyho13}.  Therefore, the higher CaT/\feiiq\ ratio observed in the ISAAC sample  could possibly reflect  the coevolution of black hole and host galaxy (in the framework of massive black hole formation and bulge growth after a wet merger) that is expected to occur   at intermediate and high $z$\  and to ultimately power the most luminous quasars. 
 
\subsection{A possible geometry of the BLR}

The narrower line width of \fe2\ and \mgii\ indicates that the distance from the central continuum source could be larger for these lines than for the Balmer lines \citep{SUL06, MAR13a}, if the velocity field associated with line broadening is predominantly virial. According to the physical condition inference from the CaT, a natural site of emission for LILs is the accretion disk that provides a high column density environment \citep{COL87,COL88, FERPER89}. Physical conditions may not be strictly the same for \oi\ and \ca2\ emission. \oi\ is more likely emitted in a region with physical conditions similar to \hb\ (\S \ref{pho}).

For example, we may envision a configuration of plane parallel distribution of BLR clouds above and below the accretion disk. In this configuration the clouds dynamics is dominated by gravity and the emitting gas motion is virialized. Some of these  clouds might be at the same distance from the ionization source as some accretion disk regions, and therefore they may share the same dynamics. 

{The \ca2\ and \oi\ emitting regions should share a similar ionization status since the two ions have  similar ionization potentials. This (along with the kinematical similarity) lead to the  interpretation that the CaT and \oil\  are emitted from the physically overlapping gas. However, they do not necessarily share exactly the same  density and the ionization status. From Figure \ref{fig:sim} we see that the observed \oil/\hb\ ratio leads to   poor constraints on \nh\ and $U$. If $r$\ is similar because of similar kinematics, then there is a large range in \nh\ and $U$    that will be consistent with the observed   \oil/\hb\ keeping the ionizing photon flux the same for the two lines.} { This indicates that  \oil\ and CaT may have a similar FWHM even if they are emitted in somewhat different conditions.}  A possible model involves an illuminated accretion disk in vertical hydrostatic equilibrium at each radius: \ca2\ may arise with higher efficiency than \oil\ from the deepest regions of the disk. 

The geometry of the BLR is likely dependent on the source location in the 4DE1 sequence. The 4DE1 is probably a sequence of Eddington ratio, which decreases from Pop. A to Pop. B \citep{MAR01}. Non gravitational forces such as radiation pressure   are expected to be operating at maximum strength in extreme Pop. A sources \citep{MAR08, NET10, NEG13}. Among Pop. B sources, we frequently observe strong redward asymmetries in the \hb\ profiles, and they are very prominent in  the spectra of our sources. The physical origin of these features is unclear (e.g \citealt{PUN10}), and may be associated with a non--Doppler process.  If redward  asymmetries are associated with non--virial gas motions, they may be seen as  perturbations  whose amplitude  increases toward the line base of the \hb\ profile, and thus toward the central continuum source. In the profile interpretation we adopt, we isolate  a VBC  in \hb\ profiles that should represent this innermost part of the Hydrogen emitting region, the VBLR.  The VBLR 
cannot  emit \fe2\ or \ca2\ because ionization level is too high.

{As mentioned, however, Hydrogen and Oxygen have very similar ionization potentials { but not necessarily the same ionization status}, and \oil\ and \hb\ could be both produced in the inner region exposed to a strong radiation field} Some \oil\ emission could be associated with the VBLR. \oil\ emission could be favoured also in an accretion disk plus wind scenario, if the innermost regions of line emission are the densest and of highest optical depth \citep{FLO12}. The wind/disk scenario is also  consistent with the idea of a mostly virial BLR for Pop. B, in which an ionization gradient with distance is responsible for differences in line profiles and differences in response times  found by reverberation mapping (the BC/VBC can be seen as a crude phenomenological model  of such ionization gradient).  It remains to be seen how much \oil\ can be produced in the VBLR.  Theoretical profiles for the different lines will be computed in a companion paper in order to ascertain how strong an \oil\ VBC might be. 

\section{Conclusion}
\label{conclusion}

We have analyzed a set of \ca2\ IR triplet + \oil\ observations for a sample of high--$z$\ luminous quasars, and reported the detection of CaT emission in all sources with exception of two. We carried out multicomponent fits that included detailed modelling of all emission features -- most notably \fe2\ and high order Paschen lines -- in the range 8000 -- 9500 \AA. The fits allowed us to retrieve accurate line widths and fluxes of CaT and \oil. The new measurements indicate that \ca2\ and optical \fe2\ emission are likely closely related.  A possible systematic difference  in the CaT/\feiiq\ ratio (which may also account for an apparent correlation of W(CaT) with $z$ and $L$) has been found and indicates that recent star formation may be driving gas enrichment in  the intermediate $z$--sample.

We present a preliminary interpretation of our measurements within a photoionization context. Photoionization models show that \ca2\ triplet and \oil\ can be emitted in regions with similar physical conditions, although there is a range of ionization parameter and density where only \oil\ emission is possible and CaT\ emission is negligible. We propose a scenario where the geometry of the BLR consists of plane parallel distributions of clouds above and below an accretion disk. In this configuration cloud dynamics is dominated by gravity. Some of these clouds could be at the same distance from the ionization source as high density regions within the accretion disk. The clouds and the emitting regions of the disk would share the same dynamics (same FWHM) but differ in  physical conditions. Much CaT emission could come from the disk, while \oil\ from less dense gas located, however, at the same distance. 

In this paper we have followed a predominantly empirical approach with a detailed description of data analysis. In a companion paper we will present additional data along with a study of the ionization mechanism associated with \ca2\ triplet and \oil\ emission. \\

The authors thank reviewer for the useful comments to improve this manuscript. 
D. Dultzin acknowledges support from Grant IN107313, PAPIIT UNAM. 
M. L. Mart\'{i}nez--Aldama acknowledges support from a CONACYT scholarship.

\clearpage

\bibliographystyle{apj}
\bibliography{bibliografia}

\begin{thebibliography}{106}
\expandafter\ifx\csname natexlab\endcsname\relax\def\natexlab#1{#1}\fi

\bibitem[{{Ammons} {et~al.}(2006){Ammons}, {Robinson}, {Strader}, {Laughlin},
  {Fischer}, \& {Wolf}}]{AMM06}
{Ammons}, S.~M., {Robinson}, S.~E., {Strader}, J., {Laughlin}, G., {Fischer},
  D., \& {Wolf}, A. 2006, \apj, 638, 1004

\bibitem[{{Barth} {et~al.}(2013){Barth}, {Pancoast}, {Bennert}, {Brewer},
  {Canalizo}, {Filippenko}, {Gates}, {Greene}, {Li}, {Malkan}, {Sand}, {Stern},
  {Treu}, {Woo}, {Assef}, {Bae}, {Buehler}, {Cenko}, {Clubb}, {Cooper},
  {Diamond-Stanic}, {H{\"o}nig}, {Joner}, {Laney}, {Lazarova}, {Nierenberg},
  {Silverman}, {Tollerud}, \& {Walsh}}]{BAR13}
{Barth}, A.~J., {et~al.} 2013, \apj, 769, 128

\bibitem[{{Bentz} {et~al.}(2013){Bentz}, {Denney}, {Grier}, {Barth},
  {Peterson}, {Vestergaard}, {Bennert}, {Canalizo}, {De Rosa}, {Filippenko},
  {Gates}, {Greene}, {Li}, {Malkan}, {Pogge}, {Stern}, {Treu}, \&
  {Woo}}]{BEN13}
{Bentz}, M.~C., {et~al.} 2013, \apj, 767, 149

\bibitem[{{Bevington} \& {Robinson}(2003)}]{BEV03}
{Bevington}, P.~R., \& {Robinson}, D.~K. 2003, {Data reduction and error
  analysis for the physical sciences} (McGraw-Hill)

\bibitem[{{Boroson}(2002)}]{BOR02}
{Boroson}, T.~A. 2002, \apj, 565, 78

\bibitem[{{Boroson} \& {Green}(1992)}]{BOR92}
{Boroson}, T.~A., \& {Green}, R.~F. 1992, \apjs, 80, 109

\bibitem[{{Braito} {et~al.}(2004){Braito}, {Della Ceca}, {Piconcelli},
  {Severgnini}, {Bassani}, {Cappi}, {Franceschini}, {Iwasawa}, {Malaguti},
  {Marziani}, {Palumbo}, {Persic}, {Risaliti}, \& {Salvati}}]{BRA04}
{Braito}, V., {et~al.} 2004, \aap, 420, 79

\bibitem[{{Bressan} {et~al.}(1998){Bressan}, {Granato}, \& {Silva}}]{BGS98}
{Bressan}, A., {Granato}, G.~L., \& {Silva}, L. 1998, \aap, 332, 135

\bibitem[{{Bressan} {et~al.}(2012){Bressan}, {Marigo}, {Girardi}, {Salasnich},
  {Dal Cero}, {Rubele}, \& {Nanni}}]{BRE12}
{Bressan}, A., {Marigo}, P., {Girardi}, L., {Salasnich}, B., {Dal Cero}, C.,
  {Rubele}, S., \& {Nanni}, A. 2012, \mnras, 427, 127

\bibitem[{{Bruhweiler} \& {Verner}(2008)}]{BRU08}
{Bruhweiler}, F., \& {Verner}, E. 2008, \apj, 675, 83

\bibitem[{{Cenarro} {et~al.}(2003){Cenarro}, {Gorgas}, {Vazdekis}, {Cardiel},
  \& {Peletier}}]{CEN03}
{Cenarro}, A.~J., {Gorgas}, J., {Vazdekis}, A., {Cardiel}, N., \& {Peletier},
  R.~F. 2003, \mnras, 339, L12

\bibitem[{{Collin-Souffrin}(1986)}]{COL86}
{Collin-Souffrin}, S. 1986, \aap, 166, 115

\bibitem[{{Collin-Souffrin}(1987)}]{COL87}
---. 1987, \aap, 179, 60

\bibitem[{{Collin-Souffrin} {et~al.}(1988){Collin-Souffrin}, {Dyson},
  {McDowell}, \& {Perry}}]{COL88}
{Collin-Souffrin}, S., {Dyson}, J.~E., {McDowell}, J.~C., \& {Perry}, J.~J.
  1988, \mnras, 232, 539

\bibitem[{{Collin-Souffrin} {et~al.}(1980){Collin-Souffrin}, {Joly}, {Dumont},
  \& {Heidmann}}]{COL80}
{Collin-Souffrin}, S., {Joly}, M., {Dumont}, S., \& {Heidmann}, N. 1980, \aap,
  83, 190

\bibitem[{{Drake} {et~al.}(2009){Drake}, {Djorgovski}, {Mahabal}, {Beshore},
  {Larson}, {Graham}, {Williams}, {Christensen}, {Catelan}, {Boattini},
  {Gibbs}, {Hill}, \& {Kowalski}}]{DRA09}
{Drake}, A.~J., {et~al.} 2009, \apj, 696, 870

\bibitem[{{Dultzin-Hacyan} {et~al.}(1999){Dultzin-Hacyan}, {Taniguchi}, \&
  {Uranga}}]{DUL99}
{Dultzin-Hacyan}, D., {Taniguchi}, Y., \& {Uranga}, L. 1999, in Astronomical
  Society of the Pacific Conference Series, Vol. 175, Structure and Kinematics
  of Quasar Broad Line Regions, ed. C.~M. {Gaskell}, W.~N. {Brandt},
  M.~{Dietrich}, D.~{Dultzin-Hacyan}, \& M.~{Eracleous}, 303

\bibitem[{{Falomo} {et~al.}(2004){Falomo}, {Kotilainen}, {Pagani}, {Scarpa}, \&
  {Treves}}]{FAL04}
{Falomo}, R., {Kotilainen}, J.~K., {Pagani}, C., {Scarpa}, R., \& {Treves}, A.
  2004, \apj, 604, 495

\bibitem[{{Ferland} {et~al.}(1998){Ferland}, {Korista}, {Verner}, {Ferguson},
  {Kingdon}, \& {Verner}}]{FER98}
{Ferland}, G.~J., {Korista}, K.~T., {Verner}, D.~A., {Ferguson}, J.~W.,
  {Kingdon}, J.~B., \& {Verner}, E.~M. 1998, \pasp, 110, 761

\bibitem[{{Ferland} \& {Persson}(1989)}]{FERPER89}
{Ferland}, G.~J., \& {Persson}, S.~E. 1989, \apj, 347, 656

\bibitem[{{Ferland} {et~al.}(2013){Ferland}, {Porter}, {van Hoof}, {Williams},
  {Abel}, {Lykins}, {Shaw}, {Henney}, \& {Stancil}}]{FER13}
{Ferland}, G.~J., {et~al.} 2013, Revista Mexicana de Astronomia y Astrofisica,
  49, 137

\bibitem[{{Flohic} {et~al.}(2012){Flohic}, {Eracleous}, \&
  {Bogdanovi{\'c}}}]{FLO12}
{Flohic}, H.~M.~L.~G., {Eracleous}, M., \& {Bogdanovi{\'c}}, T. 2012, \apj,
  753, 133

\bibitem[{{Floyd} {et~al.}(2013){Floyd}, {Dunlop}, {Kukula}, {Brown}, {McLure},
  {Baum}, \& {O'Dea}}]{FLOY13}
{Floyd}, D.~J.~E., {Dunlop}, J.~S., {Kukula}, M.~J., {Brown}, M.~J.~I.,
  {McLure}, R.~J., {Baum}, S.~A., \& {O'Dea}, C.~P. 2013, \mnras, 429, 2

\bibitem[{{Garcia-Rissmann} {et~al.}(2012){Garcia-Rissmann},
  {Rodr{\'{\i}}guez-Ardila}, {Sigut}, \& {Pradhan}}]{GAR12}
{Garcia-Rissmann}, A., {Rodr{\'{\i}}guez-Ardila}, A., {Sigut}, T.~A.~A., \&
  {Pradhan}, A.~K. 2012, \apj, 751, 7

\bibitem[{{Grandi}(1980)}]{GRA80}
{Grandi}, S.~A. 1980, \apj, 238, 10

\bibitem[{{Joly}(1987)}]{JOL87}
{Joly}, M. 1987, \aap, 184, 33

\bibitem[{{Joly}(1989)}]{JOL89}
---. 1989, \aap, 208, 47

\bibitem[{{Korista} \& {Goad}(2001)}]{KOR01}
{Korista}, K.~T., \& {Goad}, M.~R. 2001, \apj, 553, 695

\bibitem[{{Kormendy} \& {Ho}(2013)}]{kormendyho13}
{Kormendy}, J., \& {Ho}, L.~C. 2013, \araa, 51, 511

\bibitem[{{Kotilainen} {et~al.}(2007){Kotilainen}, {Falomo}, {Labita},
  {Treves}, \& {Uslenghi}}]{KOT07}
{Kotilainen}, J.~K., {Falomo}, R., {Labita}, M., {Treves}, A., \& {Uslenghi},
  M. 2007, \apj, 660, 1039

\bibitem[{{Kriss}(1994)}]{KRI94}
{Kriss}, G. 1994, in Astronomical Society of the Pacific Conference Series,
  Vol.~61, Astronomical Data Analysis Software and Systems III, ed. D.~R.
  {Crabtree}, R.~J. {Hanisch}, \& J.~{Barnes}, 437

\bibitem[{{Kuehn} {et~al.}(2008){Kuehn}, {Baldwin}, {Peterson}, \&
  {Korista}}]{KUE08}
{Kuehn}, C.~A., {Baldwin}, J.~A., {Peterson}, B.~M., \& {Korista}, K.~T. 2008,
  \apj, 673, 69

\bibitem[{{Kukula} {et~al.}(2001){Kukula}, {Dunlop}, {McLure}, {Miller},
  {Percival}, {Baum}, \& {O'Dea}}]{KUK01}
{Kukula}, M.~J., {Dunlop}, J.~S., {McLure}, R.~J., {Miller}, L., {Percival},
  W.~J., {Baum}, S.~A., \& {O'Dea}, C.~P. 2001, \mnras, 326, 1533

\bibitem[{{Landt} {et~al.}(2008){Landt}, {Bentz}, {Ward}, {Elvis}, {Peterson},
  {Korista}, \& {Karovska}}]{LAN08}
{Landt}, H., {Bentz}, M.~C., {Ward}, M.~J., {Elvis}, M., {Peterson}, B.~M.,
  {Korista}, K.~T., \& {Karovska}, M. 2008, \apjs, 174, 282

\bibitem[{{Landt} {et~al.}(2011){Landt}, {Elvis}, {Ward}, {Bentz}, {Korista},
  \& {Karovska}}]{LAN11}
{Landt}, H., {Elvis}, M., {Ward}, M.~J., {Bentz}, M.~C., {Korista}, K.~T., \&
  {Karovska}, M. 2011, \mnras, 414, 218

\bibitem[{{Magorrian} {et~al.}(1998){Magorrian}, {Tremaine}, {Richstone},
  {Bender}, {Bower}, {Dressler}, {Faber}, {Gebhardt}, {Green}, {Grillmair},
  {Kormendy}, \& {Lauer}}]{MAG98}
{Magorrian}, J., {et~al.} 1998, \aj, 115, 2285

\bibitem[{{Malkan}(1983)}]{MAL83}
{Malkan}, M.~A. 1983, \apj, 268, 582

\bibitem[{{Malkan} \& {Sargent}(1982)}]{MAL82}
{Malkan}, M.~A., \& {Sargent}, W.~L.~W. 1982, \apj, 254, 22

\bibitem[{{Marziani} {et~al.}(2013{\natexlab{a}}){Marziani}, {Martnez-Aldama},
  {Dultzin}, \& {Sulentic}}]{MAR13b}
{Marziani}, P., {Martnez-Aldama}, M.~L., {Dultzin}, D., \& {Sulentic}, J.~W.
  2013{\natexlab{a}}, The Astronomical Review, 8, 4

\bibitem[{{Marziani} \& {Sulentic}(2012)}]{MARSUL12}
{Marziani}, P., \& {Sulentic}, J.~W. 2012, The Astronomical Review, 7, 33

\bibitem[{{Marziani} \& {Sulentic}(2014)}]{MAR14}
---. 2014, \mnras, 442, 1211

\bibitem[{{Marziani} {et~al.}(2008){Marziani}, {Sulentic}, \&
  {Dultzin}}]{MAR08}
{Marziani}, P., {Sulentic}, J.~W., \& {Dultzin}, D. 2008, in Revista Mexicana
  de Astronomia y Astrofisica Conference Series, Vol.~32, Revista Mexicana de
  Astronom{\'{\i}}a y Astrof{\'{\i}}sica Conference Series, 69--73

\bibitem[{{Marziani} {et~al.}(2010){Marziani}, {Sulentic}, {Negrete},
  {Dultzin}, {Zamfir}, \& {Bachev}}]{MARZ10}
{Marziani}, P., {Sulentic}, J.~W., {Negrete}, C.~A., {Dultzin}, D., {Zamfir},
  S., \& {Bachev}, R. 2010, \mnras, 409, 1033

\bibitem[{{Marziani} {et~al.}(2013{\natexlab{b}}){Marziani}, {Sulentic},
  {Plauchu-Frayn}, \& {del Olmo}}]{MAR13a}
{Marziani}, P., {Sulentic}, J.~W., {Plauchu-Frayn}, I., \& {del Olmo}, A.
  2013{\natexlab{b}}, \aap, 555, A89

\bibitem[{{Marziani} {et~al.}(2009){Marziani}, {Sulentic}, {Stirpe}, {Zamfir},
  \& {Calvani}}]{MAR09}
{Marziani}, P., {Sulentic}, J.~W., {Stirpe}, G.~M., {Zamfir}, S., \& {Calvani},
  M. 2009, \aap, 495, 83

\bibitem[{{Marziani} {et~al.}(2003{\natexlab{a}}){Marziani}, {Sulentic},
  {Zamanov}, {Calvani}, {Dultzin-Hacyan}, {Bachev}, \& {Zwitter}}]{MAR03a}
{Marziani}, P., {Sulentic}, J.~W., {Zamanov}, R., {Calvani}, M.,
  {Dultzin-Hacyan}, D., {Bachev}, R., \& {Zwitter}, T. 2003{\natexlab{a}},
  \apjs, 145, 199

\bibitem[{{Marziani} {et~al.}(2001){Marziani}, {Sulentic}, {Zwitter},
  {Dultzin-Hacyan}, \& {Calvani}}]{MAR01}
{Marziani}, P., {Sulentic}, J.~W., {Zwitter}, T., {Dultzin-Hacyan}, D., \&
  {Calvani}, M. 2001, \apj, 558, 553

\bibitem[{{Marziani} {et~al.}(2003{\natexlab{b}}){Marziani}, {Zamanov},
  {Sulentic}, \& {Calvani}}]{MAR03b}
{Marziani}, P., {Zamanov}, R.~K., {Sulentic}, J.~W., \& {Calvani}, M.
  2003{\natexlab{b}}, \mnras, 345, 1133

\bibitem[{{Mathews} \& {Ferland}(1987)}]{MATFER87}
{Mathews}, W.~G., \& {Ferland}, G.~J. 1987, \apj, 323, 456

\bibitem[{{Matsuoka} {et~al.}(2007){Matsuoka}, {Oyabu}, {Tsuzuki}, \&
  {Kawara}}]{MAT07}
{Matsuoka}, Y., {Oyabu}, S., {Tsuzuki}, Y., \& {Kawara}, K. 2007, \apj, 663,
  781

\bibitem[{{Matsuoka} {et~al.}(2005){Matsuoka}, {Oyabu}, {Tsuzuki}, {Kawara}, \&
  {Yoshii}}]{MAT05}
{Matsuoka}, Y., {Oyabu}, S., {Tsuzuki}, Y., {Kawara}, K., \& {Yoshii}, Y. 2005,
  \pasj, 57, 563

\bibitem[{{Matsuoka} {et~al.}(2008){Matsuoka}, {Peterson}, {Oyabu}, {Kawara},
  {Asami}, {Sameshima}, {Ienaka}, {Nagayama}, \& {Tamura}}]{MAT08}
{Matsuoka}, Y., {et~al.} 2008, \apj, 685, 767

\bibitem[{{Matteucci}(2003)}]{MAT03}
{Matteucci}, F. 2003, \apss, 284, 539

\bibitem[{{Matteucci}(2012)}]{MATT12}
---. 2012, {Chemical Evolution of Galaxies} (Springer-Verlag Berlin Heidelberg)

\bibitem[{{Merloni} {et~al.}(2010){Merloni}, {Bongiorno}, {Bolzonella},
  {Brusa}, {Civano}, {Comastri}, {Elvis}, {Fiore}, {Gilli}, {Hao}, {Jahnke},
  {Koekemoer}, {Lusso}, {Mainieri}, {Mignoli}, {Miyaji}, {Renzini}, {Salvato},
  {Silverman}, {Trump}, {Vignali}, {Zamorani}, {Capak}, {Lilly}, {Sanders},
  {Taniguchi}, {Bardelli}, {Carollo}, {Caputi}, {Contini}, {Coppa}, {Cucciati},
  {de la Torre}, {de Ravel}, {Franzetti}, {Garilli}, {Hasinger}, {Impey},
  {Iovino}, {Iwasawa}, {Kampczyk}, {Kneib}, {Knobel}, {Kova{\v c}},
  {Lamareille}, {Le Borgne}, {Le Brun}, {Le F{\`e}vre}, {Maier}, {Pello},
  {Peng}, {Perez Montero}, {Ricciardelli}, {Scodeggio}, {Tanaka}, {Tasca},
  {Tresse}, {Vergani}, \& {Zucca}}]{MER10}
{Merloni}, A., {et~al.} 2010, \apj, 708, 137

\bibitem[{{Monet} {et~al.}(2003){Monet}, {Levine}, {Canzian}, {Ables}, {Bird},
  {Dahn}, {Guetter}, {Harris}, {Henden}, {Leggett}, {Levison}, {Luginbuhl},
  {Martini}, {Monet}, {Munn}, {Pier}, {Rhodes}, {Riepe}, {Sell}, {Stone},
  {Vrba}, {Walker}, {Westerhout}, {Brucato}, {Reid}, {Schoening}, {Hartley},
  {Read}, \& {Tritton}}]{MON03}
{Monet}, D.~G., {et~al.} 2003, \aj, 125, 984

\bibitem[{{Moorwood} {et~al.}(1998){Moorwood}, {Cuby}, {Biereichel}, {Brynnel},
  {Delabre}, {Devillard}, {van Dijsseldonk}, {Finger}, {Gemperlein},
  {Gilmozzi}, {Herlin}, {Huster}, {Knudstrup}, {Lidman}, {Lizon}, {Mehrgan},
  {Meyer}, {Nicolini}, {Petr}, {Spyromilio}, \& {Stegmeier}}]{MOO98}
{Moorwood}, A., {et~al.} 1998, The Messenger, 94, 7

\bibitem[{{Negrete} {et~al.}(2012){Negrete}, {Dultzin}, {Marziani}, \&
  {Sulentic}}]{NEG12}
{Negrete}, C.~A., {Dultzin}, D., {Marziani}, P., \& {Sulentic}, J.~W. 2012,
  \apj, 757, 62

\bibitem[{{Negrete} {et~al.}(2013){Negrete}, {Dultzin}, {Marziani}, \&
  {Sulentic}}]{NEG13}
---. 2013, \apj, 771, 31

\bibitem[{{Netzer}(2013)}]{NET13}
{Netzer}, H. 2013, {The Physics and Evolution of Active Galactic Nuclei}
  (Cambridge University Press)

\bibitem[{{Netzer} \& {Marziani}(2010)}]{NET10}
{Netzer}, H., \& {Marziani}, P. 2010, \apj, 724, 318

\bibitem[{{Osterbrock} \& {Ferland}(2006)}]{OF06}
{Osterbrock}, D.~E., \& {Ferland}, G.~J. 2006, {Astrophysics of gaseous nebulae
  and active galactic nuclei} (University Science Books)

\bibitem[{{Persson}(1988)}]{PER88}
{Persson}, S.~E. 1988, \apj, 330, 751

\bibitem[{{Peterson} \& {Wandel}(1999)}]{PET99}
{Peterson}, B.~M., \& {Wandel}, A. 1999, \apjl, 521, L95

\bibitem[{{Pickles}(1998)}]{PICK98}
{Pickles}, A.~J. 1998, \pasp, 110, 863

\bibitem[{{Press} {et~al.}(2002){Press}, {Teukolsky}, {Vetterling}, \&
  {Flannery}}]{PRESS}
{Press}, W.~H., {Teukolsky}, S.~A., {Vetterling}, W.~T., \& {Flannery}, B.~P.
  2002, {Numerical recipes in C++ : the art of scientific computing} (Cambridge
  University)

\bibitem[{{Punsly} \& {Zhang}(2010)}]{PUN10}
{Punsly}, B., \& {Zhang}, S. 2010, \apj, 725, 1928

\bibitem[{{Ranade} {et~al.}(2004){Ranade}, {Gupta}, {Ashok}, \&
  {Singh}}]{RAN04}
{Ranade}, A., {Gupta}, R., {Ashok}, N.~M., \& {Singh}, H.~P. 2004, Bulletin of
  the Astronomical Society of India, 32, 311

\bibitem[{{Ranade} {et~al.}(2007){Ranade}, {Singh}, {Gupta}, \&
  {Ashok}}]{RAN07}
{Ranade}, A.~C., {Singh}, H.~P., {Gupta}, R., \& {Ashok}, N.~M. 2007, ArXiv
  e-prints

\bibitem[{{Rayner} {et~al.}(2009){Rayner}, {Cushing}, \& {Vacca}}]{RAY09}
{Rayner}, J.~T., {Cushing}, M.~C., \& {Vacca}, W.~D. 2009, \apjs, 185, 289

\bibitem[{{Richards} {et~al.}(2011){Richards}, {Kruczek}, {Gallagher}, {Hall},
  {Hewett}, {Leighly}, {Deo}, {Kratzer}, \& {Shen}}]{RICH11}
{Richards}, G.~T., {et~al.} 2011, \aj, 141, 167

\bibitem[{{Rodr{\'{\i}}guez-Ardila}
  {et~al.}(2002{\natexlab{a}}){Rodr{\'{\i}}guez-Ardila}, {Viegas}, {Pastoriza},
  \& {Prato}}]{ROD02}
{Rodr{\'{\i}}guez-Ardila}, A., {Viegas}, S.~M., {Pastoriza}, M.~G., \& {Prato},
  L. 2002{\natexlab{a}}, \apj, 565, 140

\bibitem[{{Rodr{\'{\i}}guez-Ardila}
  {et~al.}(2002{\natexlab{b}}){Rodr{\'{\i}}guez-Ardila}, {Viegas}, {Pastoriza},
  {Prato}, \& {Donzelli}}]{ROD02b}
{Rodr{\'{\i}}guez-Ardila}, A., {Viegas}, S.~M., {Pastoriza}, M.~G., {Prato},
  L., \& {Donzelli}, C.~J. 2002{\natexlab{b}}, \apj, 572, 94

\bibitem[{{Rudy} {et~al.}(2000){Rudy}, {Mazuk}, {Puetter}, \&
  {Hamann}}]{RUD2000}
{Rudy}, R.~J., {Mazuk}, S., {Puetter}, R.~C., \& {Hamann}, F. 2000, \apj, 539,
  166

\bibitem[{{Sameshima} {et~al.}(2011){Sameshima}, {Kawara}, {Matsuoka}, {Oyabu},
  {Asami}, \& {Ienaka}}]{SAM11}
{Sameshima}, H., {Kawara}, K., {Matsuoka}, Y., {Oyabu}, S., {Asami}, N., \&
  {Ienaka}, N. 2011, \mnras, 410, 1018

\bibitem[{{S{\'a}nchez} {et~al.}(2004){S{\'a}nchez}, {Jahnke}, {Wisotzki},
  {McIntosh}, {Bell}, {Barden}, {Beckwith}, {Borch}, {Caldwell},
  {H{\"a}ussler}, {Jogee}, {Meisenheimer}, {Peng}, {Rix}, {Somerville}, \&
  {Wolf}}]{SAN04}
{S{\'a}nchez}, S.~F., {et~al.} 2004, \apj, 614, 586

\bibitem[{{S{\'a}nchez-Bl{\'a}zquez}
  {et~al.}(2006{\natexlab{a}}){S{\'a}nchez-Bl{\'a}zquez}, {Gorgas}, \&
  {Cardiel}}]{SAN06c}
{S{\'a}nchez-Bl{\'a}zquez}, P., {Gorgas}, J., \& {Cardiel}, N.
  2006{\natexlab{a}}, \aap, 457, 823

\bibitem[{{S{\'a}nchez-Bl{\'a}zquez}
  {et~al.}(2006{\natexlab{b}}){S{\'a}nchez-Bl{\'a}zquez}, {Gorgas}, {Cardiel},
  \& {Gonz{\'a}lez}}]{SAN06a}
{S{\'a}nchez-Bl{\'a}zquez}, P., {Gorgas}, J., {Cardiel}, N., \& {Gonz{\'a}lez},
  J.~J. 2006{\natexlab{b}}, \aap, 457, 787

\bibitem[{{S{\'a}nchez-Bl{\'a}zquez}
  {et~al.}(2006{\natexlab{c}}){S{\'a}nchez-Bl{\'a}zquez}, {Gorgas}, {Cardiel},
  \& {Gonz{\'a}lez}}]{SAN06b}
---. 2006{\natexlab{c}}, \aap, 457, 809

\bibitem[{{Sani} {et~al.}(2010){Sani}, {Lutz}, {Risaliti}, {Netzer}, {Gallo},
  {Trakhtenbrot}, {Sturm}, \& {Boller}}]{SAN10}
{Sani}, E., {Lutz}, D., {Risaliti}, G., {Netzer}, H., {Gallo}, L.~C.,
  {Trakhtenbrot}, B., {Sturm}, E., \& {Boller}, T. 2010, \mnras, 403, 1246

\bibitem[{{Schlegel} {et~al.}(1998){Schlegel}, {Finkbeiner}, \&
  {Davis}}]{SFD98}
{Schlegel}, D.~J., {Finkbeiner}, D.~P., \& {Davis}, M. 1998, \apj, 500, 525

\bibitem[{{Shapovalova} {et~al.}(2012){Shapovalova}, {Popovi{\'c}}, {Burenkov},
  {Chavushyan}, {Ili{\'c}}, {Kova{\v c}evi{\'c}}, {Kollatschny}, {Kova{\v
  c}evi{\'c}}, {Bochkarev}, {Valdes}, {Torrealba}, {Le{\'o}n-Tavares},
  {Mercado}, {Ben{\'{\i}}tez}, {Carrasco}, {Dultzin}, \& {de la
  Fuente}}]{SHA12}
{Shapovalova}, A.~I., {et~al.} 2012, \apjs, 202, 10

\bibitem[{{Shin} {et~al.}(2013){Shin}, {Woo}, {Nagao}, \& {Kim}}]{SHIN13}
{Shin}, J., {Woo}, J.-H., {Nagao}, T., \& {Kim}, S.~C. 2013, \apj, 763, 58

\bibitem[{{Sigut} \& {Pradhan}(1998)}]{SIG98}
{Sigut}, T.~A.~A., \& {Pradhan}, A.~K. 1998, \apjl, 499, L139

\bibitem[{{Sigut} \& {Pradhan}(2003)}]{SIG03}
---. 2003, \apjs, 145, 15

\bibitem[{{Sigut} {et~al.}(2004){Sigut}, {Pradhan}, \& {Nahar}}]{SIG04}
{Sigut}, T.~A.~A., {Pradhan}, A.~K., \& {Nahar}, S.~N. 2004, \apj, 611, 81

\bibitem[{{Snedden} \& {Gaskell}(2007)}]{SNE07}
{Snedden}, S.~A., \& {Gaskell}, C.~M. 2007, \apj, 669, 126

\bibitem[{{Spolaor} {et~al.}(2009){Spolaor}, {Proctor}, {Forbes}, \&
  {Couch}}]{SPO09}
{Spolaor}, M., {Proctor}, R.~N., {Forbes}, D.~A., \& {Couch}, W.~J. 2009,
  \apjl, 691, L138

\bibitem[{{Sulentic} {et~al.}(2007){Sulentic}, {Bachev}, {Marziani}, {Negrete},
  \& {Dultzin}}]{SUL07}
{Sulentic}, J.~W., {Bachev}, R., {Marziani}, P., {Negrete}, C.~A., \&
  {Dultzin}, D. 2007, \apj, 666, 757

\bibitem[{{Sulentic} {et~al.}(2014){Sulentic}, {Marziani}, {del Olmo},
  {Dultzin}, {Perea}, \& {Alenka Negrete}}]{SUL14}
{Sulentic}, J.~W., {Marziani}, P., {del Olmo}, A., {Dultzin}, D., {Perea}, J.,
  \& {Alenka Negrete}, C. 2014, \aap, 570, A96

\bibitem[{{Sulentic} {et~al.}(2000{\natexlab{a}}){Sulentic}, {Marziani}, \&
  {Dultzin-Hacyan}}]{SMD2000a}
{Sulentic}, J.~W., {Marziani}, P., \& {Dultzin-Hacyan}, D. 2000{\natexlab{a}},
  \araa, 38, 521

\bibitem[{{Sulentic} {et~al.}(2002){Sulentic}, {Marziani}, {Zamanov}, {Bachev},
  {Calvani}, \& {Dultzin-Hacyan}}]{SUL02}
{Sulentic}, J.~W., {Marziani}, P., {Zamanov}, R., {Bachev}, R., {Calvani}, M.,
  \& {Dultzin-Hacyan}, D. 2002, \apjl, 566, L71

\bibitem[{{Sulentic} {et~al.}(2012){Sulentic}, {Marziani}, {Zamfir}, \&
  {Meadows}}]{SUL12}
{Sulentic}, J.~W., {Marziani}, P., {Zamfir}, S., \& {Meadows}, Z.~A. 2012,
  \apjl, 752, L7

\bibitem[{{Sulentic} {et~al.}(2006){Sulentic}, {Repetto}, {Stirpe}, {Marziani},
  {Dultzin-Hacyan}, \& {Calvani}}]{SUL06}
{Sulentic}, J.~W., {Repetto}, P., {Stirpe}, G.~M., {Marziani}, P.,
  {Dultzin-Hacyan}, D., \& {Calvani}, M. 2006, \aap, 456, 929

\bibitem[{{Sulentic} {et~al.}(2004){Sulentic}, {Stirpe}, {Marziani}, {Zamanov},
  {Calvani}, \& {Braito}}]{SUL04}
{Sulentic}, J.~W., {Stirpe}, G.~M., {Marziani}, P., {Zamanov}, R., {Calvani},
  M., \& {Braito}, V. 2004, \aap, 423, 121

\bibitem[{{Sulentic} {et~al.}(2000{\natexlab{b}}){Sulentic}, {Zwitter},
  {Marziani}, \& {Dultzin-Hacyan}}]{SUL2000}
{Sulentic}, J.~W., {Zwitter}, T., {Marziani}, P., \& {Dultzin-Hacyan}, D.
  2000{\natexlab{b}}, \apjl, 536, L5

\bibitem[{{Trager} {et~al.}(2000){Trager}, {Faber}, {Worthey}, \&
  {Gonz{\'a}lez}}]{TRA00}
{Trager}, S.~C., {Faber}, S.~M., {Worthey}, G., \& {Gonz{\'a}lez}, J.~J. 2000,
  \aj, 120, 165

\bibitem[{{van Dokkum} \& {Conroy}(2012)}]{VAN12}
{van Dokkum}, P.~G., \& {Conroy}, C. 2012, \apj, 760, 70

\bibitem[{{Verner} {et~al.}(2003){Verner}, {Bruhweiler}, {Verner}, {Johansson},
  \& {Gull}}]{VER03}
{Verner}, E., {Bruhweiler}, F., {Verner}, D., {Johansson}, S., \& {Gull}, T.
  2003, \apjl, 592, L59

\bibitem[{{Verner} {et~al.}(1999){Verner}, {Verner}, {Korista}, {Ferguson},
  {Hamann}, \& {Ferland}}]{VER99}
{Verner}, E.~M., {Verner}, D.~A., {Korista}, K.~T., {Ferguson}, J.~W.,
  {Hamann}, F., \& {Ferland}, G.~J. 1999, \apjs, 120, 101

\bibitem[{{V{\'e}ron-Cetty} {et~al.}(2001){V{\'e}ron-Cetty}, {V{\'e}ron}, \&
  {Gon{\c c}alves}}]{VER01}
{V{\'e}ron-Cetty}, M.-P., {V{\'e}ron}, P., \& {Gon{\c c}alves}, A.~C. 2001,
  \aap, 372, 730

\bibitem[{{Wang} {et~al.}(2011){Wang}, {Wang}, {Zhou}, {Liu}, {Wang}, {Yuan},
  \& {Dong}}]{WANG11}
{Wang}, H., {Wang}, T., {Zhou}, H., {Liu}, B., {Wang}, J., {Yuan}, W., \&
  {Dong}, X. 2011, \apj, 738, 85

\bibitem[{{Wilcoxon}(1945)}]{WIL45}
{Wilcoxon}, F. 1945, Biometrics Bull., 80

\bibitem[{{Wills} {et~al.}(1985){Wills}, {Netzer}, \& {Wills}}]{WIL85}
{Wills}, B.~J., {Netzer}, H., \& {Wills}, D. 1985, \apj, 288, 94

\bibitem[{{Wyse} \& {Gilmore}(1988)}]{WYS88}
{Wyse}, R.~F.~G., \& {Gilmore}, G. 1988, \aj, 95, 1404

\bibitem[{{Zamfir} {et~al.}(2010){Zamfir}, {Sulentic}, {Marziani}, \&
  {Dultzin}}]{ZAM10}
{Zamfir}, S., {Sulentic}, J.~W., {Marziani}, P., \& {Dultzin}, D. 2010, \mnras,
  403, 1759

\end{thebibliography}

\begin{sidewaystable}[h]
\setlength{\tabcolsep}{5pt}
\begin{center}
	\caption{Basic properties of sources and log of observations}
	\begin{tabular}{c c c c c c c c c c c c c}\\ \hline \hline
	\textbf{Object$^{a}$} & \textbf{z$^b$} & \textbf{m$_{B}^{c}$} & \textbf{M$_{B}^{d}$} & \textbf{log R$_{K}^{e}$} & \textbf{Population$^f$} & \textbf{Date$^g$} & \textbf{Band$^h$} & \textbf{DIT$^i$} & \textbf{N$_{exp}^{j}$} & \textbf{Seeing$^k$} & \textbf{S/N$^l$} \\ 
	\hline
	HE0005--2355  & 1.4120 & 16.9  & --27.6 & 2.56 & B1 & 07/30/10 & K & 120 & 6 & 1.054 & 20 \\ 
	HE0035--2853  & 1.6377 & 17.0 & --28.1 & $<$ 0.21 & B2 & 07/27/10 & K & 150 & 4 & 0.840 & 55 \\ 
	HE0043--2300  & 1.5402 & 17.1  & --27.9 & 2.03 & A1 & 07/07/10 & K & 120 & 6 & 1.078 & 45 \\ 
	HE0048--2804  & 0.8467 & 17.3 & --26.0 & $...$ & B1 & 07/27/10 & H & 150 & 8 & 0.863 & 35 \\ 
	HE0058--3231  & 1.5821 & 17.1 & --27.9 & $<$ 0.24 & B1 & 07/27/10 & K & 150 & 4 & 0.759 & 10 \\ 
	HE0203--4627 & 1.4381 & 17.3 & --27.5 & 2.07 & B1 & 07/02/10 & K & 180 & 6 & 1.184 & 25 \\ 
	HE0248--3628  & 1.5355 & 16.6 & --28.2 & 0.55 & A1 & 07/22/10 & K & 120 & 6 & 0.965 & 50 \\ 
	HE1349+0007  & 1.4442 & 16.8 & --28.0 & -0.18 & B1 & 04/15/10 & K & 150 & 6 & 0.749 & 20 \\
	HE1409+0101  & 1.6497 & 16.9 & --28.3 & 0.4 & B1 & 04/15/10 & K & 150 & 6 & 0.775  & 25 \\ 
	HE2147--3212  & 1.5432 & 16.8 & --28.2 & $<$ 0.14 & B2 & 06/12/10 & K & 120 & 6 & 0.847 & 20 \\ 
	HE2202--2557  & 1.5347 & 16.7 & --28.1 & 1.8 & B1 & 07/23/10 & K & 120 & 6 & 0.606 & 40 \\ 
	HE2259--5524  & 0.8549 & 17.1 & --26.1 & $...$ & A2 & 05/25/10 & H & 150 & 8 & 1.863 & 10 \\ 
	HE2340--4443  & 0.9216 & 17.1 & --26.3 & $...$ & A1 & 07/23/10 & H & 120 & 6 & 0.568 & 25 \\ 
	HE2349--3800  & 1.6040 & 17.5 & --27.4 & 1.93 & B1 & 06/12/10 & K & 120 & 6 & 0.789 & 20 \\ 
	HE2352--4010 & 1.5799 & 16.1 & --28.8 & $...$ & A1 & 07/23/10 & K & 150 & 6 & 0.639 & 35 \\ \hline
	\end{tabular}	
    \label{table:obs}   
\end{center}
	\small 
	      $^a$ Hamburg/ESO Survey coordinate name.\\
	      $^b$ Heliocentric redshift. \\
	      $^c$ Apparent  Johnson $B$\ magnitude\\
	      $^d$ Absolute $B$\ magnitude.\\
	      $^e$ Decimal logarithm of the ratio between specific flux at 6cm and 4440 \AA. References for redshift uncertainty, apparent, absolute magnitude and ratio radio--to optical: 1: \citet{SUL04}; 2: \citet{MAR09} \\
	      $^f$ Spectral according to Eigenvector 1 scheme \citep{SUL02}\\  
	      $^g$ Date of  observation.\\
	      $^h$ Photometric band of the covered spectra range. \\
	      $^i$ Detector Integration Time (DIT)  in seconds.\\
	      $^j$ Number of exposure with single exposure time equal to DIT.\\
	      $^k$ Average seeing.\\
	      $^l$ S/N (1--$\sigma$) at continuum level.\\
\end{sidewaystable}

\newpage

\begin{sidewaystable}[ht]
\scriptsize
\begin{center}
\caption{Measurements of equivalent width, flux and FWHM of H$\beta$ and Pa9 lines.}
\scalebox{0.62}[0.62]{
\begin{tabular}{ccccccccccccccc}
    \multicolumn{15}{c}{}                                                                \\
    \hline \hline
    & & \multicolumn{3}{c}{BC}& &\multicolumn{3}{c}{NC}& &\multicolumn{3}{c}{VBC} & & Full profile \\
    \cline{3-5} 
    \cline{7-9} 
    \cline{11-13}
    \cline{15-15}
    Object name& $f_{\lambda}$&  W& $F$ &FWHM& & W & $F$ & FWHM & & W & $F$ & FWHM & FWHM \\
    & (erg s$^{-1}$ cm$^{-2}$ \AA$^{-1}$) & (\AA) & (erg s$^{-1}$ cm$^{-2}$) & (\kms) & & (\AA) & (erg s$^{-1}$ cm$^{-2}$)& (\kms) & & (\AA) & (erg s$^{-1}$ cm$^{-2}$) & (\kms) & & (\kms) \\
    \cline{1-15} 
    \\
    \multicolumn{15}{c}{\hb}\\
    \\
    \cline{1-15} 
    HE0005--2355& 2.5 $\pm$ 0.1 & 21.8 $\pm$ 2 & 54.6 $\pm$ 5 & 4780 $\pm$ 710 & & 3.7 $\pm$ 0.4 & 9.3 $\pm$ 1.0 & 790 $\pm$ 120 & & 28.7 $\pm$ 9 & 70.5 $\pm$ 21 & 10490 $\pm$ 1550 & & 5810 $\pm$ 340\\
    HE0035--2853& 2.1 $\pm$ 0.1 & 30.3 $\pm$ 1 & 65.0 $\pm$ 2 & 5140 $\pm$ 390 & & 0.7 $\pm$ 0.1 & 1.5 $\pm$ 0.1 & 620 $\pm$ 50 & & 30.2 $\pm$ 5 & 63.4 $\pm$ 9 & 9690 $\pm$  740 & & 6200 $\pm$ 190\\
    HE0043--2300& 3.5 $\pm$ 0.2 & 69.3 $\pm$ 5 & 247.7  $\pm$ 12 & 3500 $\pm$ 110 & & ... & ... & ... & & -- & -- & -- & & 3500 $\pm$ 110\\
    HE0048--2804& 0.6 $\pm$ 0.03 & 40.1 $\pm$ 4 & 25.4 $\pm$ 2 & 5500 $\pm$ 470 & & 1.4 $\pm$ 0.2 & 0.9 $\pm$ 0.1 & 620 $\pm$ 90 & & 32.4 $\pm$ 6 & 20.3 $\pm$ 4 & 10000 $\pm$ 1480 & & 6610 $\pm$ 380 \\
    HE0058--3231& 2.1 $\pm$ 0.1 & 55.4 $\pm$ 2 & 116.7 $\pm$ 3 & 5130 $\pm$ 160 & & 0.8 $\pm$ 0.1 & 1.7 $\pm$ 0.1 & 390 $\pm$ 30 & & 39.9 $\pm$ 7 & 81.3 $\pm$ 13 & 10010 $\pm$ 390 & & 6200 $\pm$ 190\\
    HE0203--4627& 2.1 $\pm$ 0.1 & 26.1 $\pm$ 3 & 54.4 $\pm$ 5 & 5490 $\pm$ 810  & & 0.4 $\pm$ 0.1 & 0.9 $\pm$ 0.1 & 550 $\pm$ 80 & & 34.1 $\pm$ 10 & 70.0 $\pm$ 21 & 10490 $\pm$ 900 & & 6630 $\pm$ 380\\
    HE0248--3628& 4.1 $\pm$ 0.1 & 40.7 $\pm$ 2 & 170.0 $\pm$ 7 & 3800 $\pm$ 150 & & 0.6 $\pm$ 0.1 & 2.3 $\pm$ 0.4 & 600 $\pm$ 50  & & 2.4 $\pm$ 1* & 10.0 $\pm$ 3* & 6000 $\pm$ 460* & & 4460 $\pm$ 170* \\
    HE1349+0007& 2.3 $\pm$ 0.1 & 33.5 $\pm$ 4 & 75.5 $\pm$ 7 & 5030 $\pm$ 430 & & 0.8 $\pm$ 0.1 & 1.9 $\pm$ 0.2 & 530 $\pm$ 80 & & 36.2 $\pm$ 11 & 80.1 $\pm$ 24 & 10980 $\pm$ 430 & & 6110 $\pm$ 350 \\
    HE1409+0101& 4.9 $\pm$ 0.1 & 26.8 $\pm$ 1 & 130.5 $\pm$ 2 & 4000 $\pm$ 160 & & 0.7 $\pm$ 0.03 & 3.6 $\pm$ 0.1 & 600 $\pm$ 20 & & 53.8 $\pm$ 2 & 256.0 $\pm$ 11 & 11000 $\pm$ 110 & & 4840 $\pm$ 50\\
    HE2147--3212& 2.0 $\pm$ 0.1 & 28.4 $\pm$ 3 & 57.7 $\pm$ 6 & 4490 $\pm$ 660 & & 0.8 $\pm$ 0.1 & 1.7 $\pm$ 0.2 & 310 $\pm$ 50 & & 31.6 $\pm$ 10 & 63.1 $\pm$ 19 & 10520 $\pm$ 1560 & & 5470 $\pm$ 320\\
    HE2202--2557& 2.1 $\pm$ 0.1 & 27.4 $\pm$ 1 & 57.6 $\pm$ 2 & 7000 $\pm$ 540 & & 0.5 $\pm$ 0.04 & 1.0 $\pm$ 0.1 & 470 $\pm$ 40 & & 11.7 $\pm$ 2 & 23.9 $\pm$ 4 & 9990 $\pm$ 760 & & 8030 $\pm$ 310\\
    HE2340--4443& 2.8 $\pm$ 0.1 & 77.5 $\pm$ 3 & 224.6 $\pm$ 6 & 3200 $\pm$ 100 & & ... & ... & ... & & -- & -- & -- & & 3200 $\pm$ 100 \\
    HE2349--3800& 1.6 $\pm$ 0.03 & 27.1 $\pm$ 1 & 43.1 $\pm$ 1 & 4000 $\pm$ 160 & & 1.5 $\pm$ 0.1 & 2.4 $\pm$ 0.1 & 850 $\pm$ 30 & & 19.8 $\pm$ 3 & 31.0 $\pm$ 4 & 10060 $\pm$ 410 & & 4870 $\pm$ 150 \\
    HE2352--4010& 7.1 $\pm$ 0.1 & 45.3 $\pm$ 1 & 327.0 $\pm$ 9 & 2900 $\pm$ 90 & & ... & ... & ... & & 4.0 $\pm$ 0.1* & 30.0 $\pm$ 2* & 4000 $\pm$ 162* & & 3300 $\pm$ 30* \\
    \cline{1-15} 
    \\
    \multicolumn{15}{c}{Pa9 $\lambda$9229}\\
    \\
    \cline{1-15} 
    HE0005--2355& 0.6 $\pm$ 0.02 & 11.9 $\pm$ 5 {(5)} & 7.0 $\pm$ 3 {(3)} & 4770 $\pm$ 2690 {(3500)} & & ... & ... & ... & & 15.6 $\pm$ 3 {(5)} & 9.1 $\pm$ 2 {(3)} & 10500 $\pm$ 2410 {f(3210)} & & 5800 $\pm$ 3270 {(4360)} \\
    HE0035--2853& 0.9 $\pm$ 0.01 & 15.8 $\pm$ 2 {(2)} & 18.0 $\pm$ 2 {(2)} & 5120 $\pm$ 1020 {(1450)} & & ... & ... & ... & & 12.7 $\pm$ 1 {(1)} & 14.6 $\pm$ 1 {(1)} & 9700 $\pm$ 740 {(800)} & & 6170 $\pm$ 3070 {(3570)} \\
    HE0043--2300& 1.3 $\pm$ 0.03 & 18.0 $\pm$ 6 {(6)} & 23.5 $\pm$ 8 {(8)} & 3660 $\pm$ 1640 {(1810)} & & ... & ... & ... & & -- & -- & -- & & 3660 $\pm$ 1640 {(1810)}\\
    HE0048--2804& 0.4 $\pm$ 0.01 & 14.6 $\pm$ 2 {(2)} & 6.3 $\pm$ 1 {(1)} & 5550 $\pm$ 1250 {(2400)} & & ... & ... & ... & & 11.9 $\pm$ 1 {(9)} & 5.0 $\pm$ 0.2 {(0.3)} & 9980 $\pm$ 880 {(1690)} & & 6650 $\pm$ 1000 {(1920)} \\
    HE0058--3231& 0.6 $\pm$ 0.06 & 86.3 $\pm$ 13 {(13)} & 51.9 $\pm$ 5 {(5)} & 5130 $\pm$ 770 {(1520)} & & ... & ... & ... & & 62.0 $\pm$ 9 {(9)} & 36.3 $\pm$ 3 {(4)} & 10020 $\pm$ 1590 {(3140)} & & 6200 $\pm$ 930 {(1840)}\\
    HE0203--4627& 0.7 $\pm$ 0.03 & 5.3 $\pm$ 4 {(4)} & 3.7 $\pm$ 3 {(3)} & 5500 $\pm$ 3180 {(3670)} & & ... & ... & ... & & 6.9 $\pm$ 4 {(4)} & 4.7 $\pm$ 2 {(3)} & 10200 $\pm$ 2340 {(2700)} & & 6610 $\pm$ 3730 {(4300)}\\
    HE0248--3628& 1.1 $\pm$ 0.02 & 12.8 $\pm$ 0.4 {(0.4)} & 14.9 $\pm$ 0.3 {(0.3)} & 4900 $\pm$ 500 {(500)} & & ... & ... & ... & & -- & -- & -- & & 4900 $\pm$ 500 {(500)}\\
    HE1349+0007& 0.6 $\pm$ 0.02 & 10.6 $\pm$ 6 {(7)} & 6.3 $\pm$ 4 {(4)} & 5100 $\pm$ 3010 {(2850)} & & ... & ... & ... & & 11.4 $\pm$ 3 {(5)} & 6.7 $\pm$ 2 {(3)} & 10860 $\pm$ 720 {(730)} & & 6190 $\pm$ 3300 {(3320)}\\
    HE1409+0101& 1.1 $\pm$ 0.04 & 12.6 $\pm$ 5 {(5)} & 13.4 $\pm$ 5 {(5)} & 4030 $\pm$ 2270 {(2200)} & & ... & ... & ... & & 15.6 $\pm$ 5 {(6)} & 26.2 $\pm$ 5 {(6)} & 10990 $\pm$ 2400 {(2440)} & & 4870 $\pm$ 710 {(730)} \\
    HE2147--3212& 0.5 $\pm$ 0.02 & 6.0 $\pm$ 4 {(4)} & 3.0 $\pm$ 2 {(2)} & 4500 $\pm$ 2600 {(3450)} & & ... & ... & ... & & 6.3 $\pm$ 4 {(4)} & 3.3 $\pm$ 1 {(1)} & 10520 $\pm$ 2410 {(3190)} & & 5470 $\pm$ 3170 {(4190)}\\
    HE2202--2557& 0.2 $\pm$ 0.004 & 20.1 $\pm$ 4 {(5)} & 8.8 $\pm$ 1 {(1)} & 6990 $\pm$ 1050 {(1040)} & & ... & ... & ... & & 8.5 $\pm$ 2 {(2)} & 3.3 $\pm$ 0.3 {(0.3)} & 10000 $\pm$ 1290 {(1290)} & & 8020 $\pm$ 3020 {(3070)}\\
    HE2340--4443& 1.2 $\pm$ 0.05 & 5.5 $\pm$ 5 {(5)} & 8.6 $\pm$ 6 {(6)} & 3200 $\pm$ 1430 {(1450)} & & ... & ... & ... & & -- & -- & -- & & 3200 $\pm$ 1430 {(1450)}\\
    HE2349--3800& 0.4 $\pm$ 0.02 & 7.7 $\pm$ 6 {(6)} & 4.0 $\pm$ 3 {(3)} & 3940 $\pm$ 2280 {(2380)} & & ... & ... & ... & & 5.2 $\pm$ 3 {(4)} & 2.9 $\pm$ 2 {(2)} & 10060 $\pm$ 2310 {(2400)} & & 4780 $\pm$ 2770 {(2880)}\\
    HE2352--4010& 2.1 $\pm$ 0.04 & 11.2 $\pm$ 3 {(3)} & 25.0 $\pm$ 6 {(6)}& 2900 $\pm$ 1110 {(1070)} & & ... & ... & ... & & 4.3 $\pm$ 0.4 {(0.4)}* & 10.0 $\pm$ 1 {(1)} & 4020 $\pm$ 450 {(430)} & & 3300 $\pm$ 470 {(490)}\\
    \cline{1-15}
    \label{tab:hb+pa9}   
\end{tabular}}
\end{center}
	      \footnotesize
	      {Notes. $^{1}$The asterisk ($*$) indicates Pop. A sources with a BC blueshifted component. $^{2}$The  component that was not observed is marked by ellipsis ($...$). $^{3}$The dashed line (--) indicates that the component is not appropriated for this kind of source. $^{4}$In parenthesis, we report the uncertainty taking into account the \fe2\ template effect. See Appendix \ref{error}.}
\end{sidewaystable}

\newpage

\begin{table}[ht]
\scriptsize
\begin{center}
\caption{Measurements of equivalent width, fluxes and FWHM of \oil\ and \ca2 triplet} 
\begin{tabular}{ccccc}
    \multicolumn{5}{c}{}                                                           \\
    \hline \hline
    & & \multicolumn{3}{c}{BC}\\
    \cline{3-5} 
    Object name& $f_{\lambda}$&  W& $F$ &FWHM\\  
    & (erg s$^{-1}$ cm$^{-2}$ \AA$^{-1}$) & (\AA) & (erg s$^{-1}$ cm$^{-2}$) & (\kms)\\      
    \cline{1-5} 
    \\
    \multicolumn{5}{c}{ {O {\sc i}} $\lambda$8446}\\
    \\
    \cline{1-5} 
    HE0005--2355& 0.7 $\pm$ 0.03 & 7.8 $\pm$ 2 {(2)} & 5.2 $\pm$ 1 {(1)} & 3500 $\pm$ 820 {(1080)} \\
    HE0035--2853& 1.0 $\pm$ 0.02 & 9.4 $\pm$ 2 {(2)} & 12.1 $\pm$ 2 {(2)} & 5000 $\pm$ 370  {(410)}\\
    HE0043--2300& 1.5 $\pm$ 0.03 & 24.5 $\pm$ 3 {(3)} & 37.1 $\pm$ 4 {(5)} & 4000 $\pm$ 360 {(390)}\\
    HE0048--2804& 0.5  $\pm$ 0.01 & 18.4 $\pm$ 1 {(2)} & 10.1 $\pm$ 1 {(1)} & 4990 $\pm$ 270 {(500)}\\
    HE0058--3231& 0.8 $\pm$ 0.08 & 44.4 $\pm$ 8 {(8)} & 35.0 $\pm$ 5 {(5)} & 4960 $\pm$ 100 {(100)}\\
    HE0203--4627& 0.8 $\pm$ 0.03 & 13.6 $\pm$ 3 {(4)} & 11.1 $\pm$ 3 {(3)} & 6000 $\pm$ 250 {(250)}\\
    HE0248--3628& 1.4 $\pm$ 0.02  & 12.2 $\pm$ 2 {(2)} & 17.3 $\pm$ 3 {(3)} & 3490 $\pm$ 260 {(260)}\\
    HE1349+0007& 0.7 $\pm$ 0.03 & 19.9 $\pm$ 4 {(5)} & 14.7 $\pm$ 3 {(4)} & 4580 $\pm$ 680 {(850)}*\\
    HE1409+0101& 1.3 $\pm$ 0.05 & 15.3 $\pm$ 2 {(2)} & 19.3 $\pm$ 3 {(3)} & 3100 $\pm$ 310 {f(310)}\\
    HE2147--3212& 0.6 $\pm$ 0.02 & 22.1 $\pm$ 6 {(11)} & 13.6 $\pm$ 4 {(7)} & 4300 $\pm$ 860 {(1450)}\\
    HE2202--2557& 0.3 $\pm$ 0.01 & 3.6 $\pm$ 1 {(3)} & 2.0 $\pm$ 0.4 {(1)} & 5810 $\pm$ 1060 {(1370)}\\
    HE2340--4443& 1.3 $\pm$ 0.05 & 15.9 $\pm$ 1 {(2)} & 22.0 $\pm$ 2 {(2)} & 3430 $\pm$ 220 {(220)}\\
    HE2349--3800& 0.6 $\pm$ 0.02 & 17.6 $\pm$ 4 {(5)} & 11.3 $\pm$ 2 {(3)} & 3480 $\pm$ 520 {(520)}\\
    HE2352--4010& 2.6 $\pm$ 0.05 & 16.6 $\pm$ 2 {(3)} & 43.5 $\pm$ 4 {(4)} & 1930 $\pm$ 110 {(110)}\\
    \cline{1-5} 
    \\
    \multicolumn{5}{c}{{Ca {\sc ii}} $\lambda$8498, $\lambda$8542, $\lambda$8662}\\
    \\
    \cline{1-5} 
    HE0005--2355& 0.7 $\pm$ 0.03 & 13.9 $\pm$ 8 (11) & 9.1 $\pm$ 5 {(7)} & 4600 $\pm$ 1070 {(1690)}\\
    HE0035--2853& 1.0 $\pm$ 0.02 & 27.9 $\pm$ 2 (2) & 28.8 $\pm$ 2 {(2)} & 4540 $\pm$ 170 {(180)} \\
    HE0043--2300& 1.4 $\pm$ 0.03 & 36.1 $\pm$ 3 {(4)} & 53.3 $\pm$ 4 {(6)} & 4000 $\pm$ 150 {(160)}\\
    HE0048--2804& 0.5 $\pm$ 0.01 & 4.9 $\pm$ 3 {(4)} & 2.6 $\pm$ 2 {(2)} & 5170 $\pm$ 2400 {(3200)} \\
    HE0058--3231& 0.8 $\pm$ 0.08 & 59.6 $\pm$ 14 {(21)} & 45.0 $\pm$ 10 {(15)} & 4910 $\pm$ 740 {(750)}\\
    HE0203--4627& 0.8 $\pm$ 0.03 & 33.2 $\pm$ 3 {(4)} & 26.3 $\pm$ 3 {(3)} & 5960 $\pm$ 530 {(650)} \\
    HE0248--3628& 1.3 $\pm$ 0.02 & 31.3 $\pm$ 2 {(2)} & 42.8 $\pm$ 2 {(3)} & 3990 $\pm$ 150 {(150)} \\
    HE1349+0007& 0.7 $\pm$ 0.03 & 24.7 $\pm$ 7 {(7)} & 17.7 $\pm$ 5 {(5)} & 4530 $\pm$ 940 {(1100)} \\
    HE1409+0101& 1.2 $\pm$ 0.05 & 19.9 $\pm$ 6 {(7)} & 41.4 $\pm$ 8 {(8)} & 3550 $\pm$ 500 {(510)} \\
    HE2147--3212& 0.6 $\pm$ 0.02 & 47.4 $\pm$ 4 {(5)} & 28.1 $\pm$ 2 {(3)} & 3990 $\pm$ 150 {(150)}\\
    HE2202--2557& 0.3 $\pm$ 0.01 & 13.0 $\pm$ 3 {(4)} & 6.8 $\pm$ 1 {(1)} & 5900 $\pm$ 330 {(400)} \\
    HE2340--4443& 1.3 $\pm$ 0.05 & 14.8 $\pm$ 10 {(11)} & 20.1 $\pm$ 14 {(15)} & 3190 $\pm$ 1700 {(1760)}\\
    HE2349--3800& 0.6 $\pm$ 0.02 & 19.9 $\pm$ 6 {(6)} & 12.3 $\pm$ 3 {(3)} & 3520 $\pm$ 700 {(730)} \\
    HE2352--4010& 2.5 $\pm$ 0.05 & 21.7 $\pm$ 3 {(3)} & 55.7 $\pm$ 6 {(6)} & 3080 $\pm$ 180 {(180)} \\
    \cline{1-5} 
    \label{tab:oi+ca}  
\end{tabular}
\end{center}
      \footnotesize 
       Notes. { $^{1}$In parenthesis, we report the uncertainty taking into account the \fe2\ template effect. $^2$The equivalent width, flux and FWHM values for \oi\ NC are: 1.3$\pm$1.0 \AA, 1.0$\pm$0.7 erg s$^{-1}$ cm$^{-2}$ and 550$\pm$250 \kms, respectively.}
\end{table}

\newpage

\begin{table}[ht]
\begin{center}
\caption{Measurements of equivalent width and flux for optical and NIR \fe2.}
\scalebox{0.8}[0.8]{
\begin{tabular}{cccccccc}
    \multicolumn{8}{c}{}                                                                \\
    \hline \hline
    & \multicolumn{2}{c}{\fe2$_{opt}$}& &\multicolumn{2}{c}{\fe2$_{NIR}$}& & \\
    \cline{2-3} 
    \cline{5-6}
    Object name & W & $F$ & & W & $F$ & & Template used\\
    & (\AA) & (erg s$^{-1}$ cm$^{-2}$) & & (\AA) & (erg s$^{-1}$ cm$^{-2}$) & & \\ 
    \cline{1-8} 
    HE0005--2355& 17.2 $\pm$ 2 & 51.4 $\pm$ 8 & & 20.4 $\pm$ 5 & 12.1 $\pm$ 9 & &Theoretical\\
    HE0035--2853& 40.7 $\pm$ 3 & 97.0 $\pm$ 8 & & 34.2 $\pm$ 4 & 32.5 $\pm$ 8 & & Theoretical\\
    HE0043--2300& 20.0 $\pm$ 1 & 78.7 $\pm$ 6 & & 2.8 $\pm$ 0.3 & 3.5 $\pm$ 0.2 & & Theoretical\\
    HE0048--2804& 22.3 $\pm$ 2 & 15.9 $\pm$ 2 & & 18.8 $\pm$ 1 & 8.2 $\pm$ 1 & & Theoretical\\
    HE0058--3231& 27.7 $\pm$ 2 & 68.3 $\pm$ 19 & & 47.3 $\pm$ 4 & 28.9 $\pm$ 7 & & Theoretical\\
    HE0203--4627& 14.4 $\pm$ 2 & 41.5 $\pm$ 6 & & 25.1 $\pm$ 4 & 17.3 $\pm$ 4 & & Theoretical\\
    HE0248--3628& 13.6 $\pm$ 0.5 & 62.7 $\pm$ 3 & & 28.1 $\pm$ 3 & 31.1 $\pm$ 1 & & Theoretical\\
    HE1349+0007& 18.5 $\pm$ 1 & 51.6 $\pm$ 6 & & 38.5 $\pm$ 8 & 23.2 $\pm$ 6 & & Theoretical\\
    HE1409+0101& 36.2 $\pm$ 2 & 203.8 $\pm$7 9 & & 3.4 $\pm$ 0.4 & 3.6 $\pm$ 1 & & Theoretical\\
    HE2147--3212& 35.0 $\pm$ 4 & 85.4 $\pm$ 12 & & 11.3 $\pm$ 2 & 5.6 $\pm$ 1 & & Semi--empirical\\
    HE2202--2557& 12.7 $\pm$ 1 & 31.1 $\pm$ 3 & & 8.1 $\pm$ 2 & 3.6 $\pm$ 1 & & Theoretical\\
    HE2340--4443& 15.4 $\pm$ 1 & 50.4 $\pm$ 3 & & 13.0 $\pm$ 1 & 15.2 $\pm$ 4 & & Theoretical\\
    HE2349--3800& 19.9 $\pm$ 1 & 35.7 $\pm$ 2 & & 8.6 $\pm$ 1 & 4.5 $\pm$ 1 & & Semi--empirical\\
    HE2352--4010& 17.6 $\pm$ 0.3 & 147.6 $\pm$ 5 & & 12.8 $\pm$ 1 & 26.8 $\pm$ 3 & &Semi--empirical\\
    \cline{1-8} 
     \label{tab:fe2}  
\end{tabular}}
\end{center}
\label{tab:multicol}
\end{table}

\newpage

\begin{table}[ht]
\caption{Observed flux ratios for the BC of \hb, Pa9, \oil\ and \ca2 triplet}
\begin{center} 
\scalebox{0.9}[0.9]{
\begin{tabular}{cccccc}
    \hline \hline
    Object name & $\log$(CaT$/$\hb) & $\log$(\oi$/$\hb) & $\log$(CaT$/$Pa9) & $\log$(\oi$/$Pa9) & $\log$(CaT$/$\oi) \\
    \hline
\\   
\multicolumn{6}{c}{No \oi\ VBC}    \\  \\   \cline{1-6}
    HE0005--2355& -0.78 & -1.02 & 0.11 & -0.13 & 0.24\\
    HE0035--2853& -0.35 & -0.73 & 0.28 & -0.09 & 0.38\\
    HE0043--2300& -0.67 & -0.82 & 0.35 & 0.20 & 0.16\\
    HE0048--2804& -0.99 & -0.40 & -0.39 & 0.20 & -0.59\\
    HE0058--3231& -0.41 & -0.52 & -0.06 & -0.17 & 0.11\\
    HE0203--4627& -0.32 & -0.69 & 0.86 & 0.48 & 0.37\\
    HE0248--3628& -0.60 & -0.99 & 0.46 & 0.07 & 0.39\\
    HE1349+0007& -0.63 & -0.71 & 0.45 & 0.37 & 0.08\\
    HE1409+0101& -0.50 & -0.83 & 0.49 & 0.16 & 0.33\\
    HE2147--3212& -0.31 & -0.63 & 0.97 & 0.66 & 0.32\\
    HE2202--2557&  -0.93 & -1.46 & -0.07 & -0.60 & 0.53 \\
    HE2340--4443& -1.05 & -1.01 & 0.37 & 0.41 & -0.04\\
    HE2349--3800& -0.54 & -0.58 & 0.49 & 0.45 & 0.04\\
    HE2352--4010& -0.77 & -0.88 & 0.35 & 0.24 & 0.11\\
    \cline{1-6} 
    \label{tab:ratios}  
\\
  \multicolumn{6}{c}{\oi\ VBC}    \\  \\   \cline{1-6} 
   HE0005--2355& -0.94 & -1.10 & -0.05 & -0.21 & 0.17\\
    HE0035--2853& -0.58 & -0.79 & -0.02 & -0.23 & 0.21\\
    HE0043--2300& -0.67 & -0.82 & 0.35 & 0.20 & 0.16\\
    HE0048--2804& -1.14 & -0.52 & -0.52 & 0.10 & -0.62\\
    HE0058--3231& -0.74 & -0.52 & -0.40 & -0.17 & -0.22\\
    HE0203--4627& -0.70 & -0.75 & 0.37 & 0.32 & 0.05\\
    HE0248--3628& -0.60 & -0.99 & 0.46 & 0.07 & 0.39\\
    HE1349+0007& -0.74 & -0.89 & 0.34 & 0.19 & 0.15\\
    HE1409+0101& -0.67 & -0.96 & 0.36 & 0.08 & 0.29\\
    HE2147--3212& -0.45 & -0.71 & 0.83 & 0.57 & 0.26\\
    HE2202--2557& -0.96 & -1.36 & -0.01 & -0.41 & 0.40\\
    HE2340--4443& -1.05 & -1.01 & 0.37 & 0.41 & -0.04\\
    HE2349--3800& -0.70 & -0.65 & 0.34 & 0.39 & -0.05\\
    HE2352--4010& -0.77 & -0.88 & 0.35 & 0.24 & 0.11\\
    \cline{1-6} 
\end{tabular}}
\end{center}
\end{table}

\newpage
\eject
\clearpage

\begin{figure}[ht]
  \begin{center}
      \includegraphics[width=6cm,keepaspectratio=true]{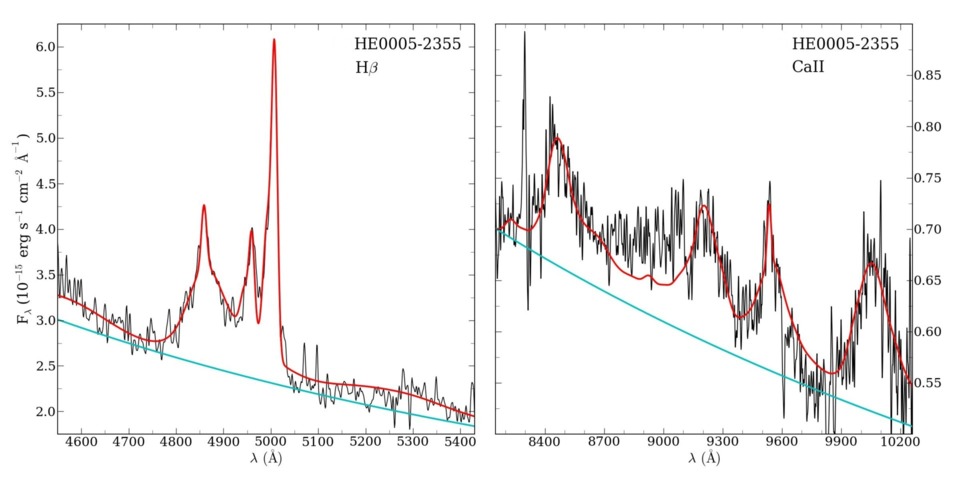} 
      \includegraphics[width=6cm,keepaspectratio=true]{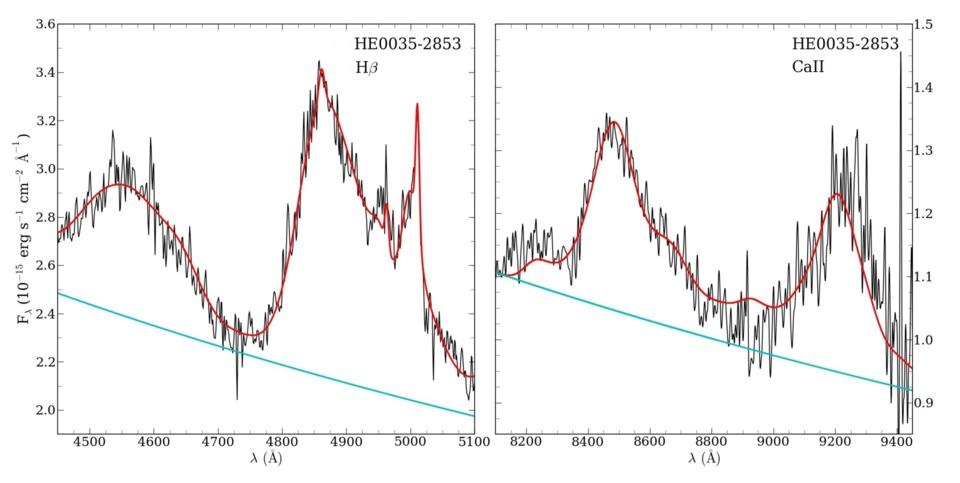} \\
      \includegraphics[width=6cm,keepaspectratio=true]{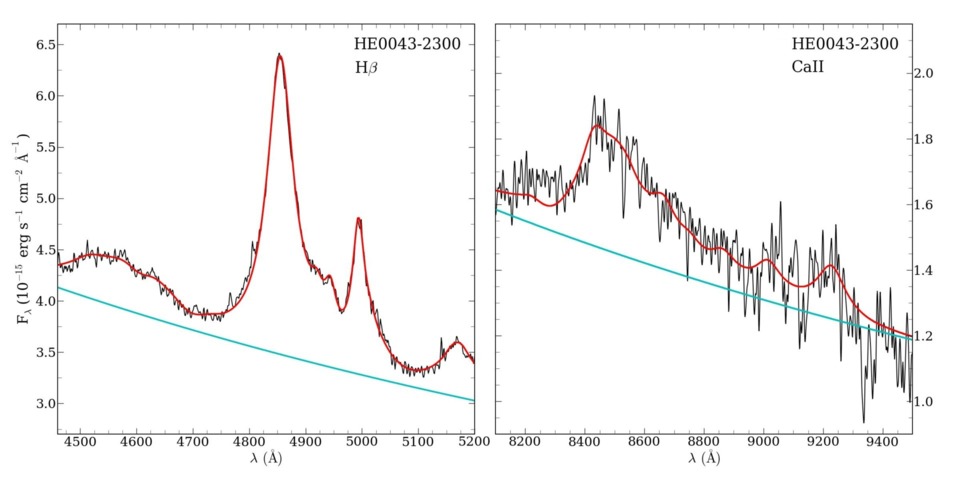} 
      \includegraphics[width=6cm,keepaspectratio=true]{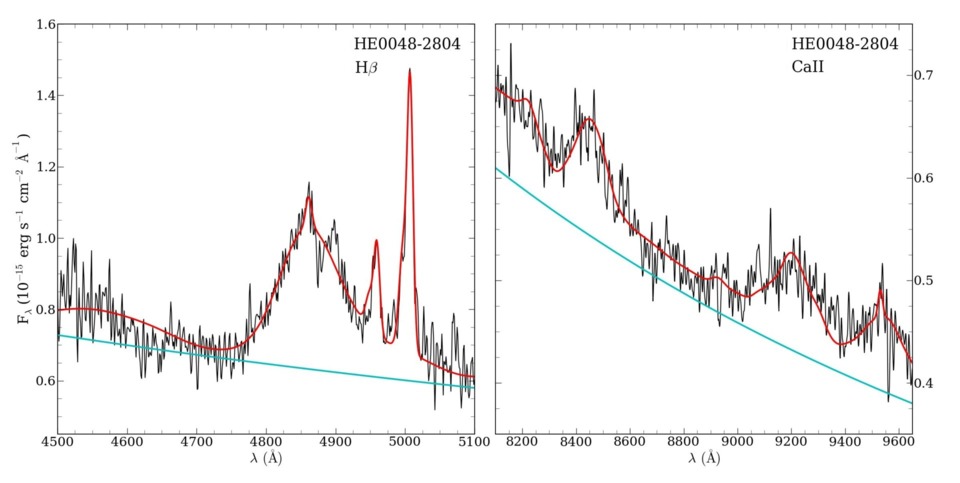} \\
      \includegraphics[width=6cm,keepaspectratio=true]{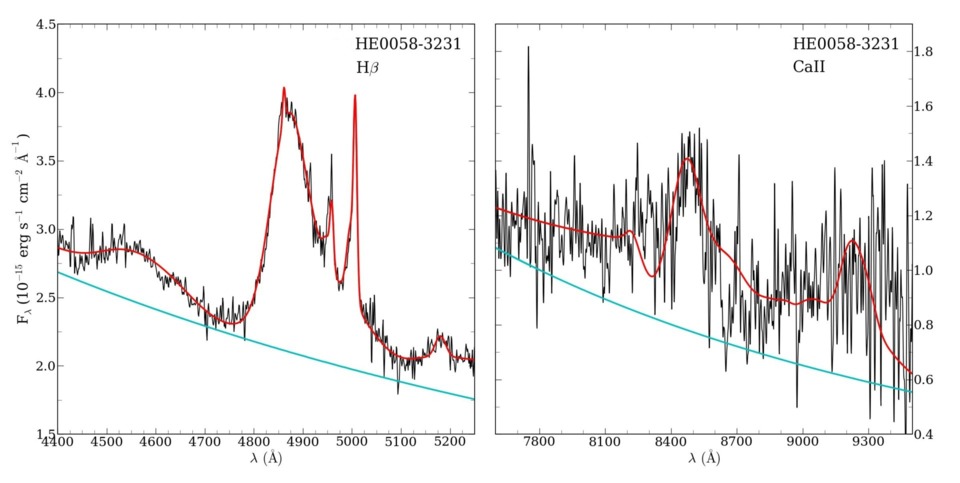} 
      \includegraphics[width=6cm,keepaspectratio=true]{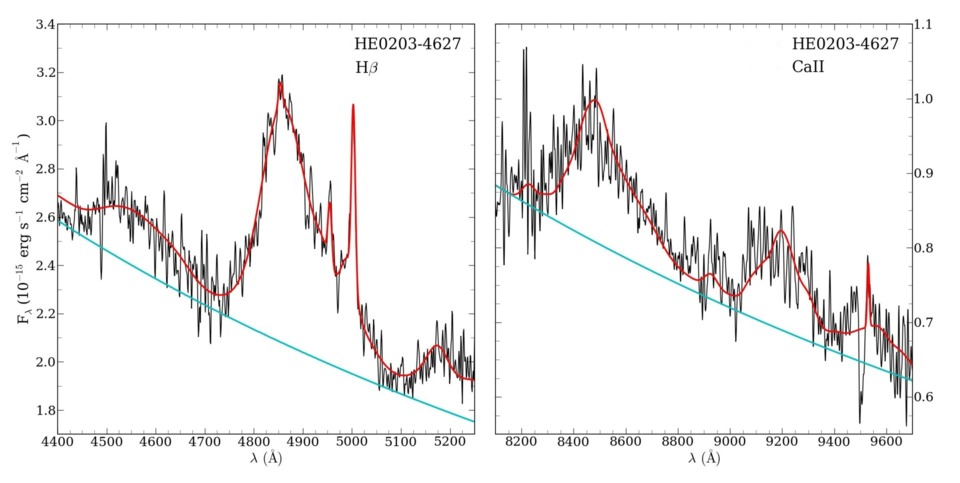} \\
      \includegraphics[width=6cm,keepaspectratio=true]{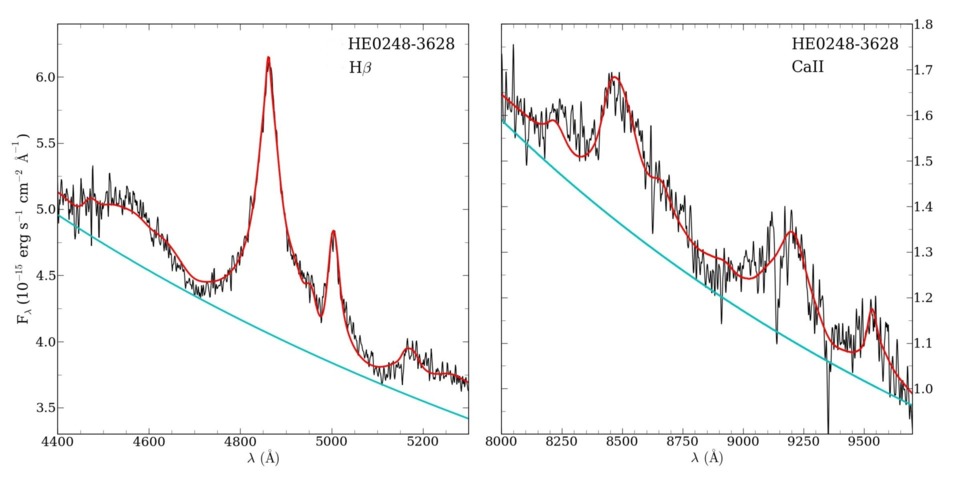} 
     \includegraphics[width=6cm,keepaspectratio=true]{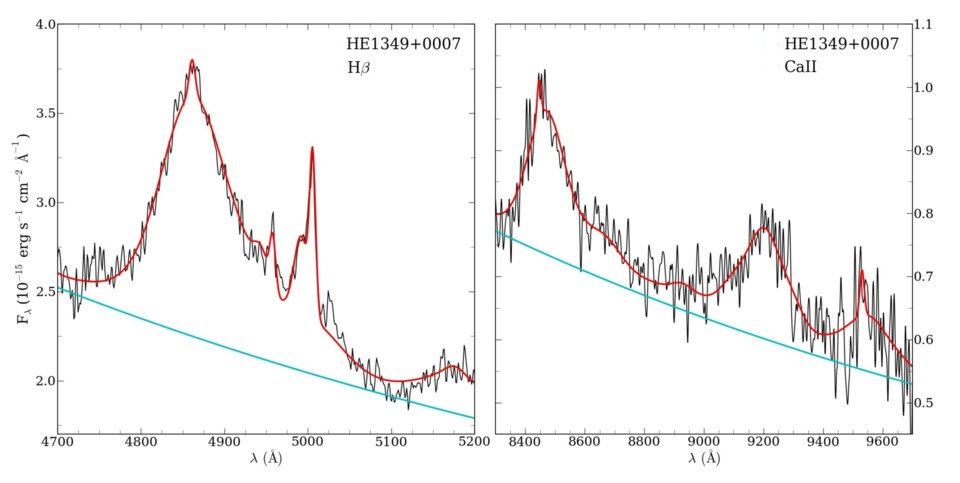} 
      \caption{Calibrated spectra. Pairs of close panels show the \hb\ (left) and \ca2\ (right) spectral regions before continuum subtraction for each object of our sample. In both panels the best fit is marked by the red color line and the continuum level is marked with the cyan line. Abscissae are rest--frame wavelength in \AA, ordinates are rest--frame specific flux in units of 10$^{-15}$ erg s$^{-1}$ cm$^{-2}$ \AA$^{-1}$.}
     \label{fig:cont}
  \end{center}
\end{figure}

\addtocounter{figure}{-1}
\begin{figure}[ht]
  \begin{center}
      \includegraphics[width=6cm,keepaspectratio=true]{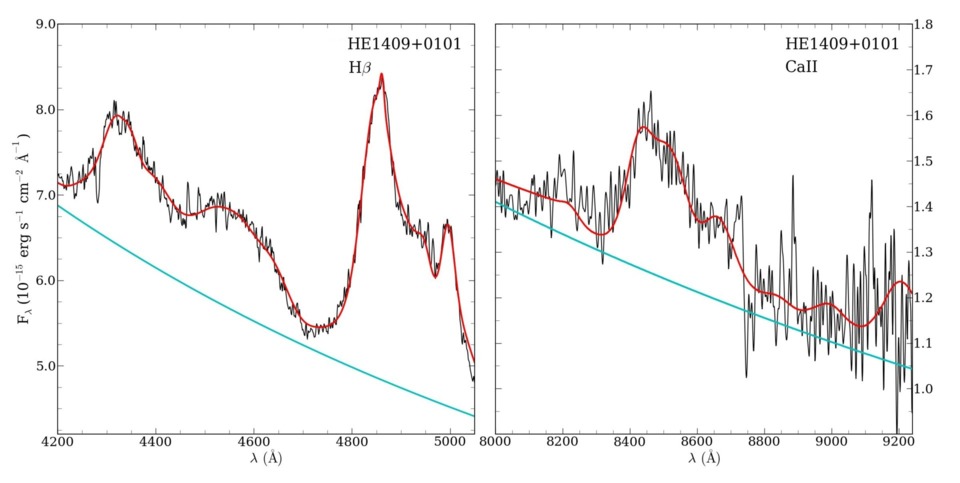} 
      \includegraphics[width=6cm,keepaspectratio=true]{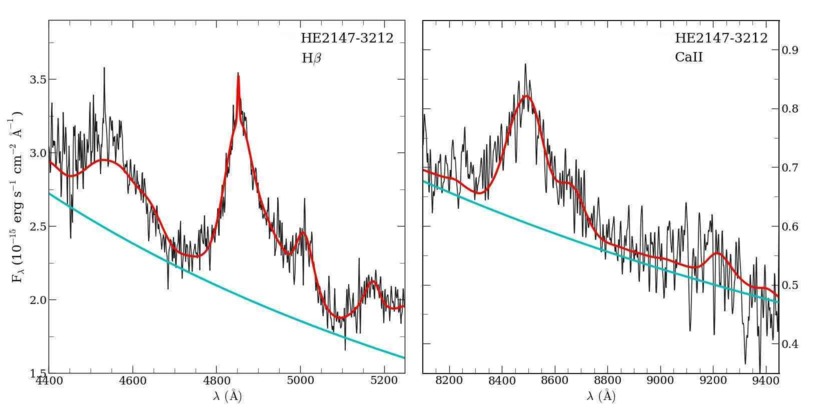} \\
      \includegraphics[width=6cm,keepaspectratio=true]{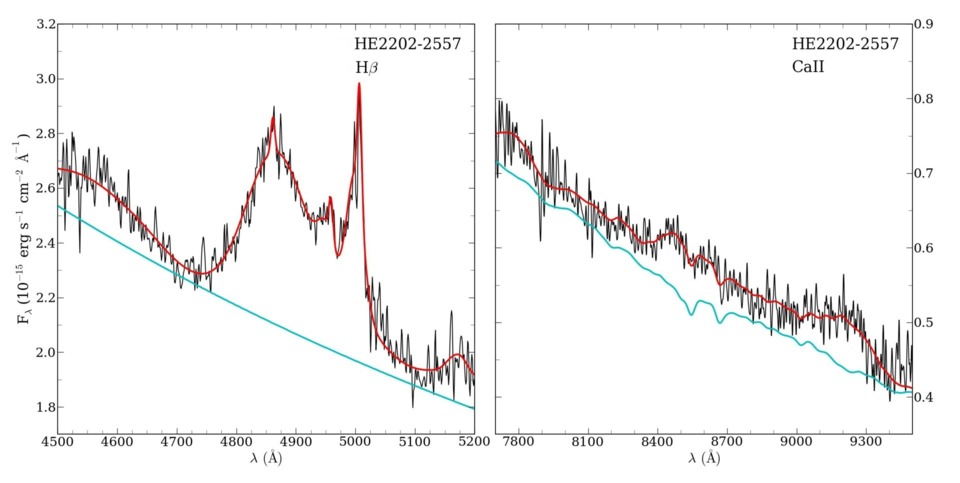} 
      \includegraphics[width=6cm,keepaspectratio=true]{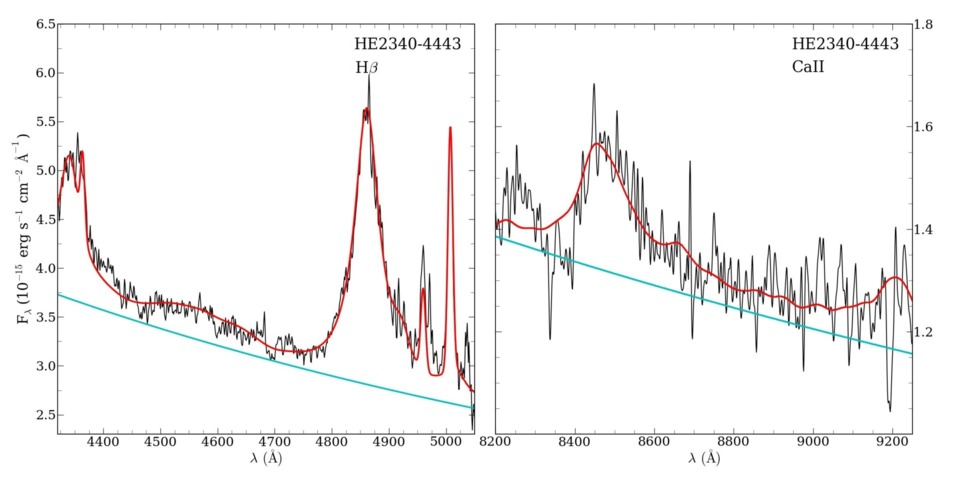} \\
      \includegraphics[width=6cm,keepaspectratio=true]{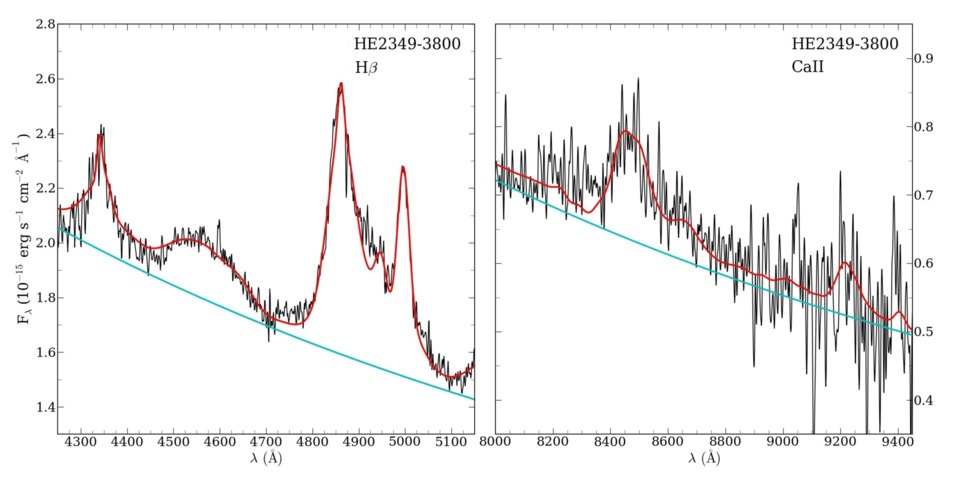} 
      \includegraphics[width=6cm,keepaspectratio=true]{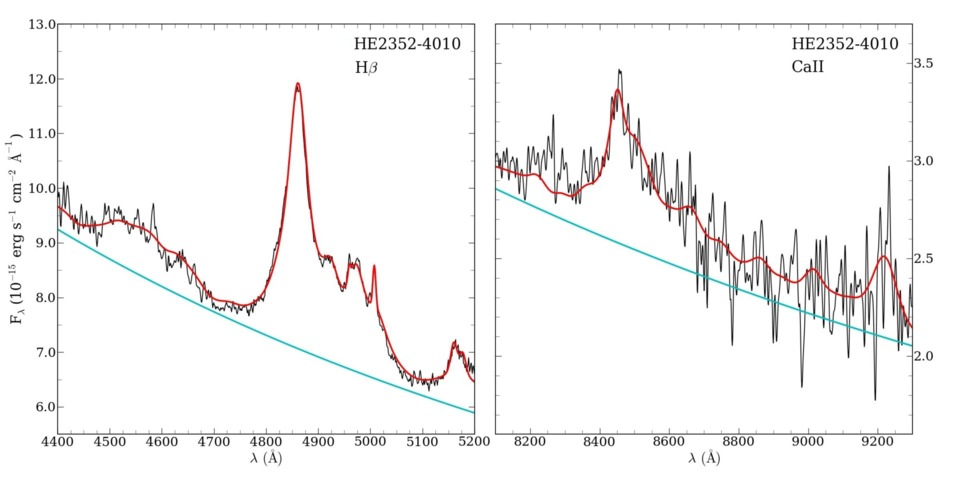} 
         \caption{Cont.}
     \end{center}
\end{figure}

\begin{figure}[ht]
  \begin{center}
      \includegraphics[width=5.5cm,keepaspectratio=true]{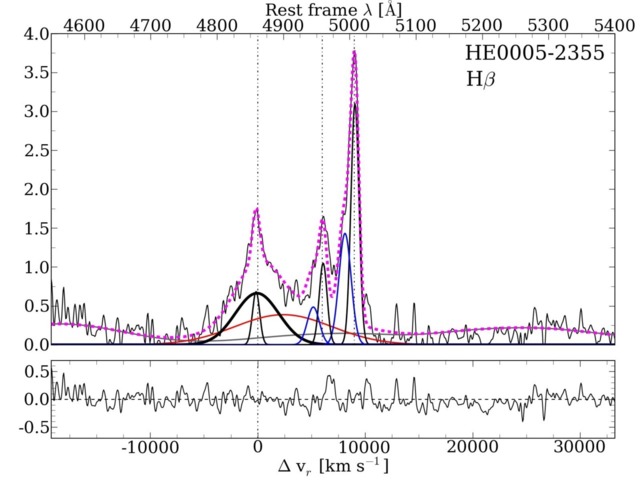}
      \includegraphics[width=5.5cm,keepaspectratio=true]{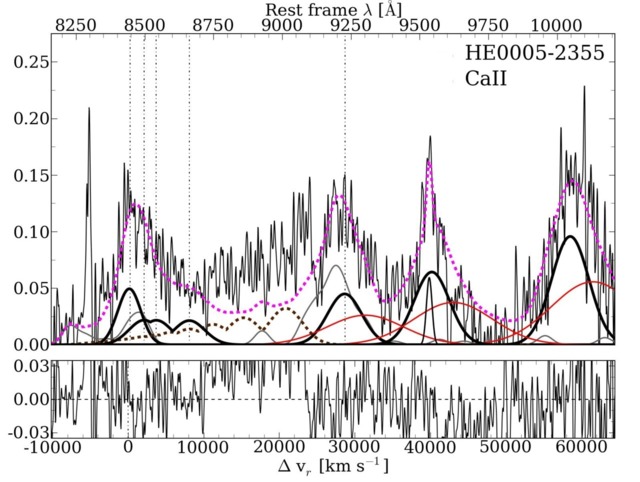} \\
      \includegraphics[width=5.5cm,keepaspectratio=true]{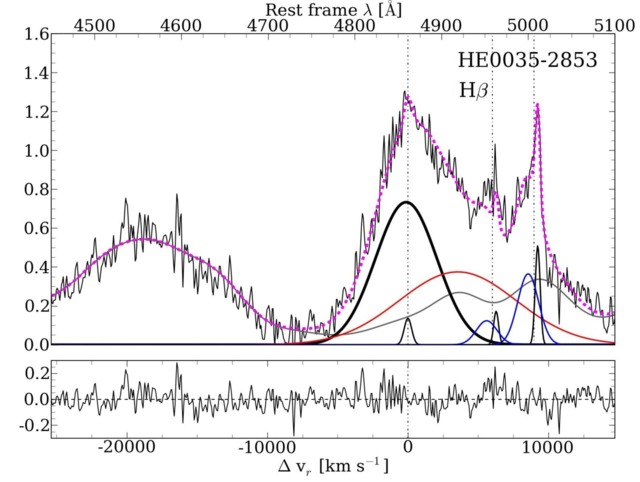} 
      \includegraphics[width=5.5cm,keepaspectratio=true]{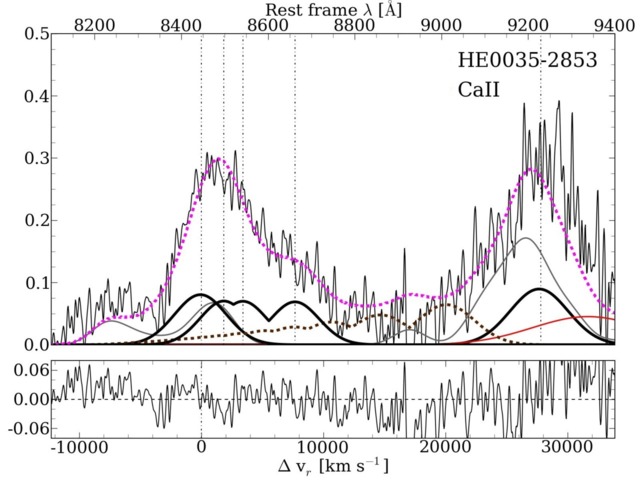} \\
      \includegraphics[width=5.5cm,keepaspectratio=true]{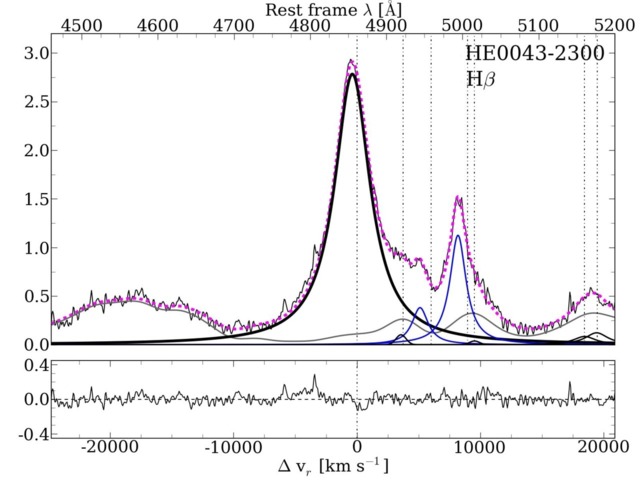} 
      \includegraphics[width=5.5cm,keepaspectratio=true]{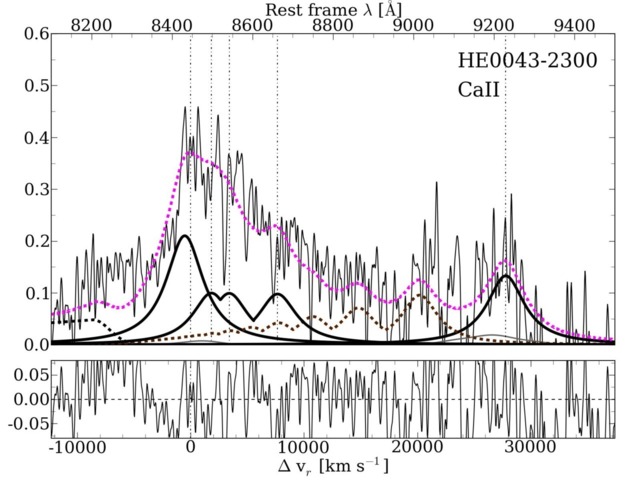}      
      \includegraphics[width=5.5cm,keepaspectratio=true]{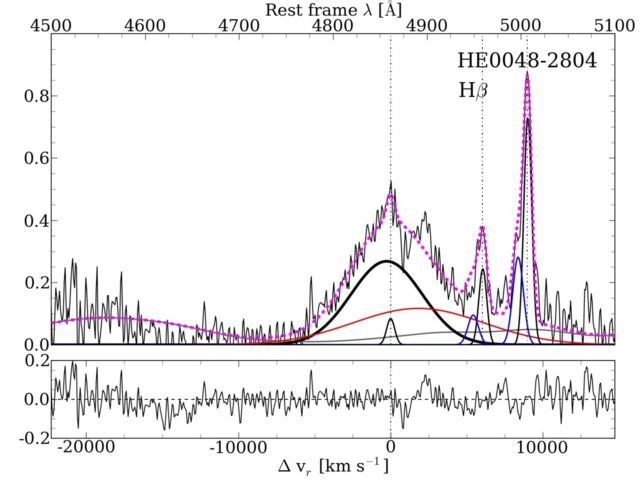} 
      \includegraphics[width=5.5cm,keepaspectratio=true]{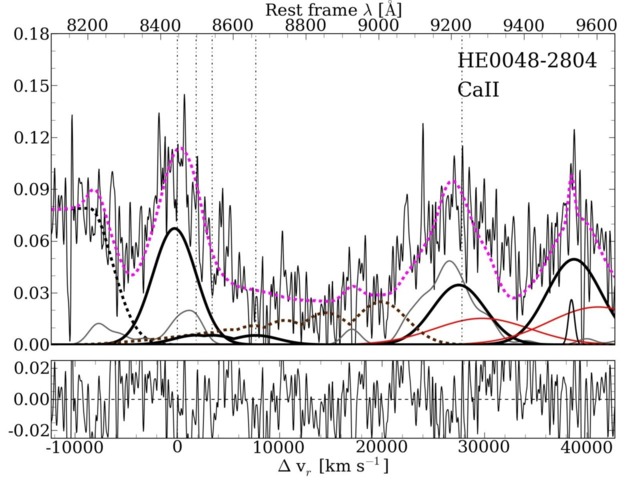} 
            \caption{Quasar spectra after continuum subtraction. Abscissae are rest--frame wavelength in \AA, ordinates are rest--frame specific flux in units of 10$^{-15}$ erg s$^{-1}$ cm$^{-2}$ \AA$^{-1}$.  The left panels show the \hb\ spectral region and right panels show the \ca2\ triplet spectral region. The vertical dashed lines are the rest frame \hb, \o3, \oil\ and \ca2\ $\lambda$8498, $\lambda$8542, $\lambda$8662. Line coding is as follows.  Thick black lines:   broad components; thin black lines: narrow components; blue lines:   blueshifted components of \hb;  red lines:  \hb\ and Pa9 VBC;   grey lines:  \fe2\ contribution;  thick dashed black lines:  Paschen continuum: thick-brown lines:  high order Paschen lines; thin--dashed pink lines: best--fit model. Note that the \ca2\ triplet and high order Paschen lines are shown as  blends of individual lines and not as their  sum.} 
     \label{fig:fits}
  \end{center}
\end{figure}

\addtocounter{figure}{-1}
\begin{figure}[ht]
  \begin{center}
 \includegraphics[width=5.5cm,keepaspectratio=true]{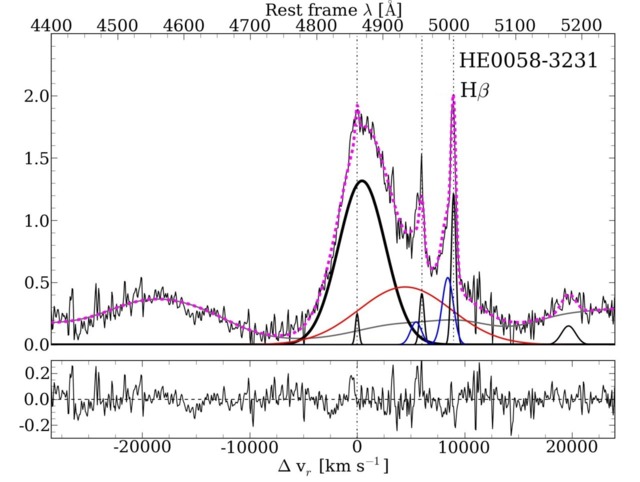}
      \includegraphics[width=5.5cm,keepaspectratio=true]{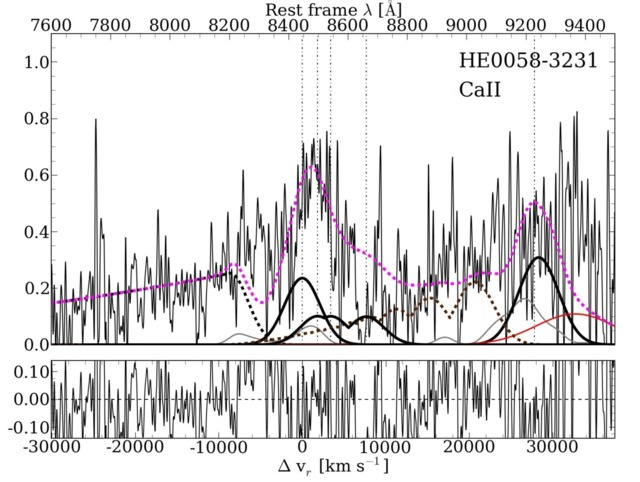}
       \includegraphics[width=5.5cm,keepaspectratio=true]{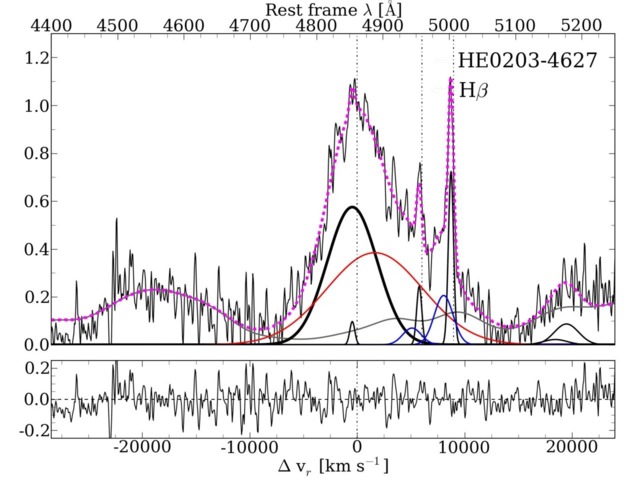} 
      \includegraphics[width=5.5cm,keepaspectratio=true]{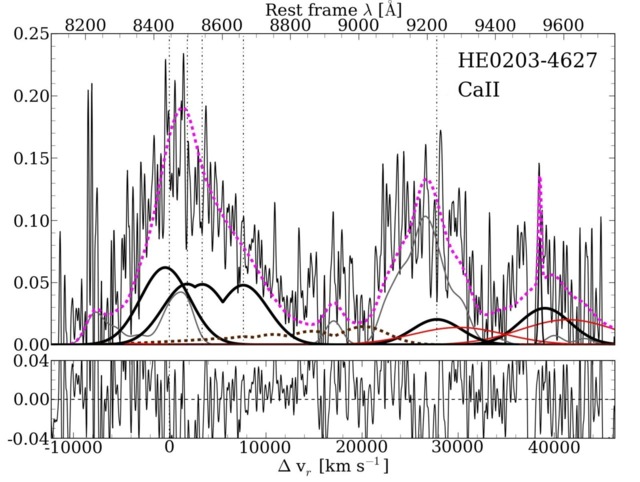} 
         \includegraphics[width=5.5cm,keepaspectratio=true]{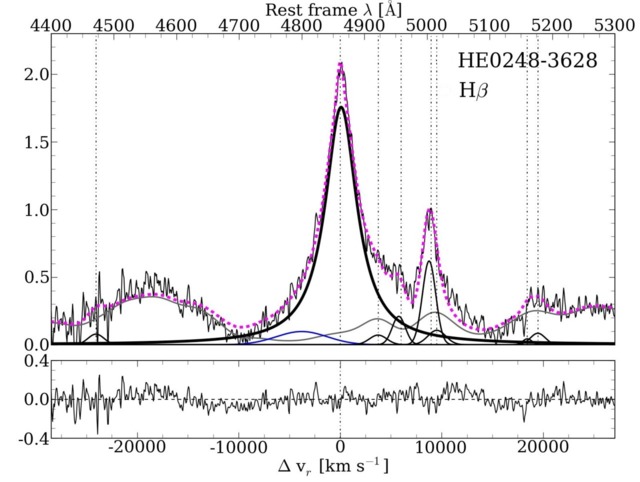}
      \includegraphics[width=5.5cm,keepaspectratio=true]{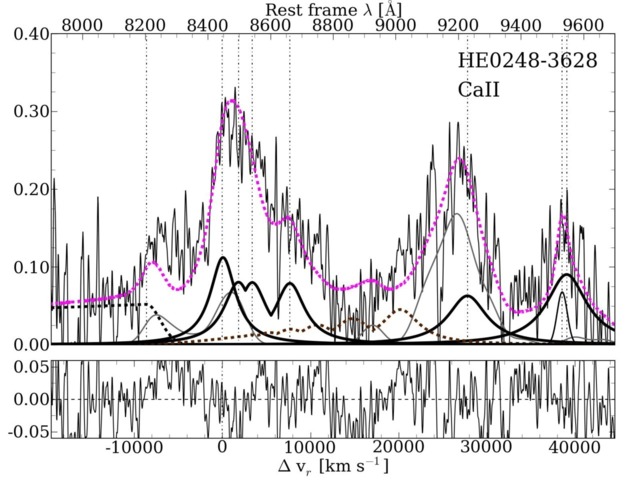}
      \includegraphics[width=5.5cm,keepaspectratio=true]{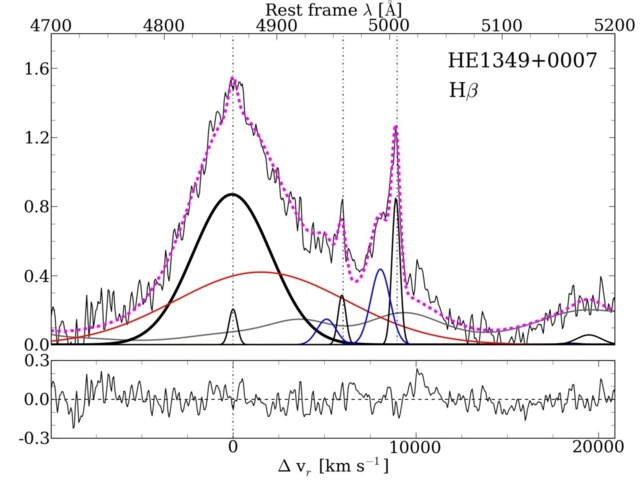} 
      \includegraphics[width=5.5cm,keepaspectratio=true]{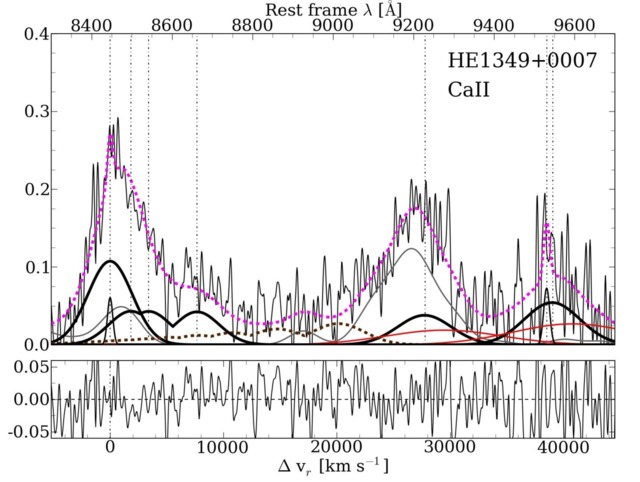} 
         \caption{Cont.}
     \end{center}
\end{figure}

\addtocounter{figure}{-1}
\begin{figure}[ht]
  \begin{center}
   \includegraphics[width=5.5cm,keepaspectratio=true]{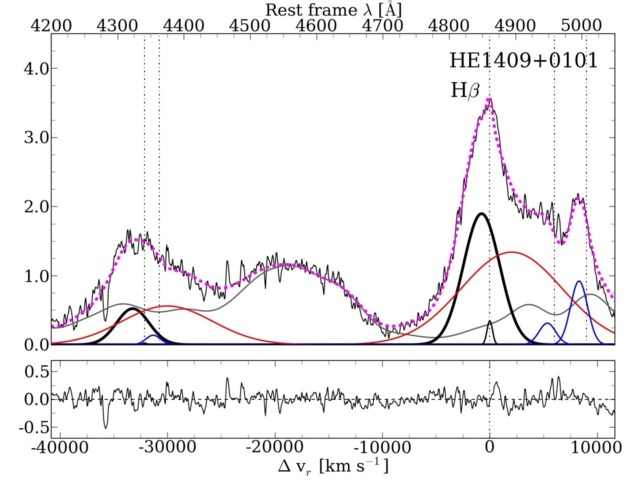}
      \includegraphics[width=5.5cm,keepaspectratio=true]{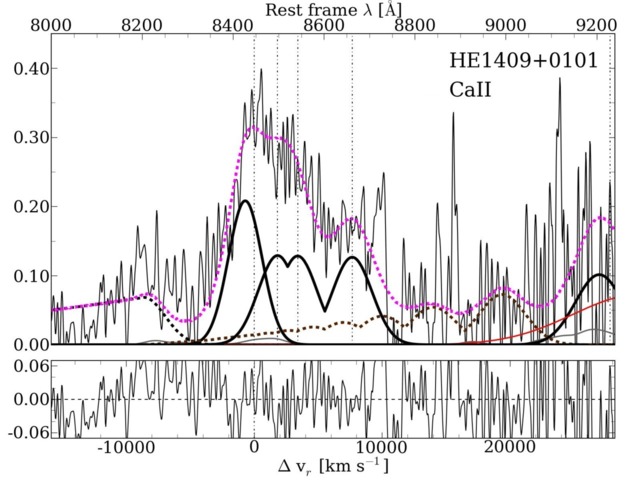}
      \includegraphics[width=5.5cm,keepaspectratio=true]{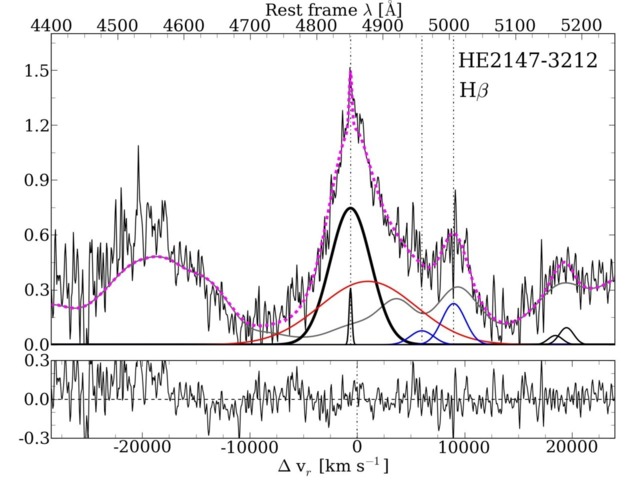} 
      \includegraphics[width=5.5cm,keepaspectratio=true]{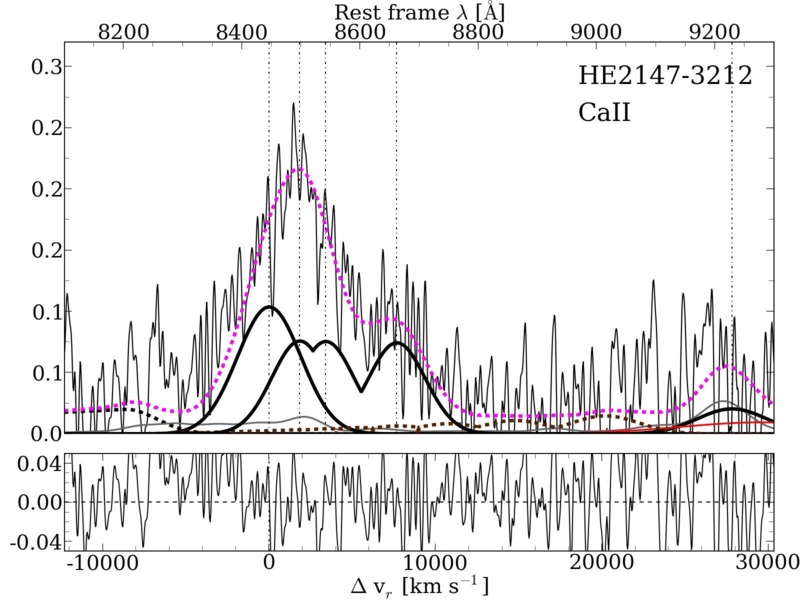}       
      \includegraphics[width=5.5cm,keepaspectratio=true]{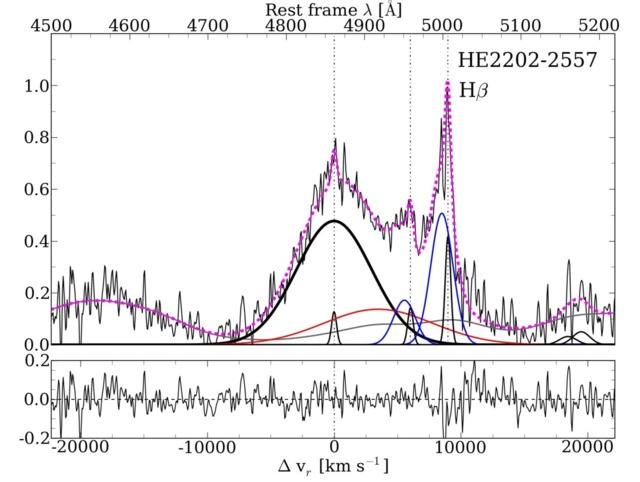}
      \includegraphics[width=5.5cm,keepaspectratio=true]{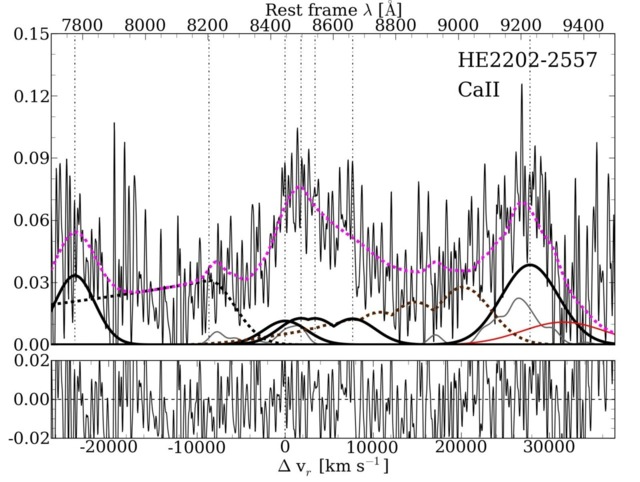}
      \includegraphics[width=5.5cm,keepaspectratio=true]{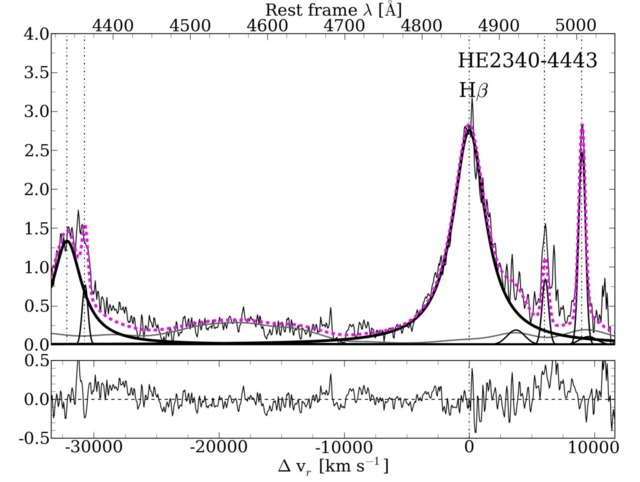}
      \includegraphics[width=5.5cm,keepaspectratio=true]{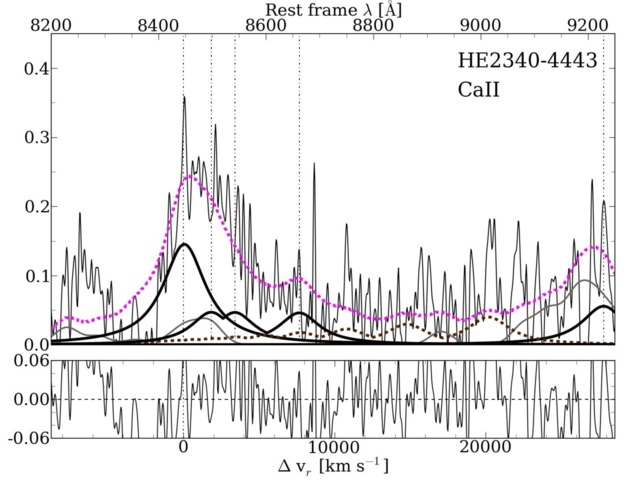}
       \caption{Cont.}
     \end{center}
\end{figure}

\addtocounter{figure}{-1}
\begin{figure}[ht]
  \begin{center}
    \includegraphics[width=5.5cm,keepaspectratio=true]{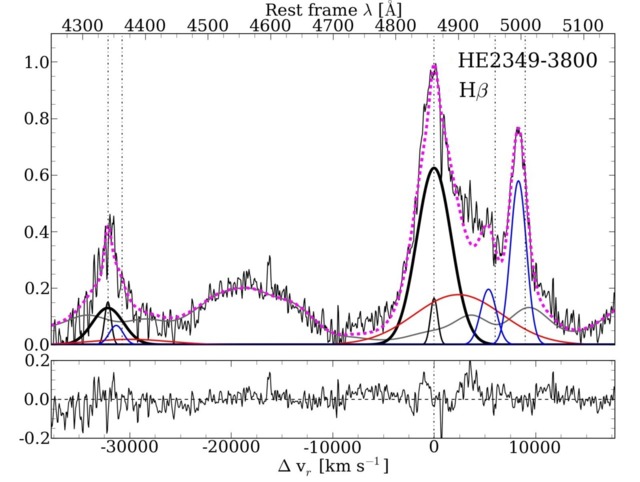}
      \includegraphics[width=5.5cm,keepaspectratio=true]{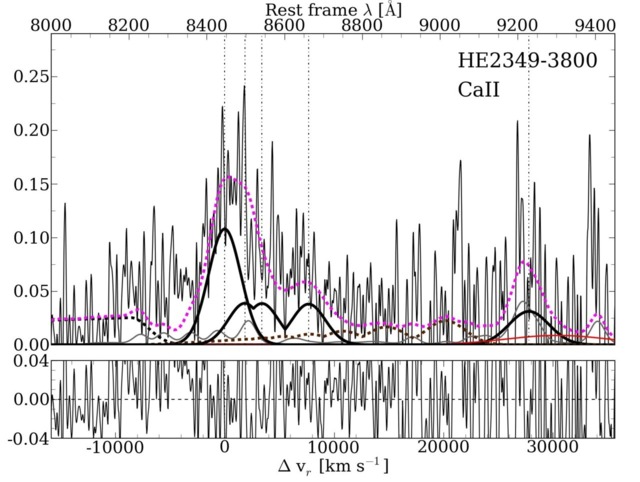}
      \includegraphics[width=5.5cm,keepaspectratio=true]{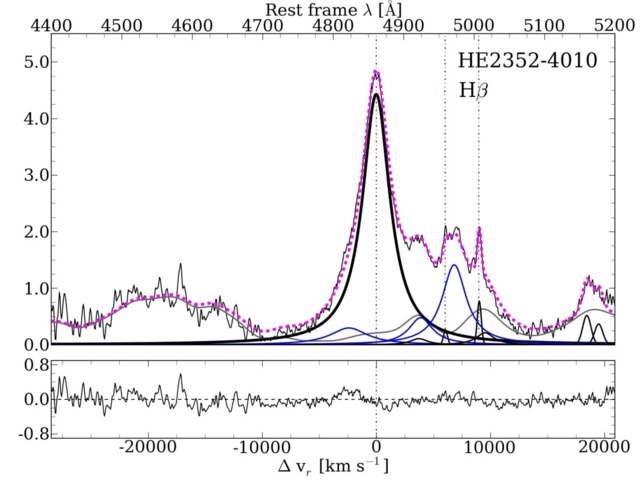}
      \includegraphics[width=5.5cm,keepaspectratio=true]{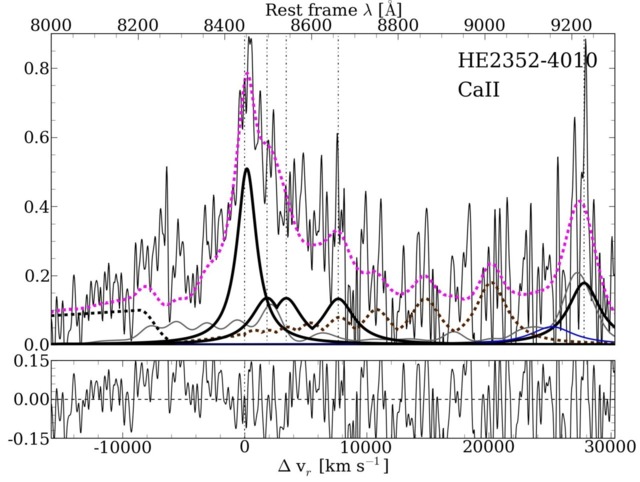}
      \caption{Cont.}
     \end{center}
\end{figure}

\eject
\clearpage

\begin{figure}[ht]
  \begin{center}
      \includegraphics[width=5.5cm,keepaspectratio=true]{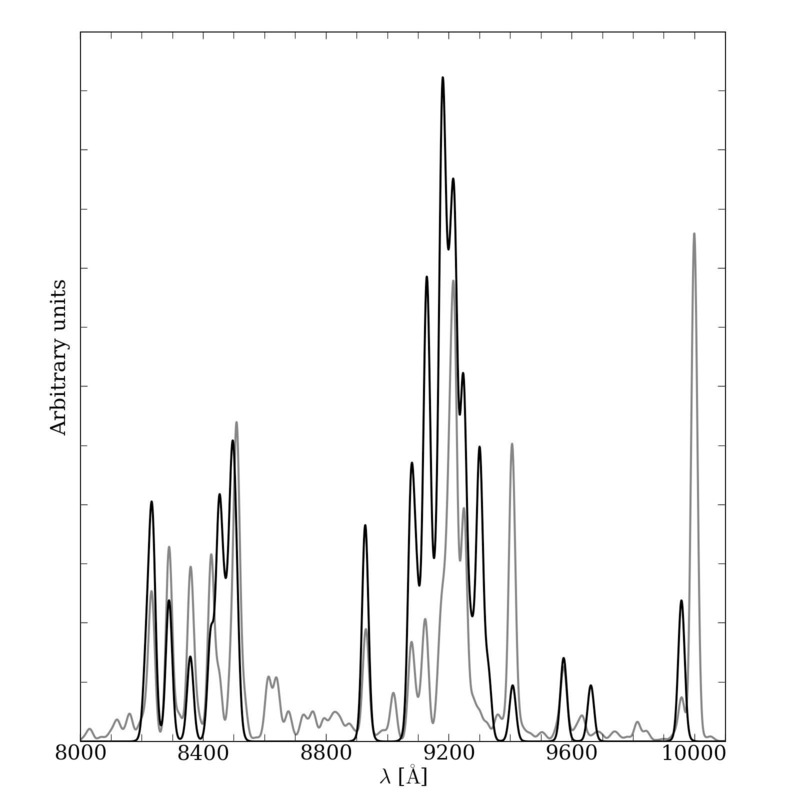}
      \includegraphics[width=5.5cm,keepaspectratio=true]{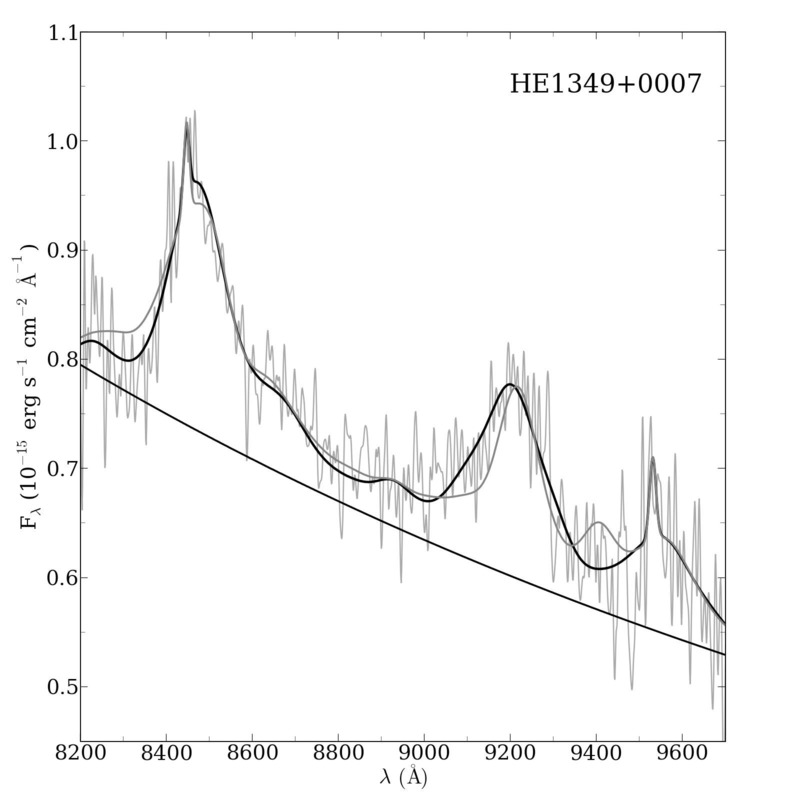}      
      \includegraphics[width=5.25cm,keepaspectratio=true]{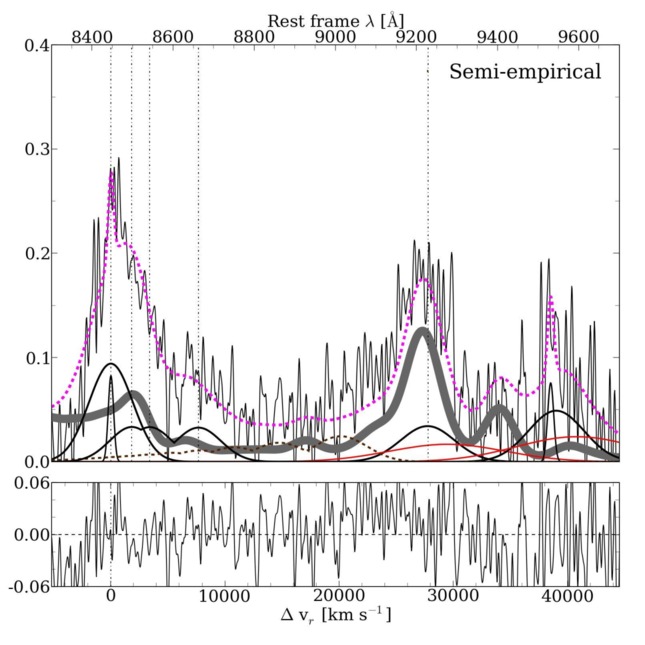}      
      \includegraphics[width=5.25cm,keepaspectratio=true]{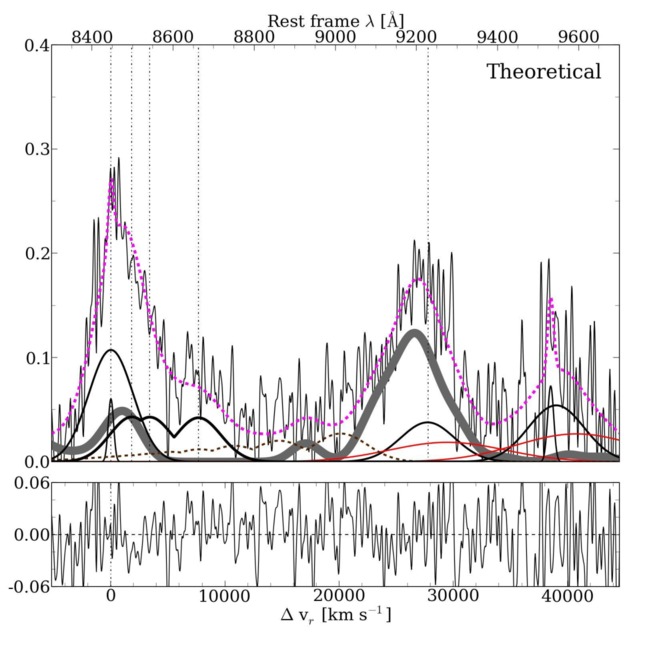}      
      \caption{Upper left panel: comparison between theoretical (black line) and semi--empirical  (grey line) template, as obtained by \citet{GAR12}. Upper right panel:  models of the NIR spectral region for  HE1349+0007,  using the two templates. The \fe2\  prominent feature at $\sim$ 9200 \AA\ is less strong in the semi--empirical template, leaving a flux deficit in the observed blend with Pa9 $\lambda$9229. The lower panels emphasise the difference between the two \fe2\ templates (thick  grey line; left: semi--empirical, right: theoretical). Line coding is the same as for the previous figure.}    
     \label{fig:fe2_templates}  
   \end{center}
\end{figure}

\begin{figure}[ht]
  \begin{center}
      \includegraphics[width=15cm,keepaspectratio=true]{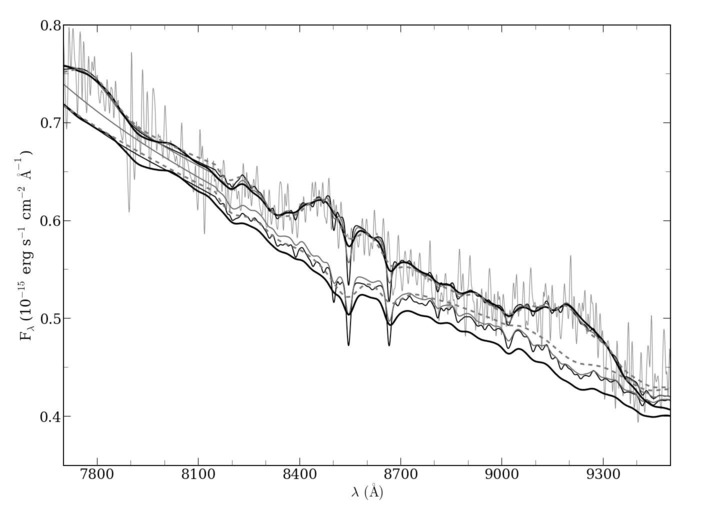}
      \caption{Stellar contribution of HE2202--2557 using different spheroid masses. The thin black and grey lines marked a host galaxy with masses of 10$^{10}$ and 10$^{11}$ M$_\odot$, respectively. The thick black line marked the host galaxy with a mass of 10$^{12}$ M$_\odot$. The grey--dashed line indicate the galaxy with the higher mass 10$^{13}$ M$_\odot$.  {Metallicity 2$Z_\odot$ and age 2.4 Gyr are assumed in all cases}. Abscissae are rest--frame wavelength in \AA, ordinates are rest--frame specific flux in units of 10$^{-15}$ erg s$^{-1}$ cm$^{-2}$ \AA$^{-1}$.}    
     \label{fig:HG}  
   \end{center}
\end{figure}

\begin{figure}[ht]
  \begin{center}
      \includegraphics[width=16cm,keepaspectratio=true]{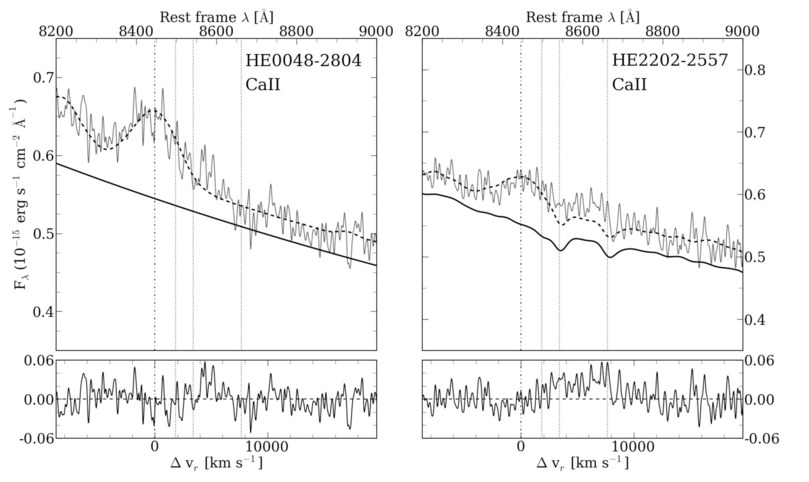}
      \caption{Quasar spectra modeled without \ca2\ triplet contribution but will all other features present. A positive residual at $\sim$ 8542--8662 \AA\ indicates the presence of the \ca2\ triplet. Vertical dash--dot and dot lines mark \oil\ and \ca2\ $\lambda$8498, $\lambda$8542, $\lambda$8662 rest--frame wavelengths. The thick black line traces the assumed continuum (for HE2202--2557 it includes the best fit host galaxy template)}.   
     \label{fig:sCa}  
\end{center}
\end{figure}

\begin{figure}[ht]
  \begin{center}
      \includegraphics[width=12cm,keepaspectratio=true]{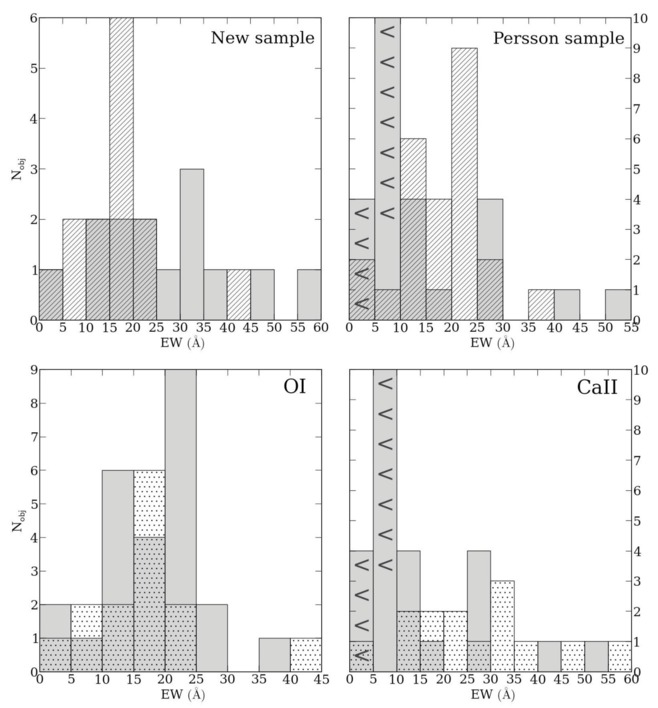}
       \includegraphics[width=12cm,keepaspectratio=true]{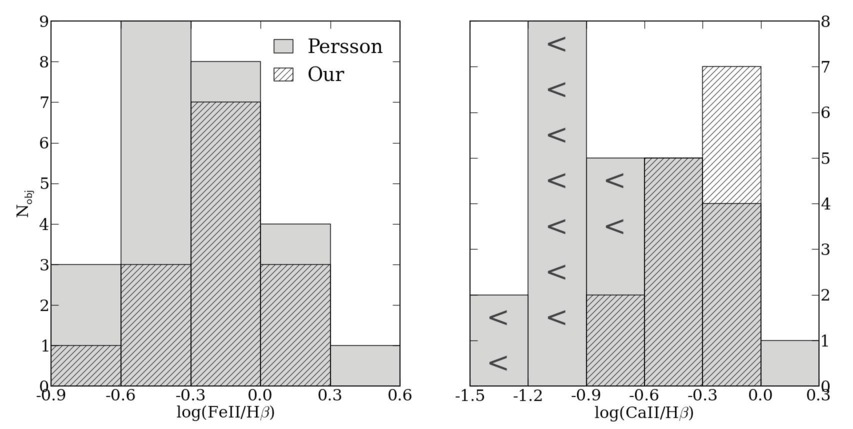}
         \caption{Top panels: Histograms of W(\oi) (hatched) and W(\ca2) (shaded) for our new ISAAC sample and \citet{PER88}. Arrowheads indicate upper limits. Middle panels: distributions of W(\oil) (left) and W(\ca2) (right) in our new sample (dotted) and from \citet{PER88} (shaded).  Bottom panel: distributions of \feiiq/\hb\ (\rfe) and CaT/\hb\ for our ISAAC sample (hatched) and \citet{PER88}'s (shaded). }
     \label{fig:histo}
  \end{center}
\end{figure}

\begin{figure}[ht]
  \begin{center}
      \includegraphics[width=7.5cm,keepaspectratio=true]{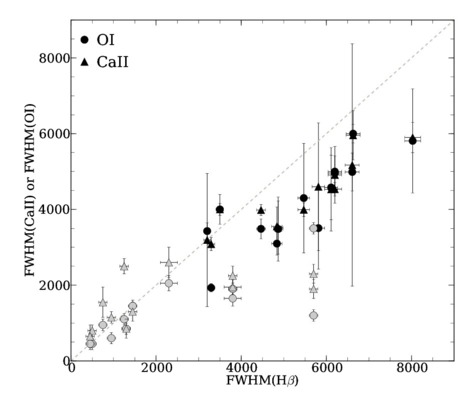}
      \includegraphics[width=7.5cm,keepaspectratio=true]{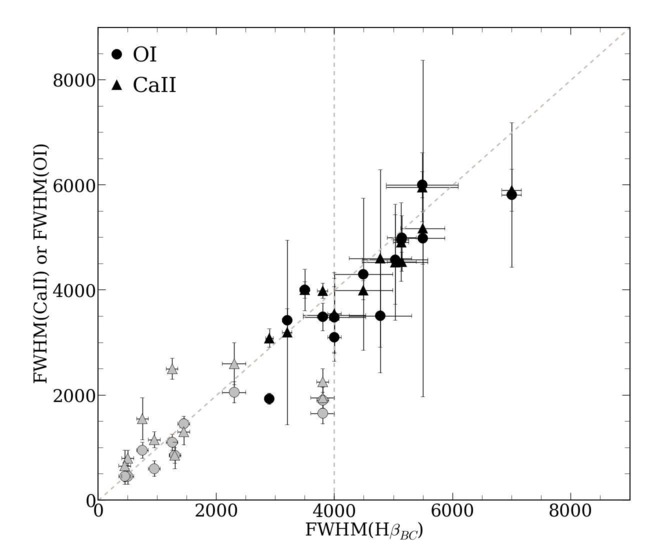}
      \includegraphics[width=7.5cm,keepaspectratio=true]{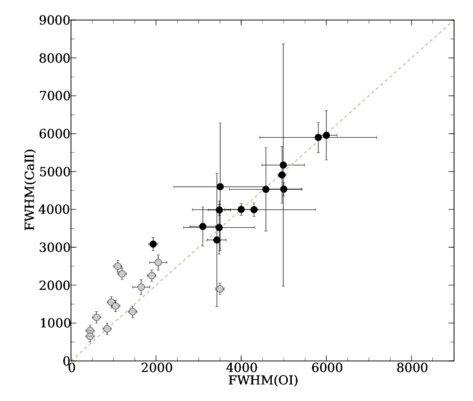}
      \caption{Top: Comparison between the \ca2 triplet, \oil\ and \hb\ FWHMs, in km s$^{-1}$. Abscissa is the H$\beta$ FWHM (full profile, i. e., BC+VBC), ordinate is the \ca2\ Triplet and OI $\lambda$8446 FWHM. Circles represents the \oi\ data, while squares are \ca2\ triplet data. Black symbols are from our sample and grey symbols represent the  Persson sample. The diagonal dashed line is the equality line. The vertical-dashed line at 4000 \kms\ indicates the separation between Pop. A and B. Middle: same as above but for FWHM \hbbc. Bottom: FWHM(CaT) vs FWHM(\oil).}
     \label{fig:fwhm}
  \end{center}
\end{figure}

\begin{figure}[ht]
  \begin{center}
      \includegraphics[width=10cm,keepaspectratio=true]{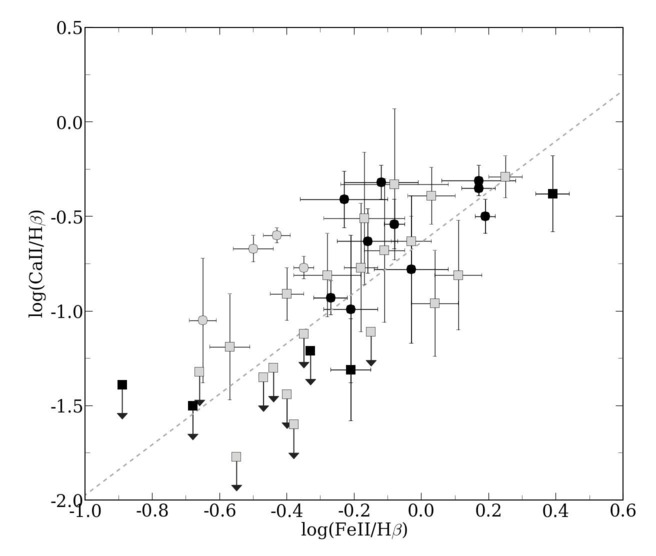}
      \includegraphics[width=10cm,keepaspectratio=true]{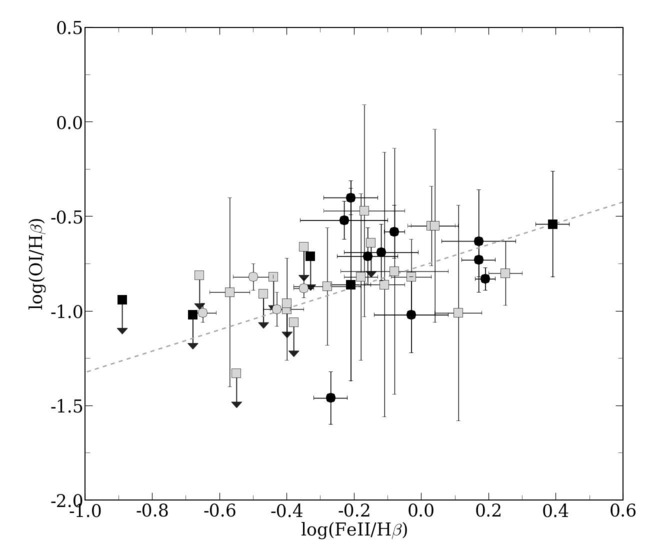}
      \caption{Top: relation between the observed ratios $\log$(\feiiq/\hb) and $\log$(CaT/\hb). Circles indicate our sample, while squares indicate Persson sample. Gray symbols indicate to Pop A. sources, while black symbols indicate to Pop. B sources. The   arrows mark    upper limits. Trendlines are represented   by  grey dashed lines. Bottom: same for $\log$(\oil/\hb$_\mathrm{BC}$) vs $\log$(\feiiq/\hb$_\mathrm{BC}$).}
     \label{fig:ratio_fe+ca}
  \end{center}
\end{figure}

\begin{figure}[ht]
  \begin{center}
      \includegraphics[width=7cm,keepaspectratio=true]{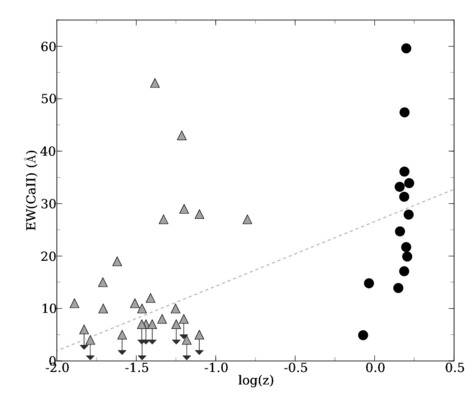}
      \includegraphics[width=7cm,keepaspectratio=true]{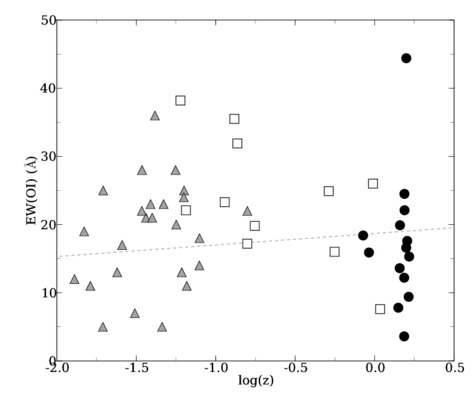}\\
      \includegraphics[width=7cm,keepaspectratio=true]{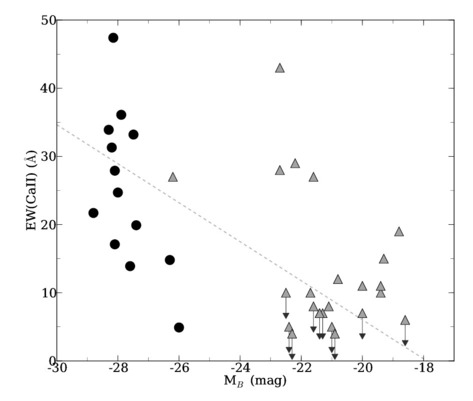}
      \includegraphics[width=7cm,keepaspectratio=true]{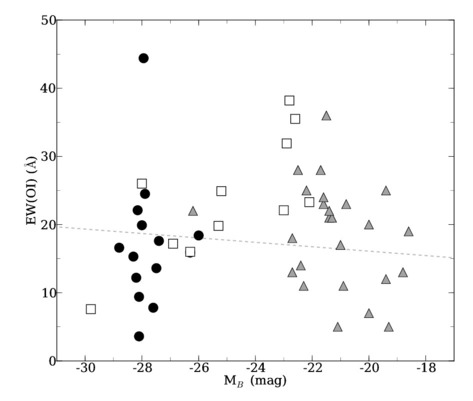}\\
      \caption{Equivalent width of CaT and \oil\ as a function of $z$\ (top) and absolute B magnitude (bottom). Filled circles: sample presented in this paper; triangles: data from \citet{PER88}; open squares: data from \citet{MAT07}. Grey dashed lines are trendlines. }
     \label{fig:lumz}
  \end{center}
\end{figure}

\begin{figure}[ht]
  \begin{center}
      \includegraphics[width=12cm,keepaspectratio=true]{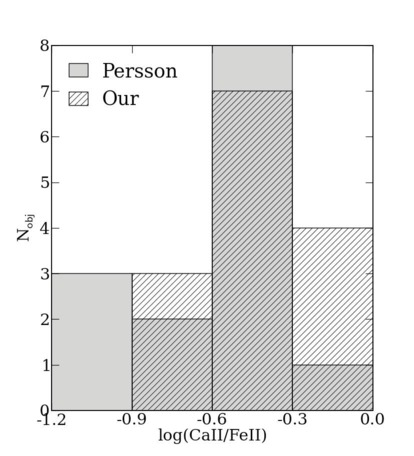}
      \caption{Distribution of the CaT/\feiiq\ for the samples presented in this paper (hatched) and in \citet{PER88}  (shaded). }
     \label{fig:ca2fe2}
  \end{center}
\end{figure}

\begin{figure}[ht]
  \begin{center}
      \includegraphics[width=13cm,keepaspectratio=true]{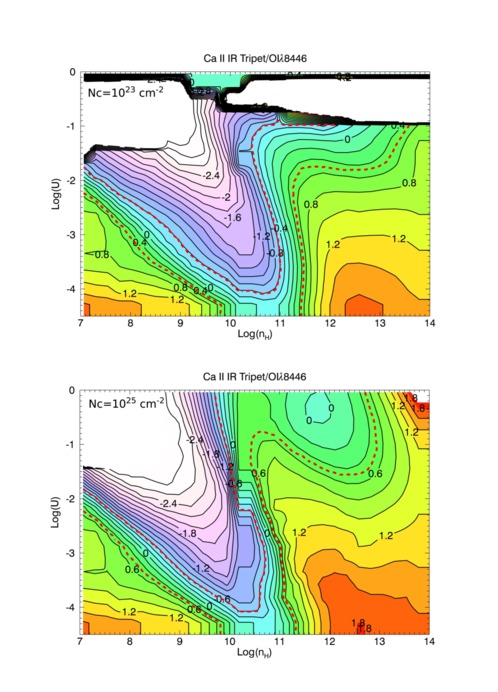}\\
      \caption{Isocontours of {\sc cloudy} simulation results for line ratio $\log$(CaT/\oil) Abscissa is Hydrogen density in cm$^{-3}$, ordinate is the ionization parameter, both in logarithmic scale. Top: models with $N_\mathrm{c}$ of 10$^{23}$ cm$^{-2}$. Bottom: models with $N_\mathrm{c}$ of 10$^{25}$ cm$^{-2}$. The dashed red line marks the maximum and the minimum value from the observations. All our measurements are inside this region. }
     \label{fig:sim_ca+oi}
  \end{center}
\end{figure}

\begin{figure}[ht]
  \begin{center}
      \includegraphics[width=14cm,keepaspectratio=true]{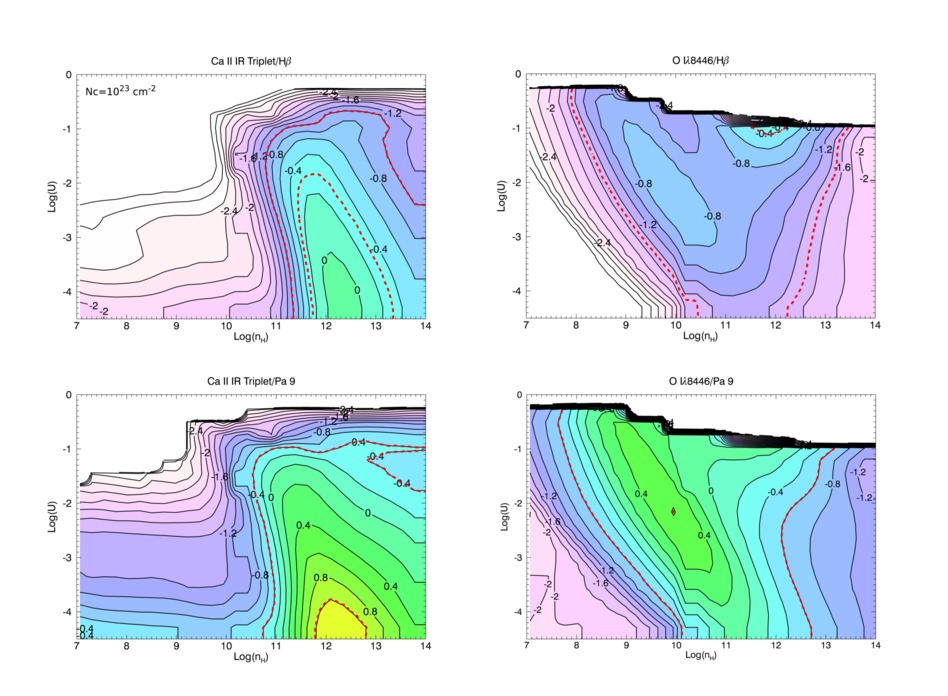}\\
      \includegraphics[width=14cm,keepaspectratio=true]{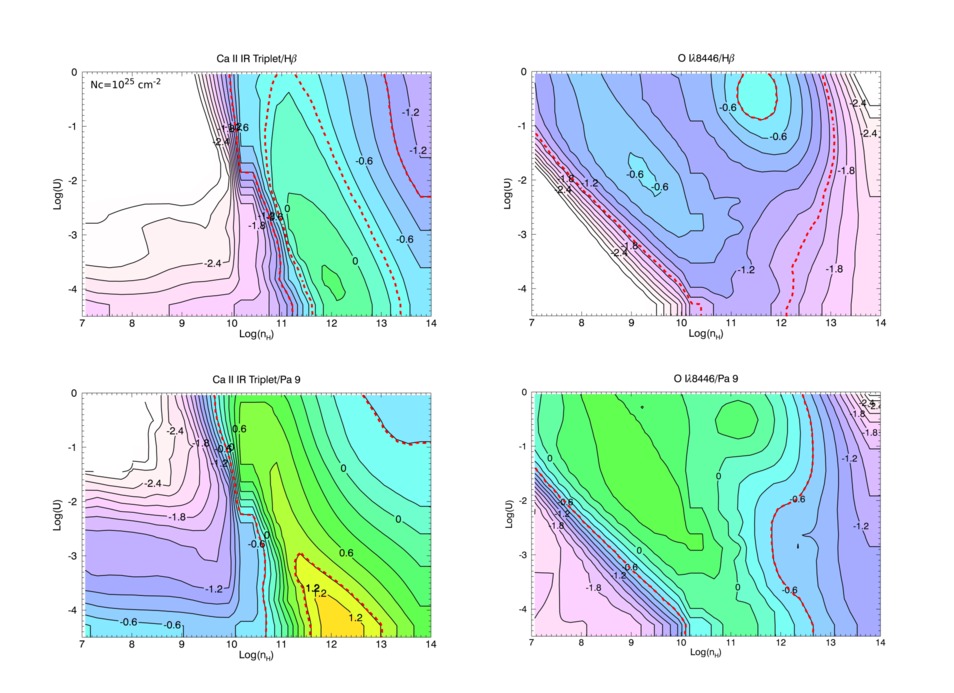}
      \caption{Isocontours of   {\sc cloudy} simulation results for   line ratios  $\log$(CaT/\hb), $\log$(\oil/\hb), $\log$(CaT/Pa9) and $\log$(\oil/Pa9). Abscissa is Hydrogen density in cm$^{-3}$, ordinate is the ionization parameter, both in logarithmic scale. Top: model with  column density \nc\ = 10$^{23}$ cm$^{-2}$. Bottom: model with  \nc\ = 10$^{25}$ cm$^{-2}$.} 
     \label{fig:sim}
  \end{center}
\end{figure}

\begin{figure}[ht]
  \begin{center}
      \includegraphics[width=15cm,keepaspectratio=true]{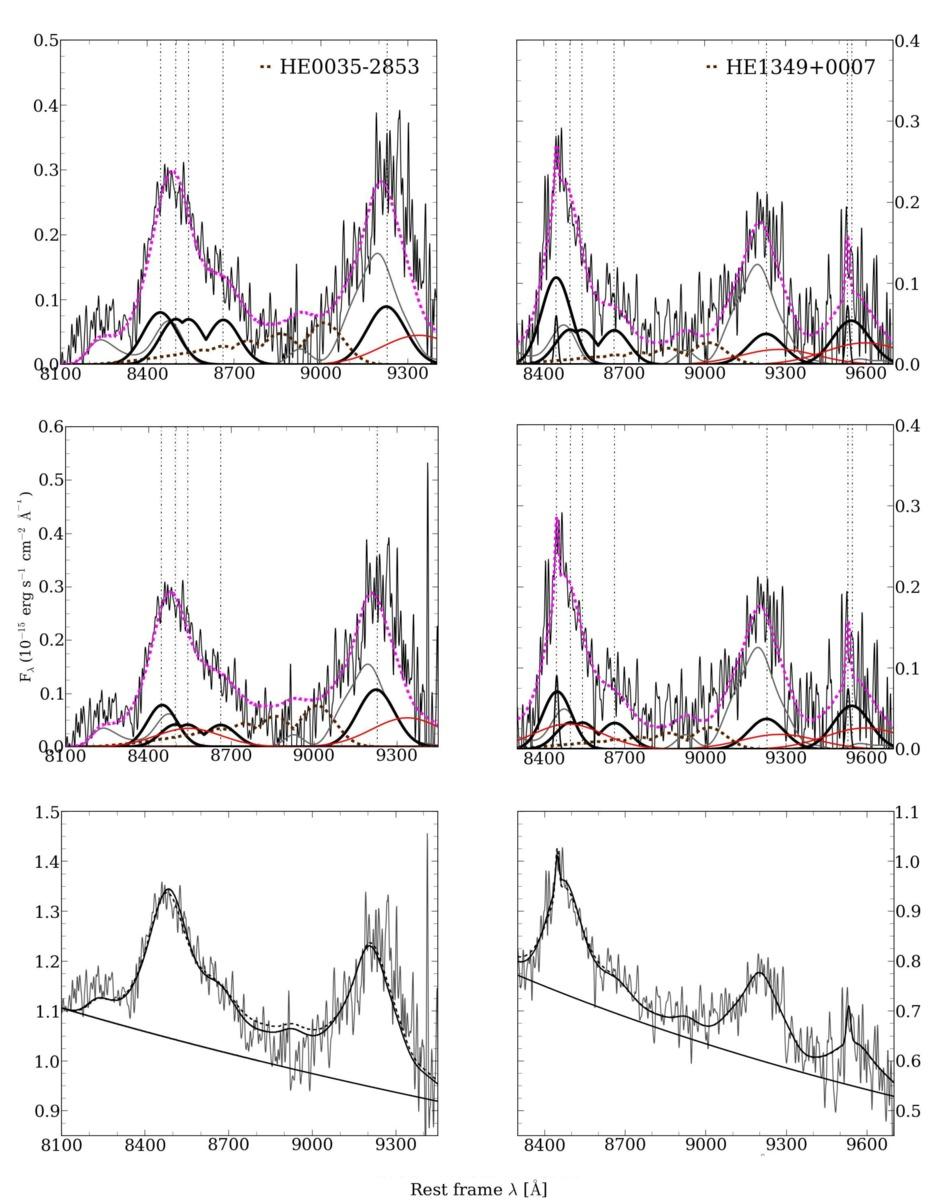}
      \caption{Fits with and without a \oil\ VBC  \oil. The upper panels show the model with only BC. The middle panels present  model fits  with \oil\ VBC. Meaning of line coding is the same of Figure \ref{fig:fits}. In the bottom panels the continuous line mark the model with \oil\ BC only, while the dashed line marks the model with BC+VBC. We don't observe a significant difference.}
     \label{fig:bc+vbc}
  \end{center}
\end{figure}

\clearpage\newpage\eject

\appendix
\section{Notes on individual sources}
\label{appex}

{\setlength{\parindent}{0.5cm} \indent}\textit{HE0005--2355} -- CaT and \oil\ are completely blended. This is the source where Pa7 is detected. Emission on the red side of the \ca2\ triplet cannot be attributed to high  order Paschen lines because the decrease in intensity should be exponential and this is not observed. \s3\ $\lambda$9531 is clearly seen.

{\setlength{\parindent}{0.5cm} \indent}\textit{HE0035--2853} -- One of the highest S/N spectra. The similarity between \hb\ and the \oil\ + CaT blend is remarkable. We see significant \fe2\ emission which is also strong in the optical region. There is a clear asymmetry in the Pa9 + \fe2\ blend at $\sim$9000 \AA\ due to \fe2. We can also see \fe2\ on the blue side of \oil. It is underestimated in both theoretical and semi--empirical \fe2\ templates.

{\setlength{\parindent}{0.5cm} \indent}\textit{HE0043--2300} -- The contribution of NIR \fe2\ emission is weak (optical \fe2\ is also weak). The whole emission feature at $\sim$ 9200 \AA\ is therefore likely to be Pa9. Although the spectrum is rather noisy we can detect high order Paschen lines (specially Pa10 and Pa11). Some excess emission on the blue side of \oil\ could not be reproduced.

{\setlength{\parindent}{0.5cm} \indent}\textit{HE0048--2804} -- This quasar shows the weakest CaT emission, if any. The blend fit is dominated by  \oil. Emission on the blend red side is well fit by  high order Paschen lines. Although \fe2\ is weak in the optical spectrum, it is reasonably strong in the NIR. \fe2\ contributes $\sim$ 70$\%$ of the observed flux in the bump at $\sim$ 9200 \AA.  Another interesting feature involves the prominent rise in intensity at $\sim$ 8200 \AA\ which may be the beginning of the Paschen continuum. An I band magnitude value is reported in the USNO-B Catalog \citep{MON03}, originating from photographic data taken $\sim$ 20 years ago. If we normalize the \hb\ spectrum to the USNO-B I magnitude, we get a significant continuum flux increase of 25$\%$, which does not produce a qualitative change in our sample statistics. Due to the large period of time between photographic observations and our data, we prefer not to re--normalize the \hb\ spectrum.

{\setlength{\parindent}{0.5cm} \indent}\textit{HE0058--3231} --  \oil\ is clearly detected but the CaT detection is uncertain. The higher order Paschen line emission could not be reproduced, especially at $\sim$ 9000 \AA. On the blue side of the spectrum the presence of emission that decreases towards shorter wavelengths is evident and may hint at the presence of  Paschen continuum.

{\setlength{\parindent}{0.5cm} \indent}\textit{HE0203--4627} -- \oil\ shows a blueshift $\sim$ -- 450 \kms\ similar to the one observed in the Hydrogen lines. \oil\ and CaT are completely blended, showing a smooth profile but with a clear red asymmetry revealing the presence of CaT emission. The low intensity of Pa9 indicates that the contribution from  higher order Paschen lines is  small. For instance, the contribution from Pa13 to the \ca2\ $\lambda$8662 profile is $<$10$\%$. For this reason the emission at $\sim$ 8900 \AA\ cannot be ascribed to  Paschen lines. The \fe2\ template underestimates the intensity of this feature. \fe2\ emission is nonetheless favored because  \fe2\ is three times stronger than  Pa9 at 9200 \AA.

{\setlength{\parindent}{0.5cm} \indent}\textit{HE0248--3628} --  The high S/N  of the spectrum allows us to detect \ca2\ $\lambda$8662, but does not allow us to distinguish \ca2\ $\lambda$8498 from $\lambda$8542. \hb\ and NIR profiles show a strong similarity. \fe2\ emission is twice as strong as Pa9 and Pa10 and can partially account for the observed emission at 8900 \AA. There is also a small feature at $\sim$ 9400 that has been identified with \fe2\ \citep{GAR12} but that is not modelled by theoretical template. Narrow \s3\ $\lambda$9531 width is comparable to the narrow optical lines. We could not reproduce emission observed on the blue side of \oil.

{\setlength{\parindent}{0.5cm} \indent}\textit{HE1349+0007} -- We found evidence for a narrow component in \oil. For many years \oil\ was thought to be produced only in the broad line region \citep{GRA80}. More recently, evidence was found that part of this line might arise in the narrow line region \citep{LAN08}. \fe2\ ($>$ 50$\%$ of flux in the Pa9 region) dominates the fit with higher order Paschen lines making a minor contribution.

{\setlength{\parindent}{0.5cm} \indent}\textit{HE1409+0101} -- The optical spectrum shows prominent \fe2\ emission while the NIR spectrum may indicate less prominent \fe2\ emission.  Some Paschen continuum emission  may be present.

{\setlength{\parindent}{0.5cm} \indent}\textit{HE2147--3212} -- Low S/N spectrum. The only feature clearly observed  is the \oil\ + CaT blend. \ca2\ $\lambda$8662 can be readily seen. We cannot fit the Pa9 line with good accuracy. 

{\setlength{\parindent}{0.5cm} \indent}\textit{HE2202--2557} -- The only source spectrum requiring subtraction of a host galaxy contribution. Host galaxy absorptions are clearly visible (especially the \ca2\ triplet). The stellar spectrum contributes 50$\%$\ of the quasar continuum luminosity.  After continuum subtraction  the \oil\ + CaT emission blend  appears similar to the  \hb\ profile. \oi\ $\lambda$7775 and \oil\  show the same width,  with \oi\ $\lambda$7775 stronger than  \oil. Emission between these lines can be well fit by the Paschen continuum.  Higher order Paschen lines are present but not unusually strong. \oil\ and CaT lines are weak, with equivalent width of a few angstroms.  The reported flux and equivalent width  errors   in Table \ref{tab:oi+ca} are estimated in a consistent way with the other sources. In the case of HE2202--2557 a large uncertainty is associated with continuum placement. Repeating the {\tt specfit} analysis indicates that $\pm$ $50$\%\ may be a more realistic estimate of 
the flux and equivalent width uncertainty.

{\setlength{\parindent}{0.5cm} \indent}\textit{HE2340--4443} -- Modest S/N at 9200 \AA\ prevents a reliable fit of the Pa9 line. High order Paschen lines and \fe2\ measurements have a large uncertainty. \oil\ and CaT are completely blended and \oil\ dominates the fit. An I band magnitude value is reported in the USNO--B Catalog \citep{MON03}.  The I magnitude would change the flux scale by 40$\%$. As for HE0048--2804 we prefer to keep the flux scale obtained from our calibration.

{\setlength{\parindent}{0.5cm} \indent}\textit{HE2349--3800} -- Low S/N makes it difficult to measure  Pa9 and \fe2. The \oil\ + CaT blend is the only distinct feature. We can isolate  the contribution of the \ca2\ $\lambda$8662 line. There is a bump on the blue side of \oi\ that we cannot reproduce.

{\setlength{\parindent}{0.5cm} \indent}\textit{HE2352--4010} --  In this case we distinguish the contribution of \ca2\ $\lambda$8662 as well as the other two lines of the triplet. We  observe a bump on the blue side of \oil\ that cannot be modeled.

\clearpage

\section{Error estimates}
\label{error}

Synthetic Gaussian line profiles were built to estimate statistical errors involved in the line parameter measures. Since we are not dealing with  isolated emission lines, we have to consider the effect of blending on error estimates. For the optical spectra, we considered the broad component (BC) and the very broad component (VBC) of \hb. For the NIR spectra, we took into account that \oil\ and CaT are blended together. Model blends were constructed using  \oil\ and the three individual components of  CaT, to which we added  Gaussian  noise. Models profiles were built considering for the typical FWHM and equivalent width measured in our spectra. For \hb\ BC we assumed FWHM = 4800 \kms, W = 24, 40, 56, 70 and S/N = 20, 30, 40, 50, 60. For  \hb\ VBC, we considered  FWHM = 10500 \kms, W = 5, 10, 20, 30, 50 and the same S/N values. For \oil, we considered  FWHM = 4500 \kms, W = 10, 15, 20, 40 and S/N = 10, 20, 30, 40, 55, and for CaT FWHM = 4300 \kms, W = 10, 20, 30, 50, 60 (sum of the three lines) and the same 
S/N values. In this way we were able to build   blended spectra with profiles similar to the ones seen in the real data of each quasars.

Let ${p_0}$ be a blended profile, with a given FWHM and W, to which noise has been added. Around it a set of profiles $p_\mathrm{0,ij}$ was built: 1) changing FWHM and keeping flux and the peak of the line fixed, and then 2) changing flux (and W) but keeping FWHM fixed. $\chi^2_\mathrm{\nu,ij}$\ values were computed for fixed noise level as a function of FWHM and flux. The original profile $p_0$\ is associated with the minimum $\chi_\nu^2$, $\chi_\mathrm{\nu, min}$. Errors are set through the ratio $F_\mathrm{ ij}$ = $\chi^2_\mathrm{\nu,ij}/\chi^2_\mathrm{\nu,min}$, and precisely by values of  $F_\mathrm{ ij}$ \ that correspond to the probability $\approx$ 0.32 (1--$\sigma$ confidence level) of reaching  a $\chi^2_\mathrm{\nu,ij}$\ value because of statistical errors (i.e., associated with noise). Values of $F_\mathrm{ ij}$\ are computed for the degrees of freedom set by the number pixels in wavelength covered by the lines, minus the number of parameters appearing in the line model. The  $F_\mathrm{ ij} $\ 
values provide confidence intervals as a function of FWHM, W, and S/N that define our uncertainty range. The FWHM error   was taken from   models where the FWHM varies and the the peak of the line is fixed. Flux can be written as the product FWHM times peak intensity.  Error propagation on FWHM and peak intensity yields the  total error on flux. The left panel of Figure \ref{fig:errors}  shows an example of blended synthetic profiles with FWHM(\oi) = 4500 \kms, FWHM(\ca2) = 4300 \kms, W(\oi) = 20 \AA, W(CaT) = 20 and S/N = 20. The right panel shows the uncertainty range on FWHM and flux estimated from the appropriate $F_\mathrm{ij}$\  value. The parameter $F = \chi_\nu^2 / \chi_\mathrm{\nu,min}^2$ is shown in Figure \ref{fig:errors} as a function of FWHM and intensity of CaT and \oil.   $F \approx 1.03$\ provides 1--$\sigma$ confidence ranges on the parameters. This is basically the classical $\chi^2_\nu$ technique for error estimates described by \citet{BEV03} and by \citet{PRESS}. 

We tested the effect of \ca2+\oi\ blending also with Monte Carlo simulations that allowed us to simultaneously fit flux and FWHM. The cases studied for the $\chi^2_\nu$ analysis were all repeated for different values of S/N ratio. The Monte Carlo simulations involved the repetition of the blend fits for a large number of trials  ($\approx$ 1500  for each case) with different realizations of Gaussian noise patterns. The uncertainty on line parameters was then retrieved by the distribution of the measurements obtained changing the random noise patterns. We found that the errors are correlated with the ones estimated through the $\chi^2_\nu$ analysis, but are  significantly smaller, between a factor 1.3 -- 1.5 and a factor $\gtrsim$\ 2 for the flux of CaT. The lower errors can be understood by considering the nature of the Monte Carlo Method and by the fact that we are dealing with spectra whose S/N is relatively high, $\gtrsim 20$\ save in one case. In these conditions, the fitting  routine converges to values 
close to the ``true'' ones i.e., the starting values in absence of noise, yielding a rather peaked parameter distribution for different noise realizations. 

In addition to Gaussian noise, another relevant source of error is continuum placement. To estimate its effect on  flux, we   considered a continuum intensity displacement of $\pm 1 \sigma$ with respect to the best fit. The W uncertainty was then estimated by propagating the errors for continuum and flux following the definition of W. 

{The uncertainty due to the inclusion of the \fe2\ template has been estimated through Monte Carlo simulations considering all components assumed to be present in the whole spectral range  8000 -- 9900 \AA.  The flux and width assumed for each line are the ones reported in  Table 2 and 3 i.e., the ones measured on the observed spectra. We run two cases for each object, in which the fitting routine  (1) considers only one template as listed  in   Table 4, and (2)   is left free to choose between the two different templates (theoretical or semiempirical).  The ratio between the errors found for  case 2 and case 1 gives us the uncertainty added by \fe2\  over the uncertainty associated with other components. We  multiplied $\chi^{2}_\nu$\ uncertainties  by this ratio (always $\gtrsim 1$)  to include the effect associated with the \fe2\ template.  The uncertainty increase  ranges from a median of $\approx$ 10 \%\ to a maximum $\approx$ 50\%.  The effect depends on  S/N and W and, in the Monte Carlo 
simulation context, should be ascribed to the inability of the fitting routine to identify which template is the best. The  uncertainties computed with and without the effect of the \fe2\ template  are reported in Tables \ref{tab:hb+pa9} and \ref{tab:oi+ca}. Figures \ref{fig:fwhm} and \ref{fig:ratio_fe+ca} error bars include \fe2\ template uncertainty. }

\setcounter{figure}{0}
\begin{figure}[ht]
  \begin{center}
      \includegraphics[width=7cm,keepaspectratio=true]{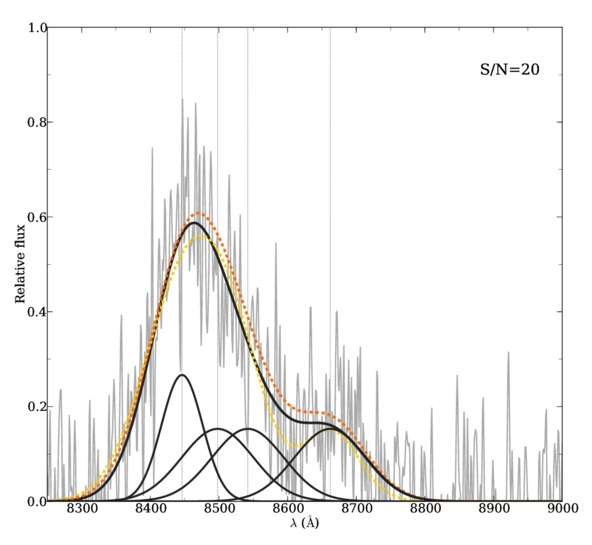}
       \includegraphics[width=7cm,keepaspectratio=true]{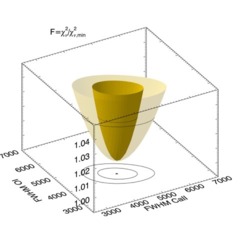}
      \includegraphics[width=7cm,keepaspectratio=true]{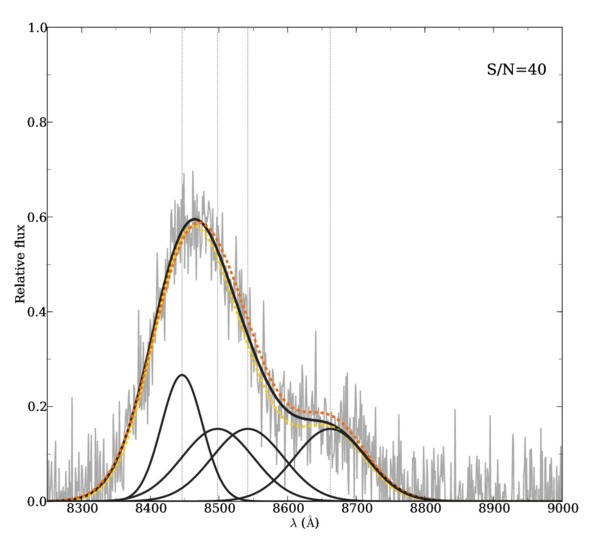}
      \includegraphics[width=7cm,keepaspectratio=true]{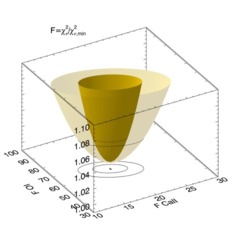}     
      \caption{In the left panels we show the synthetic profiles for the blend \oil\ + CaT with FWHM(\oil) = 4500 \kms, FWHM(\ca2) = 4300 \kms, W(\oil) = 20 \AA, W(\ca2) = 20 and S/N = 20, 40 (top and bottom, respectively). The dark line marks the model with the minimuum  $\chi^2_\nu$, while the dashed line in colour indicates the model at 1--$\sigma$ for FWHM and Flux (yellow and orange). The vertical dashed line are the rest--frame for \oil\ and \ca2\ triplet. In the right panels the behavior of the $F$  statistics is shown for the 2 different S/N values  (pale: S/N = 20, dark S/N = 40). Contour lines (traced on the 3D paraboloids and on the $xy$ plane)  trace the 1--$\sigma$ confidence ranges on FWHM and flux reached for $F\approx 1.03$.}    
     \label{fig:errors}  
   \end{center}
\end{figure}

\clearpage

\end{document}